\documentclass[aps,prx,amsmath,amssymb,nofootinbib,superscriptaddress,showpacs,floatfix,twocolumn,longbibliography]{revtex4-1} 
\usepackage{latexsym}
\usepackage{graphicx}
\usepackage{times,psfrag,subfigure}
\usepackage{enumerate}
\usepackage{amsmath}
\usepackage{dsfont}
\usepackage{dcolumn}
\usepackage{bm,bbm}
\usepackage{latexsym,amsmath,amssymb,bm,euscript}
\usepackage[normalem]{ulem}
\usepackage{dsfont}
\usepackage{textcomp}
\usepackage{tabularx}
\usepackage{setspace}
\usepackage{ctable}
\usepackage{sidecap}
\usepackage{placeins}
\usepackage{threeparttable}
\usepackage{multirow}
\usepackage[hidelinks]{hyperref}

\let \oldbm \bm
\renewcommand{\vec}[1]{\oldbm{#1}}
\newcommand{\vechat}[1]{\hat{\oldbm{#1}}}

\def\bk{{\vec k}}

\def\bn{{\vechat n}}
\def\bm{{\vec m}}
\def\br{{\vec r}}

\def\Z{\mathbb{Z}}
\def\T{\mathcal{T}}

\def\S{\mathcal{S}}
\def\I{\mathcal{I}}
\def\G{\mathcal{G}}

\def\H{\mathcal{H}}
\def\K{\mathcal{K}}

\newcommand{\beq}{\begin{equation}}
\newcommand{\eeq}{\end{equation}}
\newcommand{\beqarray}{\begin{eqnarray}}
\newcommand{\eeqarray}{\end{eqnarray}}

\newcommand{\bsigma}{\mbox{\boldmath$\sigma$}}

\allowdisplaybreaks

\begin{document}

\title{Symmetry indicators and anomalous surface states of topological crystalline insulators}

\begin{abstract}
The rich variety of crystalline symmetries in solids leads to a plethora of topological crystalline insulators (TCIs) featuring distinct physical properties, which are conventionally understood in terms of bulk invariants specialized to the symmetries at hand. While isolated examples of TCI have been identified and studied, the same variety demands a unified theoretical framework. In this work, we show how the surfaces of TCIs can be analyzed within a general surface  theory with multiple flavors of Dirac fermions, whose mass terms transform in specific ways under crystalline symmetries. We identify global obstructions to achieving a fully gapped surface, which typically lead to gapless domain walls on suitably chosen surface geometries. We perform this analysis for all 32 point groups, and subsequently for all 230 space groups, for spin-orbit-coupled electrons. We recover all previously discussed TCIs in this symmetry class, including those with ``hinge'' surface states. Finally, we make connections to the bulk band topology as diagnosed through symmetry-based indicators. We show that spin-orbit-coupled band insulators with nontrivial symmetry indicators are always accompanied by surface states that must be gapless somewhere on suitably chosen surfaces. We provide an explicit mapping between symmetry indicators, which can be readily calculated, and the characteristic surface states of the resulting TCIs. 
\end{abstract}

\author{Eslam Khalaf}
\affiliation{Max Planck Institute for Solid State Research, Heisenbergstr.\ 1, 70569 Stuttgart, Germany}
\affiliation{Department of Physics, Harvard University, Cambridge MA 02138}

\author{Hoi Chun Po}
\author{Ashvin Vishwanath}
\affiliation{Department of Physics, Harvard University, Cambridge MA 02138}

\author{Haruki Watanabe}
\affiliation{Department of Applied Physics, University of Tokyo, Tokyo 113-8656, Japan}

\maketitle
Topological phases of free fermions protected by internal symmetries (such as time reversal) feature a gapped bulk and symmetry-protected gapless surface states
\cite{Molenkamp}.  
Such bulk-boundary correspondence is an important attribute of such topological phases, and it plays a key role in their classification, which has been achieved in arbitrary spatial dimensions \cite{Kitaev09, Schnyder09, Ryu10}.
Crystals, however, frequently exhibit a far richer set of spatial symmetries, like lattice translations, rotations, and reflections, which can also protect new topological phases of matter \cite{Fu11,Hsieh12, Tanaka12, Robert-JanSlager, Teo, Lu14, Varjas15, Ken2015,  Dong16, Wang16, Ezawa16, Varjas17, Wieder17, Benalcazar17, Song17, Schindler17, Langbehn17,  Benalcazar17Aug, Fang17}.

However, several new subtleties arise when discussing topological crystalline phases. For one, spatial symmetries are often broken at surfaces, and when that happens, the surface could be fully gapped even when the bulk is topological.
Consequentially, the diagnosis of bulk band topology is a more delicate task as, a priori, the existence of anomalous surface states in this setting is only a sufficient, but not necessary, property of a topological bulk.
Furthermore, the presence of crystalline symmetries leads to a plethora of topological distinctions even when the band insulators admit atomic descriptions and are therefore individually trivial
\cite{Po17,Bradlyn17}.
For example, consider trivial (atomic) insulators, where electrons are well-localized in real space and do not support any nontrivial surface states. Suppose two atomic insulators are built from orbitals of different chemical characters, which will typically lead to different symmetry representations in the Brillouin zone (BZ). These representations cannot be modified without a bulk gap closing, and therefore the two atomic insulators are topologically distinct even though they are both derived from an atomic limit.

Since the mathematical classification of topological band structures, using K-theory, is formulated in terms of the mutual topological distinction between insulators \cite{Kitaev09, MooreFreed, Combinatorics, Ken2017}, the mentioned distinction between trivial phases is automatically incorporated. In fact, they appear on the same footing as conventional topological indices like Chern numbers. While for a physical classification one would want to discern between these different flavors of topological distinctions, they are typically intricately related.  On the one hand, the pattern of symmetry representations is oftentimes  intertwined with other topological invariants, as  epitomized by the Fu-Kane parity criterion \cite{Fu07}, which diagnoses 2D and 3D topological insulators (TIs) through their inversion eigenvalues; on the other hand, the topological distinction between atomic insulators may reside fully in their wavefunction properties and, hence, is not reflected in their symmetry eigenvalues \cite{Zak2001, Bradlyn17}.

In this work, we sidestep these issues by focusing on topological crystalline phases displaying anomalous gapless surface states, which, through bulk-boundary correspondence, are obviously topological. 
Unlike the  strong TI, these phases require crystalline symmetries for their protection.
Typically, their physical signatures are exposed on special surfaces where the protecting symmetries are preserved.
Such is the case for the 3D phases like weak TI \cite{Molenkamp}, the mirror Chern insulator \cite{Hsieh12, Tanaka12}, and the nonsymmorphic insulators featuring so-called hourglass \cite{Wang16, Ezawa16} or wallpaper \cite{Wieder17} fermions, which are respectively protected by lattice translation, reflection, and glide symmetries.
While these conventional phases feature 2D gapless surface states on appropriate surfaces, we will also allow for more delicate, 1D ``hinge'' surface states, whose existence is globally guaranteed on suitably chosen sample geometries although any given crystal facet can be fully gapped \cite{Fritz, Zhang, Teo,  Benalcazar17,  Hashimoto2017, Song17, Schindler17,  Benalcazar17Aug, Fang17, Langbehn17}. 
We will primarily focus on 3D time-reversal symmetric band structures with a bulk gap and significant spin-orbit coupling (class AII). However, our approach can be readily extended beyond this specific setting. In particular, in symmetry classes with particle-hole symmetry, the notion of hinge states can be generalized to protected gapless points on the surface \cite{Teo}.

We will base our analysis on two general techniques, which are respectively employed to  systematically study gapless surface states and identify bulk band structures associated with them. First, we will describe a surface Dirac theory approach \cite{Molenkamp, Schnyder09, Ryu10, Witten16, Fang17}. Specifically, we will consider a surface theory with multiple Dirac cones, which may transform differently under the spatial symmetries \cite{Fang17}. Such a theory can be viewed as the surface of a ``stack of strong TIs,'' which arise, for example, when there are multiple band inversions (assuming inversion symmetry) in the BZ. When an even number of Dirac cones are present at the same surface momentum, the Dirac cones can generically be locally gapped out and, naively, one expects that a trivial surface results. However, spatial symmetries can cast global constraints which obstruct the full gapping out of the Dirac cones everywhere on the surface \cite{Fang17}. When that happens, the surface remains stably gapless along certain domain walls, and hence one can infer that the bulk is topological. We will provide a comprehensive analysis of such obstructions due to spatial symmetries, and catalogue all the possible patterns of such anomalous surface states in a space group.

Next, we will turn to bulk diagnostics that informs the presence of such surface states. In particular, we will focus on those which, like the Fu-Kane criterion \cite{Fu07}, expose band topology using only symmetry eigenvalues \cite{Fu07, Ari, Hughes11, Fang12}.
Such symmetry-based indicators are of great practical values, since they can be readily evaluated without computing any wavefunction overlaps and integrals. In fact, a general theory of such indicators has been developed in Ref.\ \onlinecite{Po17}, and the indicator groups for all 230 space groups (SGs) have been computed.

Here, we provide a precise physical interpretation for all the phases captured by these indicators in class AII. In particular, we discover that the majority of them can be understood in terms of familiar indices, like the strong and weak TI $\mathbb Z_2$ indices and mirror Chern numbers. However, there exists symmetry-diagnosed band topology which persists even when all the conventional indices have been silenced. The prototypical example discussed in Ref.\ \onlinecite{Po17} is a ``doubled strong TI'' in the presence of inversion symmetry, which, as pointed out later in Ref.\ \onlinecite{Fang17}, actually features hinge-like surface states. Motivated by these developments, we relate the symmetry indicators to our surface Dirac theory and prove that, for spin-orbit coupled systems with time-reversal symmetry (TRS), any system with a nontrivial symmetry indicator is a band insulator with anomalous surface states on a suitably chosen boundary.  This is achieved by establishing a bulk-boundary correspondence between the indicators and the surface Dirac theory.
Parenthetically, we note that a nontrivial symmetry indicator constitutes a sufficient, but not necessary, condition for the presence of band topology, and we will discuss examples of topological phases which have gapless surface states predicted by our Dirac approach, although the symmetry indicator is trivial.

Although the techniques we adopted are tailored for weakly correlated materials admitting a band-theoretic description, many facets of our analysis have immediate bearing on the more general study of interacting topological crystalline phases \cite{Hermele17,Thorngren16, XiongMin, Huang17,XiongGlide}. A common theme is to embrace a real-space perspective, where ``topological crystals'' are built by repeating motifs that are by themselves topological phases of some lower dimension, and are potentially protected by internal symmetries or spatial symmetries that leave certain regions invariant (say on a mirror plane)  \cite{Hermele17, Huang17}. In particular, such picture allows one to readily deduce the interaction stability of the phases we describe \cite{Isobe15}, and we will briefly discuss how our Dirac surface theory analysis can be reconciled with such general frameworks.

\section{Symmetry indicators and bulk topology}
\label{Sec:Indicators}
In this section, we provide a precise physical meaning for the phases captured by the symmetry-based indicator groups obtained in Ref.~\onlinecite{Po17} and give explicit expressions for each of these indicators, assuming TRS and focusing on the physically most interesting case of strong spin-orbit coupling. 

We start by considering a band insulator of spinful electrons symmetric under an SG and TRS (class AII). In this setting, we can define topological indices corresponding to weak and strong TI phases protected by TRS.  For centrosymmetric SGs, the Fu-Kane formula~\cite{Fu07} allows us to compute these indices using only the parities, i.e., inversion eigenvalues, of the bands.  Namely, they can be determined just by multiplying the parities of the occupied bands without performing any sort of integrals.  This is one particular instance of the relation between the symmetry representations in momentum space and band topology.  

Recently, this relation was extended to all space groups (i.e., not restricted to inversion alone) and also to a wider class of band topology~\cite{Po17}.  By comparing the symmetry representations of the bands against that of trivial (atomic) insulators, the symmetry-based indicator group $X_{\rm BS}$ was computed for all 230 SGs~\cite{Po17} (and even for all 1651 magnetic SGs~\cite{WatanabePo17}). It was found to take the form
\begin{equation}
\label{XBS}
X_{{\rm BS}}=\mathbb{Z}_{n_1}\times\mathbb{Z}_{n_2}\times\cdots\times\mathbb{Z}_{n_N},
\end{equation}
where $N$ and $\{n_1, n_2,\cdots,n_N\}$ depend on the assumed SG.  
While $X_{{\rm BS}}$ has been exhaustively computed for all SGs in Ref.\ \onlinecite{Po17}, the concrete physical interpretation for some of the nontrivial classes was left unclear. 
We will attack this problem in the following by first recasting these topological indices into more explicit forms, akin to those in Refs.\ \onlinecite{Fu07, Fang12}, and then clarifying their physical meanings.

\subsection{Review of familiar indices}
Here, we first review some well-known symmetry-based indicators and discuss their relations with the general results in Ref.\ \onlinecite{Po17}.  Let us start with the inversion symmetry.  As mentioned above, for an SG containing the inversion symmetry $\I$, the inversion eigenvalues are related to the strong TI index $\nu_0\in\mathbb{Z}_2$ and the three weak TI indices $\nu_i\in\mathbb{Z}_2$ ($i=1,2,3$) \cite{Fu07}.  Each of the three weak TI indices gives rise to a $\mathbb{Z}_2$ factor in $X_{{\rm BS}}$ when no additional constraints are imposed.  Curiously, however, the factor in $X_{{\rm BS}}$ corresponding to the strong TI index is $\mathbb{Z}_4$, not $\mathbb{Z}_2$ \cite{Po17}.  This $\mathbb{Z}_4$ indicator will be discussed in Sec.~\ref{sec:k1}.

Next, let us discuss the mirror Chern numbers, which can be defined for every mirror symmetric plane in the momentum space \cite{Hsieh12}.  On any such plane, the single-particle Hamiltonian $\H$ can be block-diagonalized into $\H_\pm$, defined respectively in the sectors corresponding to the mirror eigenvalues $M=\pm i$.  Since $\H_\pm$ is not necessarily time-reversal symmetric, $\H_{\pm}$ can possess a Chern number $C_{\pm}\in \mathbb{Z}$, which satisfies $C_+=-C_-$ ($\equiv C_M$) due to the TRS of the total Hamiltonian $\H$.  To be more concrete, let us assume the mirror symmetry is about the $xy$ plane. 
For  primitive lattice systems, both $k_z=0$ and $k_z=\pi$ planes are mirror symmetric and can individually support $C_M^{(k_z=0)}$ and $C_M^{(k_z=\pi)}$. For body-centered systems and face-centered systems, however, $k_z=0$ is the only mirror symmetric plane and there is only one mirror Chern number $C_M^{(k_z=0)}$.

When the SG is further endowed with an $n$-fold rotation symmetry $C_{n}$ ($n=2,3,4,6$) or screw symmetry ($C_n$ rotation followed by a fractional translation along the rotation axis) whose axis is orthogonal to the mirror plane, one can diagnose the mirror Chern number $C_M$ modulo $n$ by multiplying the rotation eigenvalues~\cite{Fang12}.  
(For simplicity, in the following we will often leave the direction of a rotation or screw symmetry implicit, and, whenever necessary, label that as the $z$ axis corresponding to the third momentum coordinate.)
Naively, each $C_M$ mod $n$ produces a $\mathbb{Z}_n$ factor in $X_{{\rm BS}}$.  However, we sometimes find a ``doubled" $\mathbb{Z}_{2n}$ factor in $X_{{\rm BS}}$. The mechanism behind this enhancement is clarified in Sec.~\ref{sec:delta} and \ref{sec:k1mCh}.

Following the examination of these familiar indices, we find that $X_{{\rm BS}}$ in Eq.~\eqref{XBS} can be always factorized into ``weak factors" and a ``strong factor":
\begin{gather}
X_{{\rm BS}}=X_{\rm BS}^{(w)} \times X_{\rm BS}^{(s)}, \nonumber\\ X_{\rm BS}^{(w)}=\mathbb{Z}_{n_1}\times\cdots\times\mathbb{Z}_{n_{N-1}}, \qquad  X_{\rm BS}^{(s)}=\mathbb{Z}_{n_N}.
\label{XBS2}
\end{gather}
Every factor in $X_{\rm BS}^{(w)}$ can be completely characterized either by the weak TI indices $\nu_i$ ($i=1,2,3$) or the weak mirror Chern number $C_M^{(k_z=0)}=C_M^{(k_z=\pi)}$, both of which can be understood as arising from stacking 2D TIs.  On the other hand, for most SGs, the strong factor $X_{\rm BS}^{(s)}$ cannot be fully understood by these familiar indicators.  We will thus focus on the strong factor in the reminder of this section.

\subsection{$\mathbb{Z}_4$ index for inversion symmetry}
\label{sec:k1}
Here we show that the strong TI index $\nu_0$ can actually be promoted to a $\mathbb{Z}_4$ index in the presence of inversion symmetry. The refined index can capture topological crystalline insulators with anomalous 1D edge state, as we discuss in detail in Sec.~\ref{Sec:Surface}.

\begin{figure*}
\center
\includegraphics[width=1.3\columnwidth]{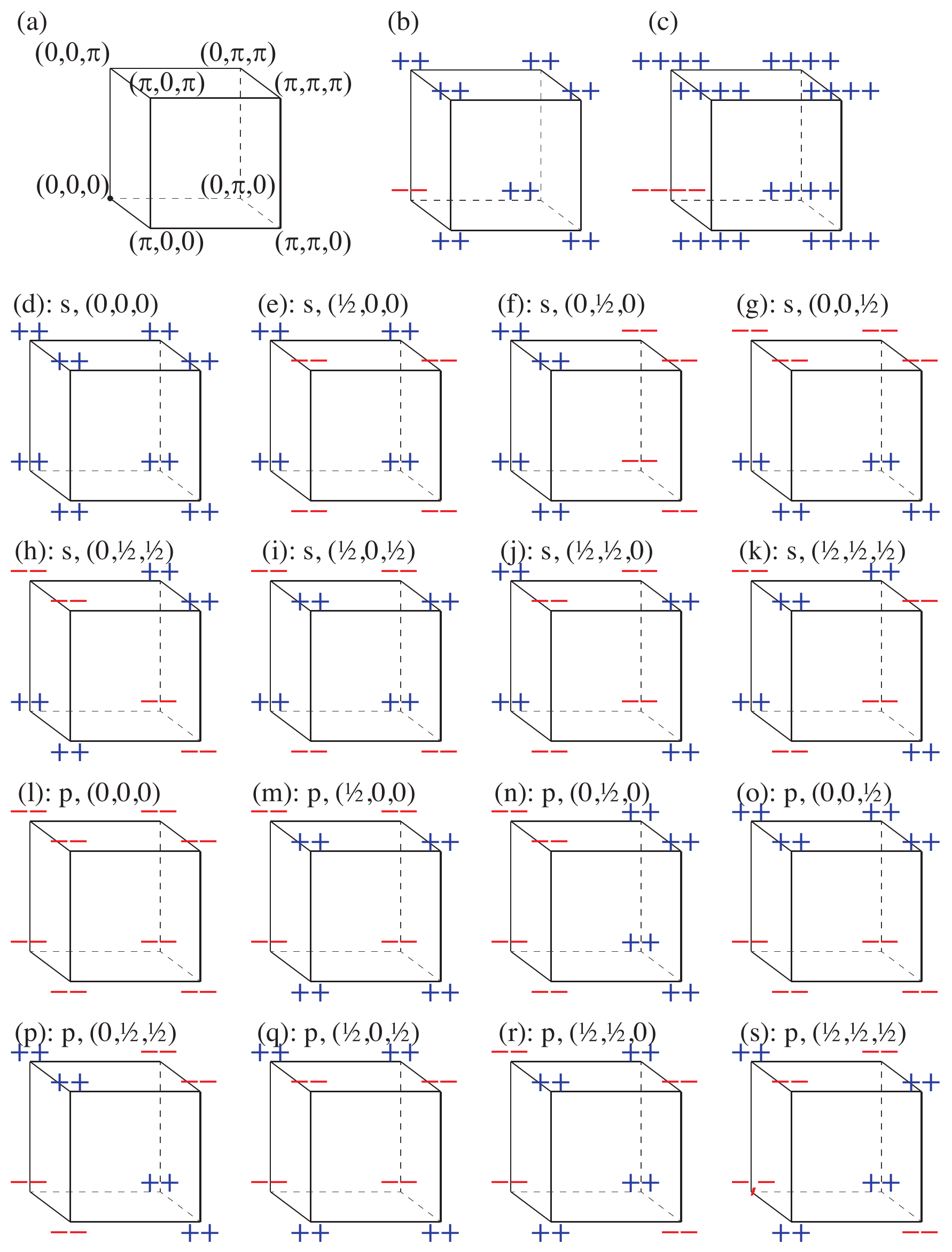}
\caption{The parity eigenvalues at high-symmetry momenta. (a): The common choice of coordinates in panels (b)--(s).  (b), (c):  An example of the parity eigenvalues for a strong topological insulator ($\kappa_1=3$, $\nu_0=1$, and $\nu_i=0$) and a higher-order topological insulator ($\kappa_1=6$ and $\nu_0=\nu_i=0$), respectively. (c) can be realized by stacking two copies of (b). (d)--(k): The parity eigenvalues of the atomic insulators constructed by placing an $s$-orbital at the position specified in each panel in every unit cell. (l)--(s): The same for (d)--(k) but a $p$-orbital is used instead of an $s$-orbital. All atomic insulators have $\kappa_1=\nu_0=\nu_i=0$ except for (d) $\kappa_1=4$, $\nu_0=\nu_i=0$ and (l) $\kappa_1=-4$, $\nu_0=\nu_i=0$.}
\label{indicator}
\end{figure*}

Let $n_{K}^{+}$ ($n_{K}^{-}$) be the number of occupied bands with even (odd) parity at each time-reversal invariant momentum (TRIM) $K$.
Due to Kramers pairing, $n_{K}^{\pm}$ is even. The Fu-Kane formula~\cite{Fu07} for the strong TI index $\nu_0\in\mathbb{Z}_2$ may be expressed as
\begin{eqnarray}
(-1)^{\nu_0}&=&\prod_{K\in\text{TRIMs}} (+1)^{\frac{1}{2}n_{K}^{+}} (-1)^{\frac{1}{2}n_{K}^{-}}\notag\\
&=&(-1)^{\frac{1}{2}\sum_{K\in\text{TRIMs}}n_{K}^{-}}.
\label{defZ2}
\end{eqnarray}
We now introduce a $\mathbb{Z}$-valued index \cite{Lu14} that is simply the sum of the inversion parities of occupied bands (up to a pre-factor):
\begin{eqnarray}
\kappa_1&\equiv&\frac{1}{4}\sum_{K\in\text{TRIMs}}(n_{K}^{+}-n_{K}^{-})\in\mathbb{Z}.
\label{defk1}
\end{eqnarray}
Using the total number of occupied bands $n\equiv n_{K}^{+}+n_{K}^{-}$, one can rewrite $\kappa_1$ as $2n-\frac{1}{2}\sum_{K\in\text{TRIMs}}n_{K}^{-}$.
Comparing this with Eq.~\eqref{defZ2}, we find
\begin{equation}
(-1)^{\nu_0}=(-1)^{\kappa_1}.
\label{nu0k1}
\end{equation}
Although Eq.~\eqref{nu0k1} suggests that $\kappa_1=\nu_0$ mod 2, $\kappa_1$ contains more information than $\nu_0$ as it is ``stable" mod $4$, in the sense that any trivial insulator has $\kappa_1=4n$ ($n\in\mathbb{Z}$) and adding or subtracting trivial bands does not alter $\kappa_1$ mod $4$ as demonstrated in Fig.~\ref{indicator}. Weak topological phases may realize $\kappa_1=2$ mod 4, but the most interesting case is when $\kappa_1=2$ mod 4 while $(\nu_1,\nu_2,\nu_3)$ all vanish.  One way of achieving this phase is to stack two copies of a strong TI as illustrated in Fig.~\ref{indicator} (b) and (c).  The surface signature of this phase was discussed in Ref.\ \onlinecite{Fang17}, and will be reintroduced in Sec.~\ref{Sec:Surface}.

The space group that contains only inversion $\I$ in addition to translations is called $P\bar{1}$ (numbered No.~2 in the standard crystallographic references \cite{ITC}). The index $\kappa_1$ can be defined for every supergroup of $P\bar{1}$ (SGs containing $P \bar{1}$ as a subgroup), i.e., for every centrosymmetric SG.

\subsection{New $\mathbb{Z}_2$ index for four-fold rotoinversion}
\label{sec:k4}
The SGs $P\bar{4}$ (No.~81) and $I\bar{4}$ (No.~82) possess neither inversion nor mirror symmetry. Hence, the indicator $X_{\text{BS}}=\mathbb{Z}_2$ found in Ref.~\onlinecite{Po17} cannot by accounted for by the Fu-Kane parity formula \cite{Fu07} or the mirror Chern number \cite{Hsieh12}. Here, we propose a new index $\kappa_4$ in terms of the eigenvalues of four-fold rotoinversion $S_4$ (the four-fold rotation followed by inversion) and show that the $\mathbb{Z}_2$ nontrivial phase is actually a strong TI.

To introduce the invariant, we note that there are four momenta in the BZ invariant under $S_4$, which we will denote by $K_4$; they are $(0,0,0)$, $(\pi,\pi,0)$, $(0,0,\pi)$ and $(\pi,\pi,\pi)$ for primitive lattice systems and $(0,0,0)$, $(0,0,2\pi)$, $(\pi,\pi,-\pi)$, and $(\pi,\pi,\pi)$ for body-centered systems.  For spinful electrons, the four possible values of $S_4$-eigenvalues are $e^{\frac{\alpha\pi}{4}i}$ ($\alpha=1,3,5,7$). Let us denote by $n_K^\alpha$ the number of occupied bands with the eigenvalue $e^{\frac{\alpha\pi}{4}i}$ at momentum $K\in K_4$.

The new $\mathbb{Z}$-valued index is (up to a pre-factor) the sum of the $S_4$-eigenvalues of occupied bands over the momenta in $K_4$ :
\begin{eqnarray}
\kappa_4&\equiv&\frac{1}{2\sqrt{2}}\sum_{K\in K_4}\sum_{\alpha}e^{\frac{\alpha\pi}{4}i}n_K^\alpha\in\mathbb{Z}.
\label{defk4}
\end{eqnarray}
This quantity is always an integer in the presence of TRS, and is stable modulo $2$ against the stacking of trivial bands.  The stability can be proven in the same way as we did for $\kappa_1$ in the previous section, i.e., by listing up all $S_4$-eigenvalues of atomic insulators.  As we show in Appendix~\ref{appk4n0}, $\kappa_4 \mod 2$ agrees with $\nu_0$.  It is, however, still useful to distinguish $\kappa_4$ from $\nu_0$, as $\kappa_4$ conveys more information than $\nu_0$ in the presence of additional symmetries, as we will discuss in the next section.  

\subsection{The combination of $\kappa_1$ and $\kappa_4$}
\label{sec:delta}
The SGs $P4/m$ (No.~83) and $I4/m$ (No.~87) contain both inversion $\I$ and four-fold rotation $C_4$ with the inversion center on the rotation axis.  The product of $\I$ and $(C_4)^2$ is a mirror $M_z$ about the $xy$ plane and one can define the mirror Chern number $C_M^{(k_z=0)}$.  The eigenvalues of the four-fold rotation determine $C_M^{(k_z=0)}$ mod 4  \cite{Fang12}.  However, $X_{\text{BS}}$ studied in Ref.~\onlinecite{Po17} contains a $\mathbb{Z}_8$ factor, which cannot be explained by the (detected) mirror Chern number alone. Here, we argue that the combination of $\kappa_1$ and $\kappa_4$ is responsible for this enhanced factor.

Since these SGs have both inversion $\I$ and rotoinversion $S_4=\I C_4$, $\kappa_1$ and $\kappa_4$ can be defined separately using Eqs.~\eqref{defk1} and~\eqref{defk4}.  Furthermore, the inversion center coincides with the rotoinversion center.  In this case, we find that the difference
\begin{equation}
\Delta=\kappa_1-2\kappa_4
\end{equation}
is stable modulo $8$, not only 4, against the stacking of trivial bands. Hence $\Delta$ mod 8 should be understood as a new bulk invariant, which can be reconciled with the strong factor $\Z_8 < X_{\rm BS}$. To see this, recall first that the definition of $\kappa_1$ and $\kappa_4$ has no ambiguity and they are perfectly well-defined as $\mathbb{Z}$-valued quantities.  Furthermore, an explicit calculation verifies that, for any trivial insulator symmetric under either $P4/m$ or $I4/m$, the two invariants $\kappa_1$ and $\kappa_4$ are not independent and $\kappa_1=2\kappa_4$ mod 8 always holds.

Even when all the mirror Chern numbers and the weak indices vanish, $\Delta$ can still be $4\neq0$ mod 8. Depending on the geometry, this phase may feature ``hinge" type 1D edge states~\cite{Song17,Fang17} which we will discuss in Sec.~\ref{Sec:Surface}.

\subsection{The combination of $\kappa_1$ and $C_M$ under six-fold rotation, screw, or rotoinversion}
\label{sec:k1mCh}
The case of SG $P6/m$ (No.~175) is similar to $P4/m$. It has a six-fold rotation $C_6$ in addition to inversion $\I$.  The product of $\I$ and $(C_6)^3$ is the mirror $M_z$ protecting the mirror Chern number $C_M^{(k_z=0,\pi)}$.  Although the eigenvalues of the six-fold rotation can only detect mirror Chern numbers mod 6, $X_{\text{BS}}^{(s)}$ contains a $\mathbb{Z}_{12}$ factor.  In this case, the combination of $C_M^{(k_z=0)}$ mod 6 and $\kappa_1$ mod 4 can fully characterize this $\mathbb{Z}_{12}$ factor (thanks to the Chinese remainder theorem).

When the rotation $C_6$ is replaced by the screw $S_{6_3}$ (i.e., $C_6$ followed by a half translation in $z$), the resulting SG is $P6_3/m$ (No.~176). In this case, the product of $\I$ and $(S_{6_3})^3$ is also a mirror symmetry.  The corresponding $X_{\text{BS}}^{(s)}$ includes the same $\mathbb{Z}_{12}$ factor.

In contrast, $P\bar{6}$ (No.~174) generated by the six-fold rotoinversion $S_6=IC_6$ does not have an inversion symmetry and $\kappa_1$ is not defined. The mirror Chern numbers $C_M^{(k_z=0,\pi)}$ protected by $M_z=(S_6)^3$ can be diagnosed by thee-fold rotation $C_3=(S_6)^2$ modulo 3 and this fully explains $X_{\text{BS}}^{(s)}=\mathbb{Z}_3$.

\subsection{Summary of $X_{\text{BS}}$}
The eight SGs studied above (SG 2, 81, 82, 83, 87, 174, 175, and 176) play the role of ``key" SGs, in the sense that they provide an anchor for understanding the  symmetry-based indicator $X_{\text{BS}}$ of any other SG.
Suppose that we are interested in one of the 230 SGs $\mathcal{G}$. To understand the nature of a nontrivial $X_{\text{BS}}^{\mathcal{G}}$, one should first identify its maximal subgroup among the eight key SGs.  Let $\mathcal{G}_0\leq\mathcal{G}$ be the key SG. Then we know that (i) the topological indices characterizing $X_{\text{BS}}^{\mathcal{G}}$ are the same as those for $X_{\text{BS}}^{\mathcal{G}_0}$, (ii) $X_{\text{BS}}^{(s),\mathcal{G}}$ and $X_{\text{BS}}^{(s),\mathcal{G}_0}$ are the same, (iii) the weak factor $X_{\text{BS}}^{(w),\mathcal{G}}$ may be reduced from $X_{\text{BS}}^{(w),\mathcal{G}_0}$ due to the constraints imposed by the additional symmetries in $\mathcal{G}\setminus\mathcal{G}_0$.  In Table~\ref{tab:XBS}, we identify the key SGs for all the 230 SGs. This table reproduces $X_{\text{BS}}$ for all 230 SGs presented in Ref.~\onlinecite{Po17}, but in a way which renders their physical properties more transparent.  

Physically, the key space groups play the role of the minimal SGs for which a certain non-trivial phase is possible. This means that a non-trivial phase in a SG $\G$ in the family of the key SG $\G_0$ will {\it remain} nontrivial even if some symmetries of $\G$ are broken, so long as those of $\G_0$ are preserved. For instance, consider SG $P4_2/m$ (No. 84) characterized by a 4-fold screw rotation $S_{4_2}$ and the inversion $\I$.  The product of $\I$ and $(S_{4_2})^2$ gives rise to a mirror symmetry, whose mirror Chern number is diagnosed mod 4 by $X_{\rm BS}^{(s)}=\mathbb{Z}_4$.  (In this case, since the inversion center and the rotoinversion ($\I S_{4_2}$) center do not agree, $\kappa_1-2\kappa_2$ does not enhance $\mathbb{Z}_4$ to $\mathbb{Z}_8$.) If we break the mirror symmetry without breaking inversion, the mirror Chern number becomes undefined, but the phases corresponding to the nontrivial $X_{\rm BS}^{(s)}$ are still protected by the inversion symmetry, and are diagnosed by the inversion index $\kappa_1$.

\begin{table*}
\begin{center}
\caption{{\bf Summary of symmetry-based indicator of band topology.} Space groups are grouped into their parental key space group in the first column. The second column lists the key topological indices characterizing nontrivial classes in $X_{\text{BS}}$. The space groups are indicated by their numbers assigned in Ref.\ \onlinecite{ITC}, and those highlighted by an underbar are the key SGs.
\label{tab:XBS}}
\begin{tabular}{cc|ccc}\hline \hline
Key Space Group	&Key Indices	&$X_{\text{BS}}^{(w)}$ & $X_{\text{BS}}^{(s)}$	&Space Groups\\\hline\hline
\multirow{11}{*}{$P\bar{1}$} & \multirow{11}{*}{$\nu_{1,2,3}$, $\kappa_1$} & $(\mathbb{Z}_2)^3$ & $\mathbb{Z}_4$ &  \underline{2}, 10, 47. \\
&&$(\mathbb{Z}_2)^2$ & $\mathbb{Z}_4$ & 11, 12, 13, 49, 51, 65, 67, 69. \\ &\\
&&\multirow{4}{*}{$\mathbb{Z}_2$}&\multirow{4}{*}{$\mathbb{Z}_4$} 
&   14, 15, 48, 50, 53, 54, 55, 57, 59, 63,\\
&&&& 64, 66, 68, 71, 72, 73, 74, 84, 85, 86, \\
&&&& 125, 129, 131, 132, 134, 147, 148, 162, \\
&&&& 164, 166, 200, 201, 204, 206, 224.\\ &\\
&& \multirow{3}{*}{0}  &\multirow{3}{*}{$\mathbb{Z}_4$} 
& 52, 56, 58, 60, 61, 62, 70, 88, 126, 130, 133,\\
&&&& 135, 136, 137, 138 141, 142, 163, 165, 167,\\
&&&& 202, 203, 205, 222, 223, 227, 228, 230.\\\hline
$P\bar{4}$ & $\kappa_4$ & 0 & $\mathbb{Z}_2$ & \underline{81}, 111--118, 215, 218. \\
$I\bar{4}$ & $\kappa_4$ & 0 & $\mathbb{Z}_2$ & \underline{82}, 119--122, 216, 217, 219, 220. \\\hline
\multirow{4}{*}{$P4/m$} && $\mathbb{Z}_2\times\mathbb{Z}_4$ &$\mathbb{Z}_8$ & \underline{83}, 123.\\
& $\nu_{1,2,3}$, $C_M^{(k_z=0,\pi)}$,  &$\mathbb{Z}_2$&$\mathbb{Z}_8$ & 124. \\
& $\kappa_1-2\kappa_4$ &$\mathbb{Z}_4$&$\mathbb{Z}_8$ & 127, 221.\\ 
&& 0 & $\mathbb{Z}_8$ & 128.\\&\\
\multirow{2}{*}{$I4/m$} & $\nu_{1,2,3}$, $C_M^{(k_z=0)}$, & $\mathbb{Z}_2$ & $\mathbb{Z}_8$ & \underline{87}, 139, 140, 229.\\
& $\kappa_1-2\kappa_4$ & 0 & $\mathbb{Z}_8$ & 225, 226\\\hline
\multirow{2}{*}{$P\bar{6}$} &\multirow{2}{*}{$C_M^{(k_z=0,\pi)}$}& $\mathbb{Z}_3$ &$\mathbb{Z}_3$ & \underline{174}, 187, 189.\\
&& 0 & $\mathbb{Z}_3$ & 188, 190.\\\hline
\multirow{2}{*}{$P6/m$} &\multirow{2}{*}{$C_M^{(k_z=0,\pi)}$, $\kappa_1$} &$\mathbb{Z}_6$ & $\mathbb{Z}_{12}$ & \underline{175}, 191. \\
&& 0 & $\mathbb{Z}_{12}$ & 192\\ & \\
$P6_3/m$ & $C_M^{(k_z=0)}$, $\kappa_1$ & 0 & $\mathbb{Z}_{12}$ & \underline{176}, 193, 194.\\
\hline \hline
\end{tabular}
\end{center}
\end{table*}

\section{Surface states}
\label{Sec:Surface}

We have seen in the previous section that some phases diagnosed by the symmetry indicator $X_{\rm BS}$ do not fall into the standard categories of strong TI, weak TIs, and mirror Chern insulators. This raises the question on the nature of the possible surface states possessed by these phases. With this question in mind, this section will be devoted to developing a  general approach to studying anomalous surface states protected by crystalline symmetries. We will argue that our approach captures most, if not all, TCIs with anomalous surface states \footnote{In fact, our approach captures any phase which admits a description in terms of K-theory/Dirac analysis, which includes all known sTCIs in class AII. While it is unlikely that there are sTCI phases which fall outside these frameworks, we leave the explicit proof of this statement as an interesting open problem.
}. In the next section, we will provide a precise relation between the surface analysis in this section and the bulk symmetry indicator discussed in the previous section.

We start by considering a sample with a specific geometry, and boundary conditions which are periodic or open along different directions. The surface is a compact manifold\footnote{For sample geometries with sharp edges, e.g. a cube, we can always round them into a (practically indistinguishable) smooth region with very large curvature.}
described by a surface Hamiltonian $h_{\br,\bk}$, which depends on the position on the surface $\br$ and the surface momentum $\bk =- i \nabla_{\br}$, a vector in the tangent space to the surface at point $\br$. The procedure of obtaining a surface Hamiltonian from a given bulk Hamiltonian is explained in detail in Appendix \ref{app:surfaceD}. To summarize, it starts by introducing some space-dependent parameter in the Hamiltonian, and changing it across the surface from its value inside the sample to a different value outside it (i.e., in the vacuum). This is then followed by projecting the bulk degrees of freedom onto the space of states localized close to the surface and rewriting the Hamiltonian in terms of these surface degrees of freedom.

We then proceed to define the notion of an ``anomalous surface topological crystalline insulator (sTCI)'' based on its surface properties:
we say a band insulator is a sTCI if there exists a geometry with some boundary conditions such that the surface Hamiltonian
\begin{enumerate}[(i)]
\item is not gapped everywhere; 
\item cannot be realized in an independent lower dimensional system with the same symmetries; and
\item can be gapped everywhere by breaking the crystalline symmetry or going through a bulk phase transition. 
\end{enumerate}
This definition includes familiar phases such as weak TIs (protected by translation) and conventional TCIs, as well as the recently discovered TCI with surface hinge modes \cite{ Song17, Schindler17, Langbehn17,  Benalcazar17Aug, Fang17}. 

Given a gapped bulk, the surface region where the gap vanishes is necessarily boundaryless, since otherwise the spatial spread of the surface states would diverge as we approach the boundary of the gapless region, necessitating the existence of gapless bulk states as in Weyl semimetals \cite{Wan11, Huang15, Xu15}.
This boundaryless region residing on the surface of the 3D bulk could be a 2D, 1D (collection of closed curves) or a 0D region (collection of points). We will show below that the latter is impossible for an insulator in class AII, but first, let us point out that designating a certain dimensionality to a gapless region generally depends not only on the symmetries at play, but also on the geometry of the sample. For example, the hinge insulators protected by rotation \cite{Song17, Fang17} possess 1D helical gapless modes when placed on a sphere, but exhibit gapless 2D Dirac cones when the surface is a plane normal to the rotation axis (with periodic boundary conditions along the two in-plane directions), as we illustrate Fig.~\ref{Planes}c. The relation between 1D and 2D surface states will be explored in more detail in Sec.~\ref{Sec:Dispersion}. 

To explore the types of possible surface states, we start by defining a (surface) high-symmetry point as a point left invariant by some crystalline symmetries. That is to say, the stabilizer group of this point contains elements apart from the identity. Let us first consider a generic point $\vec r$ (that is not a high-symmetry point). Since the only symmetry at $\vec r$ is TRS, the surface Hamiltonian at $\vec r$ can only be gapless if it is locally identical to (i) the surface of a strong TI, or (ii) a (movable) domain wall between two different quantum-spin Hall phases. This can be understood in terms of the classification of stable topological defects \cite{Teo10}, which implies that 1D and 2D defects are topologically stable in class AII (with a $\Z_2$ classification), whereas 0D defects are not. The latter typically describe vortices in superconductors and require particle-hole or chiral symmetry to be stable. 

At a high-symmetry point, one extra complication arises from the fact that the crystalline symmetry acts locally as an on-site discrete unitary symmetry, which can be used to block-diagonalize the local Hamiltonian. Each block can potentially have less symmetries than the original Hamiltonian. As a result, we need to consider the additional possibility of class A (unitary-type) defects. Only 1D defects, representing domain walls between two phases with different Chern numbers, are stable in class A \cite{Teo10} and have a $\Z$ classification. They will only exist if the symmetry leaves a line or curve on the surface invariant, which only occurs for mirror symmetry in 3D. Such line can host an arbitrary number of 1D gapless modes due to the $\Z$ classification, which can be understood in terms of mirror Chern number \cite{Hsieh12}.

Our argument can be summarized as follows: a gapless surface Hamiltonian at a certain point \emph{without fine tuning} will be a part of (i) a single domain wall, (ii) multiple intersecting domain walls (this could only happen at a high-symmetry point since an extra symmetry is required to stabilize the crossing), or (iii) the 2D gapless surface of a strong TI. The last case does not correspond to a sTCI since it does not require a crystalline symmetry for its stability, while the former two cases correspond to sTCIs with domain walls, which are either movable or are pinned to high-symmetry points or lines. Consequently, we arrive at a unified description for the surface states of all sTCIs in terms of ``globally irremovable, locally stable topological defects.'' Our approach is reminiscent to the approach of Ref.~\onlinecite{Hermele17}, which argued that a symmetry protected topological (SPT) phase protected by a point group symmetry can be understood in terms an embedded SPT of a lower dimensionality stabilized by the point-group symmetry. 

Having reduced our problem to the study of globally stable configurations of surface domain walls, the next simplification is achieved by observing that a surface domain wall in class AII can be generically constructed by first stacking together two strong TIs \footnote{We note that, throughout this work, ``stacking'' refers to direct summation of Hilbert spaces rather than real space stacking.}, and then adding a mass term which changes sign at the domain wall. This is also consistent with the analysis of Sec.~\ref{Sec:Indicators} and Refs.~\onlinecite{Fang17,Song17}, and it means that the surface theory of any sTCI can be constructed by stacking strong TIs and studying the symmetry transformation properties of the possible mass terms. Our basic building block in the following sections will be the doubled strong TI (DSTI), which is constructed by stacking two strong TIs.
In particular, we note that every even entry in the strong factor $X_{\rm BS}^{\rm (s)}$ of any space group can be viewed as the stack of two strong TIs (e.g., by rewriting $4 = 1+3$). If the order of $X_{\rm BS}^{\rm (s)}$ is even (say $X_{\rm BS}^{\rm (s)} = \mathbb Z_4$), then the DSTIs correspond precisely to the even subgroup of $X_{\rm BS}^{\rm (s)}$ (Table \ref{tab:XBS}); however, if the order is odd, then this identification does not hold, and any given entry could correspond to either a strong or a doubled strong TI. We will elaborate further on such identification ambiguity in Sec.\ \ref{Sec:BulkSurface}.

\subsection{Stacked strong TIs}
\label{Sec:DSTI}
As explained in detail in Appendix \ref{app:surfaceD}, the gapless Hamiltonian on the surface of a strong TI can be written as
\beq
\label{HTI}
h_{\br,\bk} = (\bk \times \bn_{\br}) \cdot \bsigma.
\eeq
Here, $\bn_\br$ is the normal to the surface at point $\br$ and $\bsigma =(\sigma_x, \sigma_y, \sigma_z)$ is the vector of Pauli matrices representing the spin degrees of freedom. TRS is implemented as $\T =i \sigma_y \K$, where $\K$ denotes complex conjugation, and it protects the gaplessness of $h_{\br,\bk} $.

A DSTI is constructed by stacking two strong TIs, whose surface Hamiltonian can be described by two copies of (\ref{HTI}). The only possible $\T$-symmetric mass term which can gap-out the surface has the form 
\beq
\label{Mass}
M_{\br} = m_{\br} \bn_{\br} \cdot \bsigma \otimes \tau_y,
\eeq 
with $\tau_{x,y,z}$ denoting the Pauli matrices in the orbital space of the two copies. 

To study the action of spatial symmetries on the DSTI model, we start by considering a generic spatial symmetry $g = \{ R_g ~|~ \vec \tau_g\}$, where $R_g \in {\rm O}(3)$ is a point-group operation and $\vec \tau_g$ denotes the (possibly fractional) translation in $g$. $R_g$ can be parameterized by three pieces of data:  $\theta_g$,  $\vechat n_g$, and $\det R_g$, where $\theta_g$ and $\vechat n_g$ are respectively the rotation angle and axis, and $\det R_g = \pm 1$ indicates whether $R_g$ is a proper or improper rotation. We can construct a natural spin-1/2 representation of the point-group action of $g$, given by $v_g \equiv \exp(- i \theta_g \, \vechat n_g \cdot \vec \sigma/2)$. 

The action of a spatial symmetry $g$ on the surface Hamiltonian (\ref{HTI}), derived in Appendix \ref{app:surfaceD} starting from its action on the bulk Hamiltonian, is given by 
\beq
g \cdot h_{\br,\bk} \equiv v_g h_{\br,\bk} v_g^\dagger =  h_{g \cdot \br,R_g \vec k },
\eeq
which leaves the surface Hamiltonian (as a whole) invariant for any spatial symmetry $g$. Clearly, the same conclusion holds when we generalize $v_g \mapsto u_g \equiv \eta_g v_g$ for any ${\rm U}(1)$ phase factor $\eta_g$.
Time-reversal symmetry further restricts this phase to $\eta_g = \pm 1$. 
In addition, we demand $\eta_g $ to respect the (projective) group structure of the point group. Specifically, for a pair of symmetry elements $g, g'$, we demand $u_{g g'} u_g^\dagger u_{g'}^\dagger = v_{g g'} v_g^\dagger v_{g'}^\dagger $, which fixes $\eta_{g g'} = \eta_g \eta_{g'}$
\footnote{Note that, as one can freely redefine $\theta_g \mapsto  \theta_g + 2\pi $, we have made an implicit sign choice in the definition of $v_g$. $\eta_g = u_g^\dagger v_g $ is unaffected by such ambiguity if one demands a simultaneous redefinition of both $u_g$ and $v_g$. Admittedly, however, this is a matter of convention; as we will see, the actual physical quantity of interest (called the ``signature'') is manifestly well-defined, as it involves the comparison of $\eta_g$ realized by two strong TIs in the same SG.}.

Microscopically, different values of $\eta_g$ arise naturally since the ${\rm O}(3)$ rotational symmetry is reduced to a finite, discrete point group in a crystal, and both spin and orbital angular momenta contribute to the symmetry representation.
Yet, when $u_g$ is traceless, say for $u_g = i \eta_g \sigma_z$, the sign $\eta_g$ can be removed through a basis transformation, and hence $\eta _g =\pm 1$ do not give rise to distinct representations. However, such a basis transformation will simultaneously modify the form of the surface Hamiltonian (\ref{HTI}), which in turn amounts to a redefinition of the helicity of the surface Dirac cone, as we will discuss later. In the following, we will say $u_g$ is a ``signed representation'' when we want to emphasize the importance of the sign choice $\eta_g = \pm 1$, regardless of whether this choice actually gives rise to distinct representations.

Spatial symmetries may force the mass term $m_\br$ to vanish along some curve, which creates a domain wall hosting a propagating 1D helical gapless mode. To see this, consider a spatial symmetry operation $g$ which leaves the mass term (\ref{Mass}) invariant. The action of the symmetry on the mass term can be generally written as 
\beq
g \cdot M_\br \equiv 
u^{\tau}_g M_\br (u^{\tau}_g)^{\dagger}
= M_{g \cdot \br} .
\eeq
Here, $u^{\tau}_g \equiv u_g^{(1)} \oplus u_g^{(2)}$ is the signed representation of $g$ acting on the DSTI, where $u_g^{(i)} = \eta_g^{(i)} v_g$ for the two strong TIs labeled by $i=1,2$.

The invariance of the mass $M_\br$ under symmetry action implies that
\beq
m_{g \cdot \br} = s_g m_\br, \quad (u_g^\tau)^{\dagger} \left( \bn_{g \cdot \br} \cdot \bsigma \otimes \tau_y \right) u_g^\tau = s_g  \bn_\br \cdot \bsigma \otimes \tau_y,
\eeq
with $s_g = \pm 1$. This means that, for any signed representation $u^{\tau}_g$ of the symmetry group $\G$ on the DSTI, we can define its ``signature'' as a map $s: \G \mapsto \Z_2$, which assigns a $\pm$ sign to each symmetry operation $g \in \G$ according to whether or not the mass changes sign under the action of $g$ (in the specified signed representation). The only condition which should be satisfied by $s_g = \pm 1$ is that it is a group homomorphism i.e.  $s_{g_1} s_{g_2} = s_{g_1 g_2}$ for $g_{1,2} \in \G$. The signature $s_g$ is not to be confused with the sign $\eta_g$. While the former describes a representation on the {\it DSTI} specifying the transformation properties of the mass term \eqref{Mass}, the latter describes the sign choice of the representation on a {\it single} strong TI. The two are related by Eq.~\eqref{sg} given below.
 
The properties of the mass term on the surface are fixed by specifying the signatures of the symmetry operations, which are in turn completely fixed by the parameters $\eta_g^{(1, 2)}$ and $\det R_g$ via the relation
\beq
\label{sg}
s_g = \det R_g \, \eta_g^{(1)}  \eta_g^{(2)}.
\eeq
The appearance of $\det R_g$ in this expression follows from the fact that $\bn \cdot \bsigma$ is a pseudoscalar.
In the following, we will usually use the terminology ``a $\pm$ representation for symmetry $g$'' for symmetry representations on the DSTI to indicate that the symmetry $g$ has a signature $s_g = \pm$.

Let us point out that, although our DSTI model  with two flavors of Dirac fermions is not sufficient for implementing all possible sTCIs of interest, say those with a high mirror Chern number, it is sufficient for constructing the ``root states'' which generate such states upon stacking. To see this, consider a system consisting of $n$ copies of the DSTIs. In this case, there are $k_n$ independent $\T$-preserving mass terms $m_{i,\br}$, $i=1,\dots,k_n$. The crystalline symmetries act on the vector $\bm_\br =(m_{1,\br},\dots,m_{k_n,\br})$ as orthogonal transformations leaving the length of the vector (which gives the magnitude of the gap at a given point) fixed (see appendix~\ref{app:surfaceD}). Any crystalline symmetry apart from mirror leaves at most two points on any given surface invariant. As a result, it will only protect anomalous surface states if it enforces the existence a domain wall between a point $\br$ to its image under symmetry, $g \cdot \br$. Such a domain wall will be irremovable if and only if there is no trajectory connecting $\bm$ and $O_g \bm$ on the $(k_n-1)$-sphere, establishing a correspondence between the zeroth homotopy group of the $(k_n-1)$-sphere and the stable domain walls in a model with $n$ DSTIs. Since the zeroth homotopy group of the $(k-1)$-sphere is trivial for $k>1$, we deduce that we cannot build stable domain walls whenever $n>1$. For $n=1$, $k_n = 1$ and only one mass term is possible. In this case, the fact that the  0-sphere (just two points) has two disconnected components implies the possibility of having $\Z_2$ domain walls, thereby establishing a $\Z_2$ classification for sTCIs {\it not protected by mirror symmetry} in class AII. Physically, the $\Z_2$ classification here simply descends from that of 2D TIs in class AII.
 
The only exception to the previous analysis is mirror symmetry. In this case, we need to consider the mass vector in the mirror plane. We find that it has to remain invariant under the action of mirror symmetry at any point in this plane, $O_M \bm = \bm$, which is only possible if one of the eigenvalues of $O_M$ is $+1$. In a representation which does not satisfy this condition, e.g $O_M = -1$, the mass vector $\bm$ will necessarily vanish in the mirror plane regardless of $n$. Nevertheless, the state with $n>1$ DSTIs can always be built by stacking the ``root'' state implemented using a single mirror-symmetric DSTI. 

\begin{figure}
\center
\includegraphics[width=0.35\columnwidth]{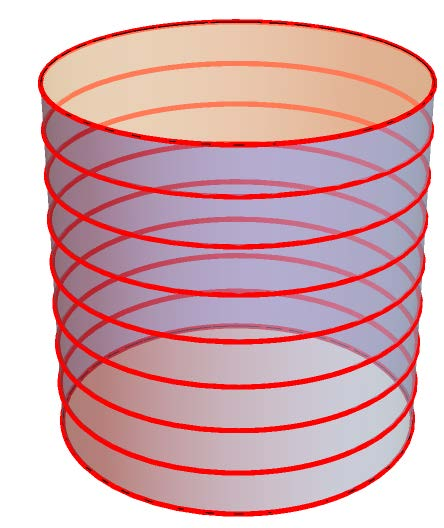}
\caption{Surface states for the weak topological insulator can be understood as choosing a ``$-$'' representation for the translation along the ``weak'' direction.}
\label{WeakTI}
\end{figure}

Notice that, up to this point, the analysis is general and applies to any crystalline symmetry. For instance, weak TIs can be understood within the DSTI model by considering a surface with periodic boundary along the ``weak'' direction, and choosing a ``$-$'' representation for translation along this direction. This will enforce the existence of a domain wall for every unit lattice translation along this direction, leading to the surface states obtained by stacking 2D quantum-spin Hall systems (Fig.~\ref{WeakTI}).
Alternatively, for a system without any weak index, i.e., all the lattice translations are assigned a ``$+$'' signature, some other elements in the SG could also be in a ``$-$'' representation and lead to other patterns of gapless modes on suitably chosen surfaces.

Having outlined our general framework, we now apply it to classify all sTCIs in class AII. We will proceed in two steps: first, we will focus on phases protected solely by point-group symmetries; second, we will discuss how to extend these results consistently to cover all the 230 space groups.

\subsection{Crystallographic point groups} 
\label{Sec:Pointgroups}
This subsection is devoted to the study of crystalline point groups, whose associated sTCIs can be understood by considering a spherical geometry with open boundary conditions in all directions. We will describe a procedure to construct all possible sTCIs in a given symmetry group, and then use it to obtain the complete classification of sTCIs protected by point group symmetries.

We begin by reviewing the natural action of the individual point-group symmetries on the physical degrees of freedom, and discussing the types of gapless surface states they can protect. 
Afterwards, we will provide a general procedure to construct all sTCIs in a given point group, and use it to obtain an exhaustive classification of sTCIs for the 32 crystallographic point groups.

\subsubsection{Crystallographic point group symmetries}

{\bf Inversion}: Inversion symmetry acts by inverting position and momentum while keeping the spin unchanged
\beq
\I: \quad \br \rightarrow -\br, \qquad \bk \rightarrow -\bk, \qquad \bsigma \rightarrow \bsigma.
\eeq
According to (\ref{sg}), inversion can be represented with negative signature by taking its action on the DSTI to be $\I^\tau = \I \oplus \I = \I \otimes \tau_0$ (corresponding to $\eta^{(1)} = \eta^{(2)}$), or with positive signature by acting on the DSTI as $\I^\tau = \I \oplus (-\I) = \I \otimes \tau_z$ (corresponding to $\eta^{(1)} = -\eta^{(2)}$).

In the ``$-$'' representation, the mass term changes sign between a point and its image under inversion, leading to a ``hinge'' phase with a helical gapless mode living on some inversion-symmetric curve on the sphere. The curve can be moved around but cannot be removed without breaking inversion. Such a phase is graphically illustrated in the $C_i$ entry in Fig.~\ref{PG}. Two copies of this state can be trivialized by adding the mass terms $m^{x,z}_\br (\bn_\br \cdot \bsigma) \tau_{x,z} \mu_y$, where $\mu$ denote the Pauli matrices in this additional orbital space of copies. 
These additional mass terms, like (\ref{Mass}), change sign under inversion; however, $m^{x,z}_\br$ can be chosen so that they do not both vanish at a point where the original mass term, $m_\br$ in Eq.\ \eqref{Mass}, also vanishes. This leads to a completely gapped surface. As a result, gapless modes protected by inversion will have a $\Z_2$ classification, as anticipated in the general discussion of the previous subsection.

{\bf $C_n$ rotation}: An $n$-fold rotation acts by rotating the position and momentum as vectors and rotating the spin as a spinor. An $n$-fold rotation about the $z$-axis is implemented as
\beq
C_{nz}: \br \rightarrow C_{nz} \br, \qquad \bk \rightarrow C_{nz} \bk \qquad \bsigma \rightarrow e^{-i \frac{\pi}{n} \sigma_z} \bsigma e^{i \frac{\pi}{n} \sigma_z},
\eeq
A positive signature representation is obtained by taking $C_n^\tau = C_n \oplus C_n = C_n \otimes \tau_0$ (corresponding to $\eta^{(1)} = \eta^{(2)}$), whereas the choice $C_n^\tau = C_n \oplus (-C_n) = C_n \otimes \tau_z$ (corresponding to $\eta^{(1)} = -\eta^{(2)}$) leads to a negative signature \footnote{These two choices obviously do not exhaust all possible choices for the different irreps in the two copies \cite{Fang17}, but they are sufficient for constructing representations with both signatures, which suffices for our analysis.}. Notice that, here, the ``$+$'' and ``$-$'' representations are opposite to the inversion case, since $\det \mathcal I = -1$ but $\det C_n = 1$ (cf.~Eq.~\eqref{sg}).

We note that the ``$-$'' representation is only possible for even $n$, since it violates the condition $C_n^n = -1$ whenever $n$ is odd. Thus, $n=3$ is only consistent with a ``$+$'' representation, whereas $n=2,4,6$ can have of either signature. A ``$-$'' representation in this case forces the mass to vanish at $n/2$ (closed) curves related by rotation and intersecting at the two points left invariant by rotation (the poles) (as shown in Fig.~\ref{PG} PGs $C_2$, $C_4$ and, $C_6$). 
We remark that these conclusions have already been drawn in Ref.\ \onlinecite{Fang17} using a slightly different language.

Similar to the case of inversion, two copies of the described DSTI can be gapped out by adding the mass term $m_\br' (\bn_\br \cdot \bsigma) \tau_z \otimes \mu_y$, which does not change sign under rotation and can be chosen to be positive everywhere. This means that rotations lead to a $\Z_2$ classification as well, consistent with our general discussion.

{\bf Mirror symmetry}: Mirror symmetry acts by inverting the position and momentum components perpendicular to the mirror plane and the spin component parallel to it. For example, the action of mirror symmetry about the $xy$ plane is implemented as
\beq
M_z: \quad (x,y,z) \rightarrow (x,y,-z), \qquad  \bsigma \rightarrow \sigma_z \bsigma \sigma_z.
\eeq
In the mirror plane, $M_z$ flips the sign of $\bn_\br \cdot \bsigma$. Thus, in the ``$-$'' representation, implemented by taking the mirror to act in the DSTI as $M_z^\tau = M_z \oplus M_z = M_z \otimes \tau_0$ (corresponding to $\eta^{(1)} = \eta^{(2)}$), the mass term $m_\br$ has to vanish in this plane. The ``$+$'' representation, on the other hand, is implemented in the DSTI as $M_z^\tau = M_z \oplus (-M_z) = M_z \otimes \tau_z$ (corresponding to $\eta^{(1)} = -\eta^{(2)}$), which does not impose a constraint on the mass term in the mirror plane and leads to a completely gapped surface. 

As we noted in the previous subsection, the ``$-$'' representation for the mirror implies that the mirror plane remains gapless regardless of the number of DSTIs  stacked together. This can be seen from the fact that, for any number of copies, any mass term will have the form $m_\br (\bn_\br \cdot \bsigma) \otimes \Gamma$, where $\Gamma$ denotes matrices in the orbital space of copies, which will always vanish in the mirror plane. As discussed previously, this implies a $\Z$ classification corresponding to the mirror Chern number in real space \cite{Hsieh12}.

{\bf $S_n$ rotoinversion}: As $S_n = \mathcal I C_n$, its action on a DSTI can be readily understood through the corresponding discussions for $\mathcal I$ and $C_n$ above. Note that a ``two-fold rotoinversion'' is simply a mirror symmetry, which leaves a plane invariant in 3D, and is differentiated from $S_{3,4,6}$, which only leave the origin invariant. It is worth noting that among the three roto-inversion groups $\bar 3$, $\bar 4$, and $\bar 6$, only $\bar 4$ is not a direct product of smaller groups.

\subsubsection{Classification of sTCIs in the 32 crystallographic point groups}
We are now in a position to perform a systematic investigation of sTCIs in the 32 crystallographic point groups. As we argued in the beginning of this section, all these states can be built from either the DSTI model considered in Sec.~\ref{Sec:DSTI}, or copies thereof. 

In Sec.~\ref{Sec:DSTI}, we proposed that there is a one-to-one correspondence between the signatures of the different symmetries and the pattern of gapless modes on the surface. One part of this correspondence is obvious since surface states corresponding to different symmetry signatures cannot be deformed into each other without changing these signatures. We conjecture that the opposite is also true: two patterns of surface modes can always be deformed into each other if they correspond to the same representation signatures for all the possible symmetries. We have checked this explicitly for several examples, where seemingly different surface state patterns corresponding to the same symmetry signatures turned out to be deformable into each other.

\begin{figure}
\center
\includegraphics[width=0.85\columnwidth]{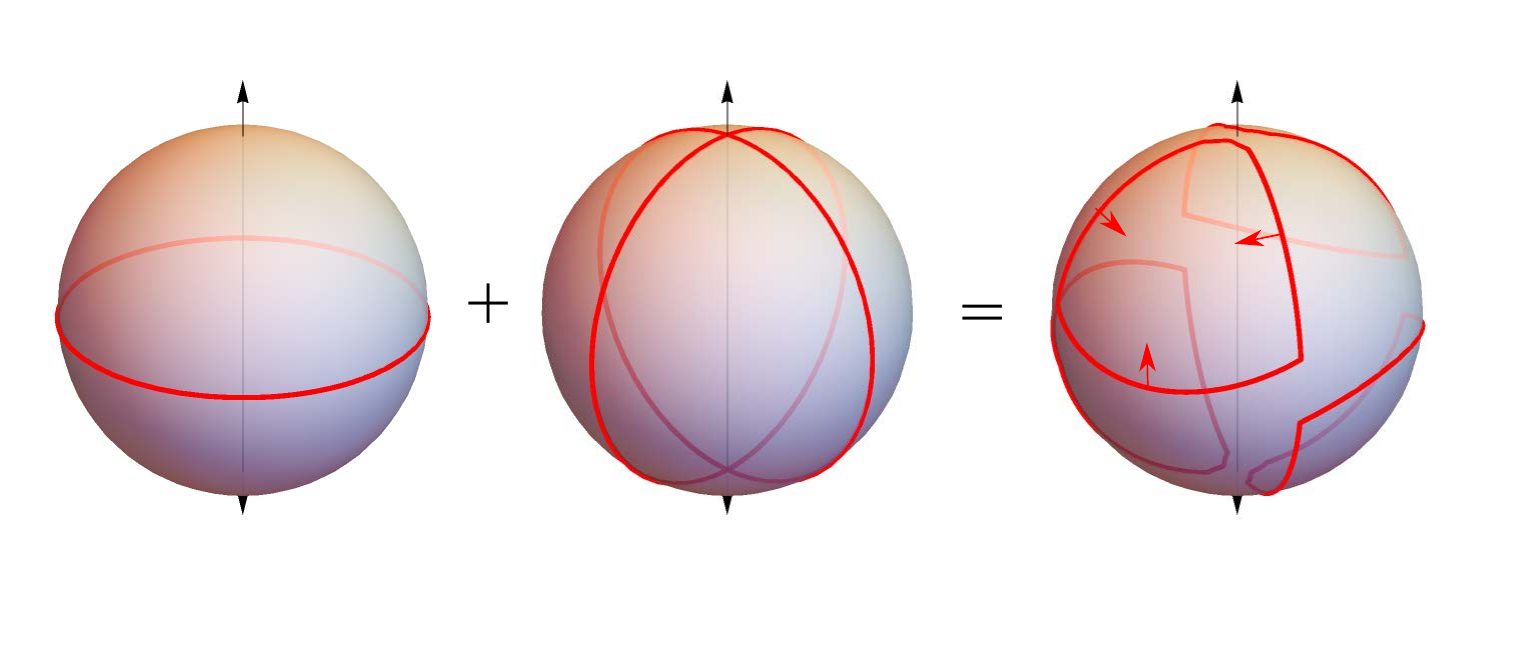}
\caption{In the presence of the four-fold rotoinversion $S_4$ only, the ``equator'' and the ``hinge'' state are deformable to each other since their sum can be trivialized as shown here.}
\label{Trivial}
\end{figure}

For example, consider the point group $\bar 4$, where the only symmetry is a 4-fold rotoinversion. In this case, there seems to be two distinct surface state patterns corresponding to the ``$-$'' representation for $S_4$, given by the ``equator'' state and the ``hinge'' state (cf.~Fig.~\ref{Trivial}). These two can, nevertheless, be deformed into each. The reason is that, unlike rotation, rotoinversion does not leave the poles fixed. Therefore, there is no symmetry constraint on the mass terms at the poles, implying the intersection of gapless modes at the poles is spurious, i.e., it is unstable against symmetry-allowed perturbations.
Thus, we can move the hinges symmetrically away from the poles, brining them to the equator. This can be seen more clearly by adding the two phases and noting that a mass term can be added to gap-out the modes at their intersection points (this is possible since rotoinversion does not enforce any local constraints on the mass). The resulting surface can be deformed to a trivial one as shown in Fig.~\ref{Trivial}. 
Alternatively, such correspondence can also be understood in a slightly more general language (i.e., beyond the Dirac theory analysis), as we elaborate in Sec.~\ref{Sec:Discussion}.

Using our conjecture that the classification of sTCIs reduces to classifying distinct signatures of the different symmetries, we now proceed by constructing these phases explicitly. We first note that two symmetries $g_{1,2} \in \G$ related by conjugation, i.e. there is an $h \in \G$ such that $g_1 =h^{-1} g_2 h$, necessarily have the same signature. In the following, we will call two symmetries independent if they are not related by conjugation, so that their signatures can be assigned independently. For two rotations (mirrors), independence means that the rotation axes (mirror planes) are not related by any symmetry operation in the symmetry group $\G$. This implies that we can classify all possible signatures by specifying the signatures for a minimal set of independent generators of the group (i.e., every group element can be written as a product of powers of the elements in the set, and the size of the set is as small as possible). Notice that the generators in a minimal set can always be chosen to be independent \footnote{If two generators $g_1$ and $g_2$ are related by $g_2 = h g_1 h^{-1}$, with $h$ not in the minimal generator set (since otherwise it is not minimal), we can always replace $g_2$ by $h$ in the generator set.}. Due to the fact that mirror and $C_3$ rotation symmetries behave differently compared to other symmetries in sTCI classification, we need the generator set to satisfy the extra conditions (while respecting the condition that the generator set is minimal):
\begin{enumerate}
\item The number of independent mirror symmetries included as generators is as large as possible.
\item The number of independent $C_3$ rotations included as generators is as large as possible .
\end{enumerate}
Note that these two conditions are consistent with each other since we cannot generally increase (decrease) the number of independent mirror symmetries by decreasing (increasing) the number of $C_3$ symmetries in a generator set \footnote{The reason is that two mirrors related by $C_3$ multiplication are also related by $C_3$ conjugation.}. Notice also that a minimal generator set with a maximal number of independent mirror symmetries actually contains all of them, which can be explicitly verified (cf.~Table \ref{Classification})
These conditions mean that we should include the maximum number of independent $C_3$ rotations in the generator set, as long as there is no other generator set with smaller size. For example, in the PG $T$, which can be generated using either two independent $C_3$ rotations or a $C_2$ and a $C_3$, we have to choose the former since it includes more $C_3$ rotations; in contrast, in the PG $C_6$, which can be generated using a single $C_6$ or a $C_3$ together with a $C_6$, we have to choose the former since it contains a smaller number of generators (the latter is not really a minimal generator set).

\begin{figure*}
\center
\includegraphics[width=0.95\textwidth]{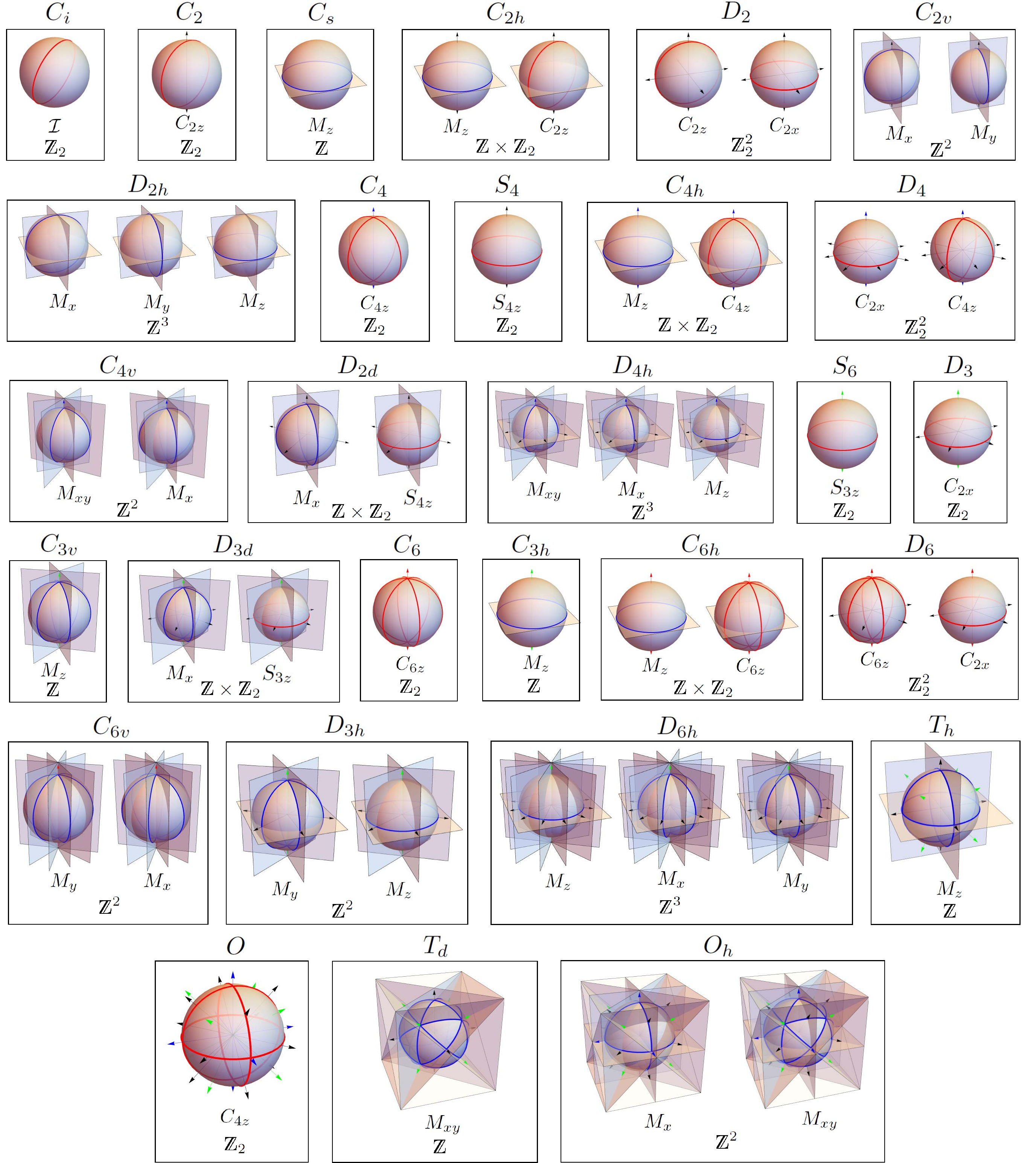}
\caption{Graphical illustration of the surface states of the ``hinge'' phases, which generate all sTCIs  for the 32 crystallographic point groups, on a sphere. For each state, the operators which need to be taken in the ``$-$'' representation are shown as well as the resulting classification. 
Red circles indicate $\Z_2$ modes which would be gapped-out if two copies of the system are stacked together, whereas blue circles indicate $\Z$ modes protected by a mirror Chern number. Rotation axes are shown with black, blue, green, or red color for two-, four-, three- and six-fold rotations respectively.}
\label{PG}
\end{figure*}

Once we have a minimal generator set satisfying these properties, we can read off the classification of sTCIs as follows:
\begin{enumerate}
\item To every mirror symmetry in the generator set, we assign a factor of $\Z$ indicating the number of gapless modes in this mirror plane. The phase which generates this factor is obtained by choosing the ``$-$'' representation for the corresponding mirror symmetry in the DSTI model.
\item To every symmetry generator other than mirror and $C_3$, we assign a factor of $\Z_2$. Each of these $\Z_2$ phases is generated by choosing the ``$-$'' representation for the corresponding symmetry generator.
\end{enumerate}
Implementing these rules leads to the classification of sTCIs in all crystallographic point groups, which we tabulate under $\S_{\rm PG}$ in Table~\ref{Classification}. The ``hinge'' states generating the different $\Z$, $\Z_2$ factors by choosing the ``$-$'' representation in the corresponding symmetry are graphically illustrated in Fig.~\ref{PG}.

\begin{table}
\center
\caption{{\bf Classification of point group sTCIs} $\S_{\rm PG}$ denotes the classification of point group sTCIs. The generators are chosen according to the criteria defined in the main text. $\bn_1$ and $\bn_2$ denotes two of the four 3-fold rotation axes in the cubic PGs. These describe systems with cubic symmetry with four 3-fold axes along the cube body-diagonal.}
\begin{tabular}{c|c|c|c}
\hline \hline
Symbol & ~~~PG~~~ & Generators & ~~~~~$\S_{\rm PG}$~~~~~\\
\hline
 1 & $C_1$ & 1 & 0\\
 $\bar{1}$ & $C_i$ & $\I$ & $\Z_2$\\
 2 & $C_2$ & $C_{2z}$ & $\Z_2$\\
 $m$ & $C_s$ & $M_z$ & $\Z$\\
 $2/m$ & $C_{2h}$ & $M_z, C_{2z}$ & $\Z \times \Z_2$\\
 $222$ & $D_2$ & $C_{2z}, C_{2x}$ & $\Z_2^2$\\
$mm2$ & $C_{2v}$ & $M_x, M_y$ & $\Z^2$\\
$mmm$ & $D_{2h}$ & $M_x, M_y, M_z$ & $\Z^3$\\
$4$ & $C_4$ & $C_{4z}$ & $\Z_2$\\
 $\bar{4}$ & $S_4$ & $S_{4z}$ & $\Z_2$\\
 $4/m$ & $C_{4h}$ & $C_{4z}, M_z$ & $\Z \times \Z_2$\\
$422$ & $D_4$ & $C_{4z}, C_{2x}$ & $\Z_2^2$\\
$4mm$ & $C_{4v}$ & $M_{x}, M_{xy}$ & $\Z^2$\\
$\bar{4}2m$ & $D_{2d}$ & $S_{4z}, M_{x}$ & $\Z \times \Z_2$\\
$4/mmm$ & $D_{4h}$ & $M_{x}, M_{z}, M_{xy}$ & $\Z^3$\\
$3$ & $C_3$ & $C_{3z}$ & 0\\
$\bar 3$ & $S_6$ & $S_{3z}$ & $\Z_2$\\
$32$ & $D_3$ & $C_{3z}, C_{2x}$ & $\Z_2$\\
$3m$ & $C_{3v}$ & $C_{3z}, M_{x}$ & $\Z$\\
$\bar 3m$ & $D_{3d}$ & $S_{3z}, M_{x}$ & $\Z \times \Z_2$\\
$6$ & $C_6$ & $C_{6z}$ & $\Z_2$\\
$\bar 6$ & $C_{3h}$ & $C_{3z}, M_z$ & $\Z$\\
$6/m$ & $C_{6h}$ & $C_{6z}, M_z$ & $\Z \times \Z_2$\\
$622$ & $D_6$ & $C_{6z}, C_{2x}$ & $\Z_2^2$\\
$6mm$ & $C_{6v}$ & $M_{x}, M_{\frac{\sqrt{3}x+y}{2}}$ & $\Z^2$\\
$\bar 6m2$ & $D_{3h}$ & $C_{3z}, M_z, M_{x}$ & $\Z^2$\\
$6/mmm$ & $D_{6h}$ & $M_z, M_{x}, M_{\frac{\sqrt{3}x+y}{2}}$ & $\Z^3$\\
$23$ & $T$ & $C_{3,\bn_1}, C_{3,\bn_2}$ & 0\\
$m\bar 3$ & $T_h$ & $C_{3,\bn_1}, M_z$ & $\Z$ \\
$432$ & $O$ & $C_{3,\bn_1}, C_{4,z}$ & $\Z_2$\\
$\bar 43m$ & $T_d$ & $C_{3,\bn_1}, M_{\bn_1}$ &  $\Z$\\
$m\bar 3 m$ & $O_h$ & $C_{3,\bn_1}, M_{x},M_{xy}$ & $\Z^2$\\
\hline \hline
\end{tabular}
\label{Classification}
\end{table}

\subsection{Space groups}
In this subsection, we extend the analysis of the previous section to space groups, which requires considering symmetries that do not fix any point in space (lattice translations and non-symmorphic symmetries). We first start by discussing the main complication arising from the inclusion of non-symmorphic symmetries, which requires a certain choice of the sample geometry and boundary conditions. We then provide a few examples of sTCIs protected by non-symmorphic symmetries. 
Next, we discuss how to consistently combine the preceding results with lattice translations, i.e., we systematically study the spatial-symmetry constraints on weak indices \cite{Varjas17}.
Finally, we will provide a complete classification of sTCIs in the 230 SGs. We note that the order followed in this section, by considering first non-symmorphic symmetries and then translations, may seem a bit incongruous given that non-symmorphic symmetries are generalizations of translation. This ordering is however natural within our present context in which the strong part $\S_{\rm SG}^{\rm (s)}$ of the classification (which does not include translations) is first generalized from the point group case to include non-symmorphic symmetries before presenting a separate discussion for the weak part $\S_{\rm SG}^{\rm (w)}$ where translation plays a major role.

\subsubsection{Non-symmorphic symmetries}
A non-symmorphic symmetry arises when a point-group symmetry is combined with an irremovable, fractional translation, which results in a symmetry that leaves no point in space invariant. 
The extension of the analysis of Sec.~\ref{Sec:Pointgroups} to include non-symmorphic symmetries is, in principle, straightforward.
As we will elaborate on later, to analyze sTCIs protected by non-symmorphic symmetries, it suffices to first assume that all the weak indices of the system are trivial, i.e., all the lattice translations are given a ``$+$'' representation. Microscopically, this is the case when we can stick with DSTI models with degrees of freedom arising form the $\Gamma$ point in the BZ, such that, in momentum space, the fractional translation associated to a non-symmorphic symmetry becomes irrelevant. This restriction is justified in Sec.\ \ref{Sec:BulkSurface} and Appendix \ref{app:XBSRep}.

As we can focus on the $\Gamma$ point in the momentum space, the main conceptual difference in understanding sTCIs protected by point-group and non-symmorphic symmetries, therefore, lies in the real space. Recall that, to expose the anomalous surface states of an sTCI, one has to consider a geometry respecting the protecting spatial symmetries.
In order to preserve a non-symmorphic symmetry on the surface, we need to consider a sample with periodic boundary conditions along the directions of the fractional translation vector of the symmetry. A non-symmorphic symmetry, which can either be a screw ($n$-fold rotation followed by a fractional lattice translation along the rotation axis) or a glide (mirror reflection followed by a fractional lattice translation along a vector in the mirror plane), does not leave any point invariant. As a result, the surface states protected by non-symmorphic symmetries will have a $\Z_2$ classification (see the discussion at the beginning of this section).

The anomalous surface states protected by non-symmorphic symmetries can be understood by considering a cylinder geometry whose axis is parallel to the fractional translation axis, along which periodic boundary condition is taken. An $n$-fold screw in a ``$-$'' representation would give rise to a state with $n$ symmetry-related hinges related, along the sides of the cylinder. This surface state is very similar to the surface state protected by rotation, in that the hinges can be moved around freely as long as they preserve the screw or rotation as a set.

\begin{figure}
\center
\includegraphics[width=0.8\columnwidth]{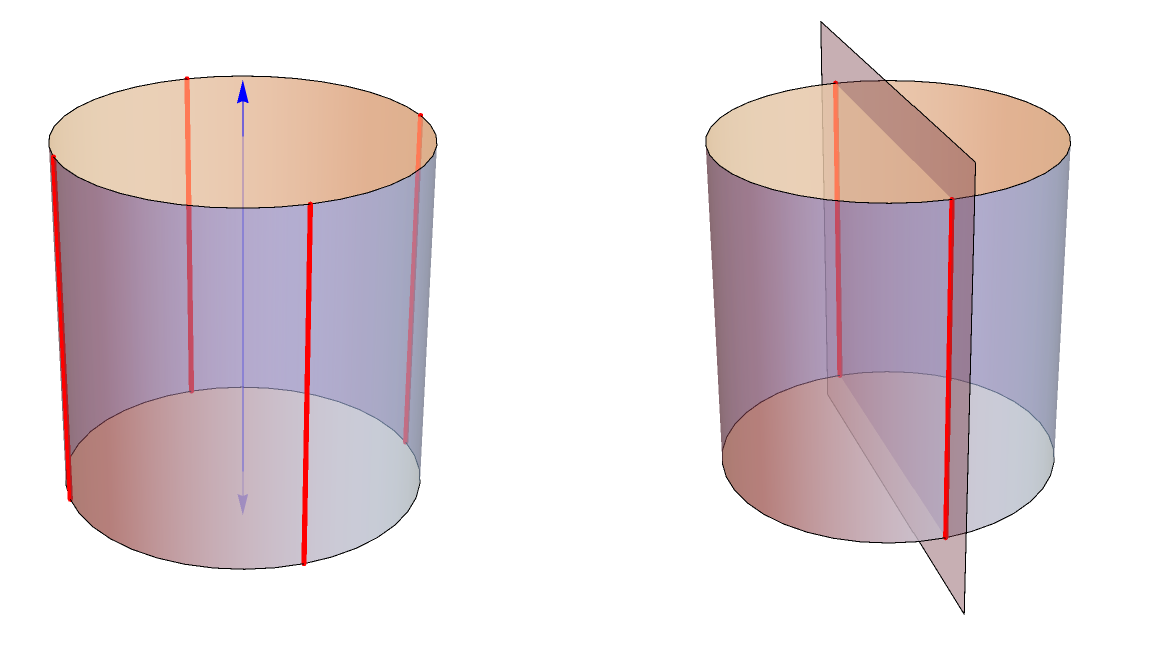}
\caption{Illustration for the hinge state protected by a 4-fold screw (left) and a glide (right) on a cylinder geometry with periodic boundary conditions along the screw direction.}
\label{Screwglide}
\end{figure}

Extending a mirror symmetry to its non-symmorphic counterpart, a glide, leads to a more drastic modification of the physics. Picking a ``$-$'' representation for a glide symmetry gives rise to a state with two hinge modes confined to the glide plane shown in Fig.~\ref{Screwglide}. Despite looking similar to surface states protected by mirror symmetry, this state differs in an essential aspect, as it becomes trivial when added to itself. The reason is that, unlike mirror symmetry, a mirror Chern number cannot be defined for glide symmetry, since it does not act as an on-site symmetry in the glide plane. One can also understand such modification from a momentum-space perspective: unlike a mirror, the band eigenvalues $\pm i$ of a glide are not invariantly defined, as they are interchanged upon the addition of a reciprocal lattice vector to the crystal momentum. Consequentially, one cannot define a $\mathbb Z$-valued Chern number using a glide symmetry. We note here that glide symmetry fits more naturally than mirror within our signed representation approach. The reason is that the approach, by itself, only allows for $\Z_2$-type hinges modes resulting from the symmetry constraints on the surface mass term. The appearance for $\Z$-type hinge modes in the mirror case is an exception due to the fact that it leaves invariant an extended (1D) region on the surface where the effective symmetry class is reduced from AII to A. Such unusual property, which is not shared by any other symmetry, is the main reason why an integer invariant (mirror Chern number) can be defined only in the mirror case.

\begin{figure}
\center
\includegraphics[width=0.95\columnwidth]{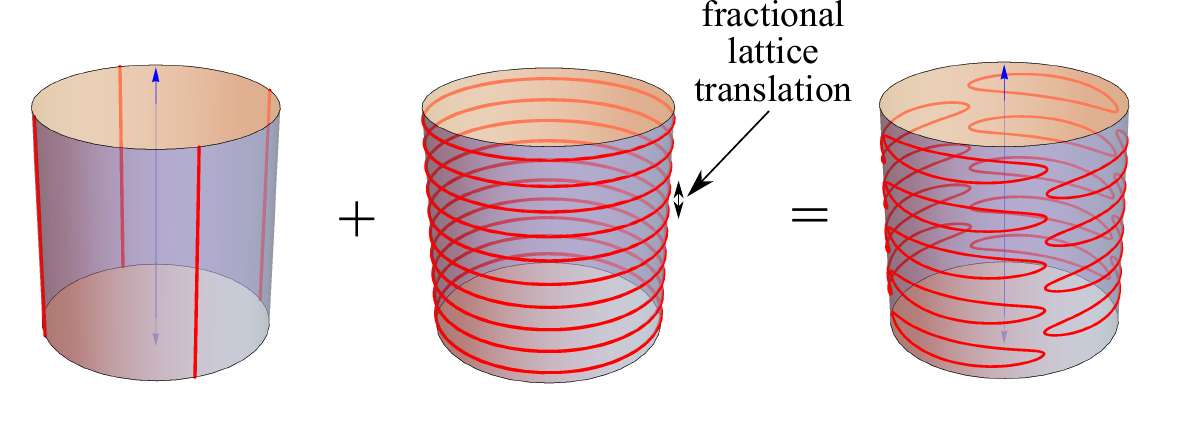}
\caption{Illustration for the two equivalent surface states corresponding to a fourfold screw. The equivalence of the two can be established by noting that their sum can be deformed to a trivial state while preserving the screw symmetry.}
\label{Screwdef}
\end{figure}

As an example of surface states protected by a non-symmorphic symmetry, let us consider SG $P4_2$ (No.~77), where the only symmetry is a 4-fold screw $4_2$. On a cylinder geometry with a periodic boundary condition taken along the screw axis, a ``$-$'' representation for the screw symmetry leads to the state with four hinges shown in Fig.~\ref{Screwglide}. We note that a state with domain walls at every half lattice translation along the screw axis is also consistent with the ``$-$'' representation for the screw symmetry. This state is gapless everywhere on the surface, and looks very similar to the surface of a weak TI shown in Fig.~\ref{WeakTI}. The main difference in this case is that such a state is unstable against being gapped out everywhere except for the four hinges. That is, the two surface state configurations shown in Fig.~\ref{Screwdef} both correspond to choosing the ``$-$'' representation for the screw symmetry and they can be symmetrically deformed into each other as shown in the figure.

Another example is SG $P6_3/m$ (No.~176), which is one of the key space groups considered in Table~\ref{tab:XBS}. This space group is generated by a 6-fold screw and a mirror about a plane normal to the screw axis. Choosing a cylinder geometry with a periodic boundary condition along the screw axis, we may consider a ``$-$'' representation for either the screw or the mirror. In the former case, we get a phase with six hinges, whereas the latter is characterized by gapless modes protected by a mirror Chern number in the mirror planes. Notice that, in this case, we have several mirror planes related by screws which all have the same mirror Chern number.
The signatures of the mentioned mirror and screw are independent, and consequentially the resulting classification is $\Z \times \Z_2$.

While we have argued quite generally that the sTCI classification for non-symmorphic symmetries can be understood from the point-group counterpart, our discussion so far has one caveat:
Unlike 3-fold rotations, a three-fold screw with a $1/3$ lattice translation along the rotation axis, $S_{3_1}$, does admit a ``$-$'' representation.  However, such a choice implies $(S_{3_1})^3$, a unit lattice translation, is also assigned a ``$-$'' representation, leading to a nontrivial weak TI index, and hence the described phase falls outside of our present discussion. This brings us to the last piece of consideration required for our classification of sTCIs: the incorporation of weak phases.

\subsubsection{Lattice translation \label{sec:transX}}
As we have briefly addressed at the end of Sec.\ \ref{Sec:DSTI}, in our framework a weak TI is characterized by the set of lattice translations taking a ``$-$'' signature. More specifically, consider a weak TI characterized by the vector $\vec G \equiv \frac{1}{2} \sum_{i=1}^{3} \nu_i \vec G_i$, where $\vec G_i$ denotes a reciprocal lattice vector and $\nu_i = 0,1$ is the associated weak index. The signature of a lattice translation, characterized by the vector $\vec t$, is then given by $e^{-i \vec G \cdot \vec t}$.

Generally, the presence of additional spatial symmetries leads to a restriction on the possible weak indices. For instance, a cubic system cannot realize a weak index which favors a particular axis, say $\nu_1 = \nu_2 = 0$ but $\nu_3=1$. As discussed in Ref.\ \onlinecite{Varjas17}, such restrictions are encoded in the SG constraints on the admissible values of $\vec G$, and originate either from a point-group or non-symmorphic symmetry. As we discuss below, the same problem can be analyzed through a complementary, though equivalent, perspective by studying the symmetry restrictions on the signature assignments to the lattice vectors $\vec t$.

Let $p$ be a symmetry in the SG, and $T_{\vec t}$ a lattice translation\footnote{We use $T_{\vec t}$ to denote the symmetry element in the SG corresponding to a translation by the lattice vector $\vec t$.}. We then see that the signatures of $T_{\vec t}$ and $p \, T_{\vec t} \,  p^{-1}$ are necessarily identical. More generally, a weak index is possible if and only if the corresponding signature assignment to lattice translations, generated by choosing a ``$-$'' signature for $T_{\vec t}$, is symmetric under the described conjugation by any symmetry in the SG. This requirement captures all the restrictions from the point-group actions.

The presence of non-symmorphic symmetries can further restrict the possible weak indices. To see this, note that, for any non-symmorphic symmetry $g$ which is not a 3-fold screw, $g^n$ is a lattice translation for some even integer $n$. Therefore, regardless of the signature chosen for $g$, the translation $g^n$ is always taking a ``$+$'' signature, and hence any weak index demanding a ``$-$'' signature for $g^n$ is forbidden.

These restrictions can be illustrated using the following examples. First, consider the (symmorphic) space group 143 whose only symmetry is a threefold rotation and suppose the lattice vectors in the plane perpendicular to the rotation axis are denoted by $\vec t_1$ and $\vec t_2$. The action of threefold rotation sends $\vec t_1$ to $\vec t_2$ and $\vec t_2$ to $- (\vec t_1 + \vec t_2)$. This means that the translations along $\vec t_{1,2}$ satisfy $C_3^{-1} T_{\vec t_1} C_3 = T_{\vec t_2}$ and $C_3^{-1} T_{\vec t_2} C_3 = T_{\vec t_1}^{-1} T_{\vec t_2}^{-1}$. The first equation implies that translations along $\vec t_1$ and $\vec t_2$ necessarily have the same signature and the second equation implies this signature is necessarily +1, thereby ruling out the possibility of any weak phases for the in-plane translations. The resulting weak classification is $\Z_2$ corresponding to translation along the rotation axis only. 

To illustrate the effect of non-symmorphic symmetries, we consider space group 4 which has a single twofold screw rotation. Since this screw squares to a lattice translation along its axis, it forces translation along this axis to have +1 signature, ruling out the possibility of a weak phase along this direction. The weak classification is then $\Z_2 \times \Z_2$ corresponding to the two independent in-plane translations. Space groups 76 and 167 provide examples in which both restrictions are present simultaneously. The former contains a single fourfold screw rotation, which in addition to ruling out a weak phase for translation along the screw axis also forces the signatures for the two orthogonal in-plane translations to be the same leading to a $\Z_2$ weak classification. The sixfold screw characterizing the latter places even more restrictions by ruling out any weak phase either in the plane (similar to the case of SG 143 considered above) or along the screw axis, leading to the absence of any weak phases.

Extra restrictions on the weak indices may arise due to the type of unit cell e.g. primitive vs body-centered. For example, SG 23 which features three orthogonal intersecting $C_2$ axes in a body-centered unit cell has a $\Z_2$ weak classification due to the fact that all translations can be expressed in terms of a single body-centered translation and the action of $C_2$ rotations on it. We point out here that our analysis is equally applicable to the two special space groups 24 and 199 which are the only non-symmorphic space groups not containing any screws or glides. Both SGs have a body-centered unit cell containing three non-intersecting $C_2$ axes and are thus very similar to SG 23 with a single independent body-centered translation leading to a $\Z_2$ weak classification. 

While we have  discussed the constraints on weak phases utilizing the group structure of the SG, it is instructive to connect it to the more microscopic DSTI model we described. To this end, observe that, if the surface Dirac cone originates from degrees of freedom around a TRIM $\vec k$, the signed representation for the lattice translation $T_{\vec t}$ is given by $ \eta_{T_{\vec t} } = e^{-i \vec k \cdot \vec t} = \pm 1$. According to Eq.\ \eqref{sg}, a DSTI built from strong TIs with effective degrees of freedom coming from the TRIMs $\vec k^{(1)}$ and $\vec k^{(2)}$ would then realize a weak index of $\vec G = \vec k^{(2)} - \vec k^{(1)}$.

\subsubsection{Classification of sTCIs in the 230 space groups}
Having separately described how to extend our point-group results to sTCIs protected by a lattice translation or non-symmorphic symmetry, which does not fix any point in space, we now tackle the problem of classifying all sTCI phases built from stacking strong TIs in all the 230 space groups. 
From the discussion on point groups, we see that the desired classification can be reconciled with the different ways to assign the signature $\pm 1$ to a generating set of the space group, which we denote by $\S_{\rm SG}$.

We first focus on a subgroup $\S_{\rm SG}^{\rm (s)} \leq \S_{\rm SG}$ classifying sTCIs which are ``not weak,'' i.e., those for which the signature $s_{T}=+1$ for all lattice translations $T$.
We  now argue that the sTCIs described by an element of $\S_{\rm SG}^{(s)}$ can be readily studied via $\S_{\rm PG}$. Recall that, given a SG $\G$, a corresponding point group $\G_{\rm p}$ is defined by ``modding out'' the translation part of $\G$, i.e. $\G_{\rm p} \equiv \G/T_\G$, where $T_\G$ is the translation subgroup of $\G$ (which is always a normal subgroup).
This procedure reduces screws/glides in $\G$ to rotations/mirrors in $\G_{\rm p}$,
The classification of sTCIs in any given SG $\G$ is the same as the classification of $\G_{\rm p}$, except for the fact that every mirror in $\G_{\rm p}$ which derive from a glide in $\G$ should be assigned a $\Z_2$ rather than $\Z$ factor. More concretely, we note that, for any consistent signature assignment on $\G$ satisfying $s_T = +1$, any two symmetries in $\G$ with the same point-group action will be given the same signature. This induces a consistent signature assignment on $\G_{\rm p}$. Conversely, for any consistent assignment on $\G_{\rm p}$ one can define an assignment on $\G$ by composing with the canonical projection $\G \rightarrow \G_{\rm p}$. This demonstrates the stated one-to-one correspondence, up to the modification required for mirror vs.\ glide\footnote{We remark that, in the case that a mirror and a glide both exist in a given SG, they will always be independent, since a mirror plane and a glide plane can never be related by a space group symmetry. This means that the procedure described above is well-defined, and does not depend on the choice of the generator set (which satisfies the conditions of Sec.~\ref{Sec:Pointgroups}). }.

Next, we incorporate weak phases into our discussion. Recall that the computation of $\S_{\rm SG}$ amounts to the identification of a minimal generating set of the SG, paying a special attention to the presence of mirror and three-fold rotation symmetries (conditions outlined in Sec.\ \ref{Sec:Pointgroups}). 
Consider a  weak index, generated by assigning a ``$-$'' signature to a lattice vector $\vec t$, which satisfies all the previously outlined SG constraints.
This implies the signature of $\vec t$ is not fixed by the part of indices contained in $\S_{\rm SG}^{\rm (s)}$, and therefore $\vec t$ must must be incorporated into the generating set for the SG. In addition, if the SG possesses a mirror $m$ about a plane normal to $\vec t$, the mentioned weak TI is  ``promoted'' to a weak mirror Chern insulator, i.e., the classification is modified from $\Z_2$ to $\Z$.
When this happens, we should append the shifted mirror, $m\, T_{\vec t}$,  instead of the translation, $T_{\vec t}$, to our generator set.
Following this procedure, one incorporates all the needed lattice translations or shifted mirrors into the generator set. Each of such additional generator for the SG then corresponds to either a $\Z_2$-valued weak TI index or a $\Z$-valued weak mirror-Chern index, which we append to $\S_{\rm SG}^{\rm (s)}$ to arrive at the full classification  $\S_{\rm SG}^{\rm (s)}$.
Implementing this rule leads to Table \ref{tab:SurfaceXBS_Full} in Appendix \ref{app:XBSRep}, which provides the sTCI classification $\S_{\rm SG}$ for all the 230 SGs.

Finally, we comment on the meaning of adding two phases in $\S_{\rm SG}$, which is an abelian group representing the equivalence classes of distinct sTCI phases. The subtlety arises if there is no geometry for which both phases exhibit anomalous surface states. This will never be the case for SGs containing only symmorphic elements, but is possible in the presence of non-symmorphic symmetries. However, we have to remember elements of $\S_{\rm SG}$ are distinct {\it bulk} phases despite being physically defined by their surface signatures. This means that the possibility of anomalous states on any given surface is completely fixed by the bulk, as will be discussed in detail in Sec.~\ref{Sec:BulkSurface}. Moreover, it is always possible to distinguish two distinct sTCI phases by considering them on many different geometries. For instance, given two phases 1 and 2 and two geometries $G_{1}$ and $G_2$, such that 1 (2) exhibits surface states on $G_{1}$ ($G_{2}$) but not on $G_{2}$ ($G_{1}$), we can distinguish the sum of the two phases from either phase by the fact that it exhibits surfaces states on {\it both} geometries $G_1$ and $G_2$.

\subsection{Surface dispersion at special planes}
\label{Sec:Dispersion}
Although the 1D surface modes shown in Fig.~\ref{PG} can be, in principle, detected by means of transport measurements, the most experimentally accessible tool to investigate such surface states is provided by the angle-resolved photoemission spectroscopy (ARPES), which probes the energy dispersion at a given surface. The dispersion measured by ARPES can be readily understood from the surface state pattern given in real space in Fig.~\ref{PG}. In brief, we consider what happens when we attach a tangent plane to the sphere at a given point, i.e., when we consider a flat, macroscopic crystal facet with the same surface normal as the considered point on the sphere.

\begin{figure}
\center
\includegraphics[width=0.7\columnwidth]{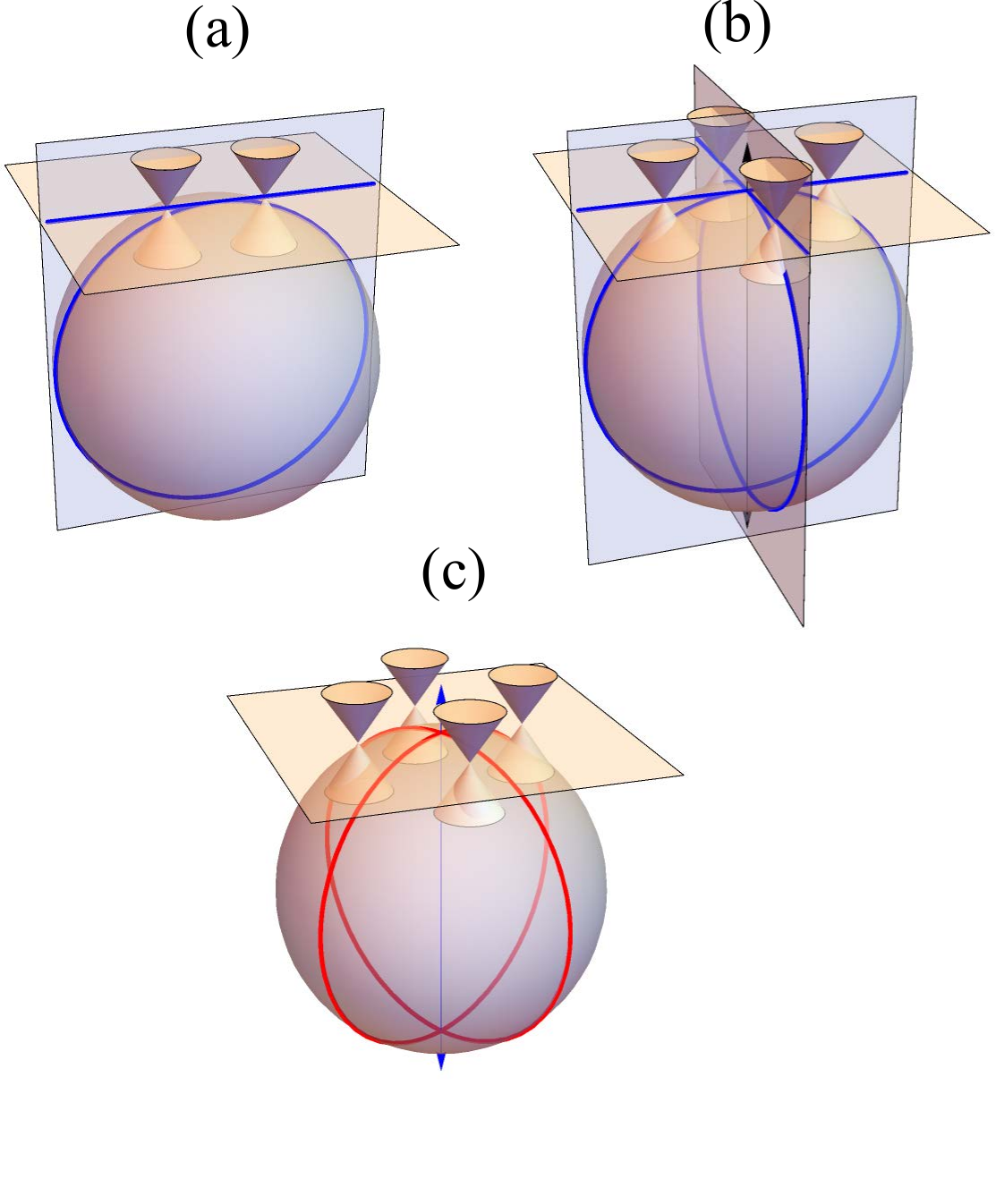}
\caption{Illustration of the surface dispersion in planes tangent to the sphere whose normal is (a) in a single mirror plane, (b) in two mirror planes, or (c) parallel to a 4-fold rotation axis.}
\label{Planes}
\end{figure}

We note that the surface modes can generally be freely moved on the surface except when they pass through a rotation-invariant point or lie in a mirror/glide plane. As a result, a tangent plane at a point that is not rotation-invariant or lies in a mirror/glide plane will generically have gapped dispersion. Therefore, the analysis of surfaces where 2D gapless modes may exist boils down to considering the cases where the normal to the surface (i) lies in a single mirror/glide plane, (ii) lies at the intersection of multiple mirror/glide planes, or (iii) is parallel to a rotation axis. These three cases are illustrated in Fig.~\ref{Planes}, and we elaborate on them below.

\subsubsection{Single mirror/glide plane}
Here, we investigate the dispersion in a tangent plane whose normal lies in a single mirror plane.  The dispersion in the plane close to the $\Gamma$ point can be written as
\beq
\label{HS}
h_\bk = k_y \sigma_x - k_x \sigma_y 
\eeq
with reflection acting as $M_x \cdot h_{k_x,k_y} \equiv \sigma_x h_{k_x,k_y} \sigma_x=   h_{-k_x,k_y} $. Reflection symmetry implies that any mass term has the form $m_\bk \sigma_z  \otimes \Gamma$, which necessarily vanishes at $k_x=0$ since it satisfies $m_{-k_x,k_y} = -m_{k_x,k_y}$. It is, therefore, not possible to gap-out the surface but it is possible to move the two Dirac cones apart by adding the $\T$-symmetric term $m_1 \sigma_x \tau_y$, which shifts the Dirac cones to $\bk = (0,\pm m_1)$ \cite{Hsieh12}. The glide case can be considered very similarly. A surface glide symmetry has the form $M_x e^{i \bk \cdot \vec t}$ for some fractional translation $\vec t$. Thus, close to the $\Gamma$ point, mirror and glide act in the same way.

\subsubsection{Rotation axis}
This case was considered in Ref.~\onlinecite{Fang17}. We start with the surface Hamiltonian (\ref{HS}), with 2-,4- or 6-fold rotation acting as $C_n \cdot h_\bk \equiv  e^{i \frac{\pi}{n} \sigma_z} \tau_z h_\bk \tau_z e^{-i \frac{\pi}{n} \sigma_z} =   h_{C_n \bk}  $. The only mass term consistent with $C_n$ and time-reversal symmetry is $m_\bk \sigma_z \tau_x$ for $n=2, 6$ or $m_\bk \sigma_z \tau_y$ for $n=4$, both satisfying $m_{C_n \bk} = -m_{\bk}$. This, in particular, implies that the mass $m_\bk$ vanishes at $\bk=(0,0)$, so the Dirac cones cannot be gapped. We can, however, add the symmetry-allowed term $m_1 \tau_z$, which will move the Dirac cones away from $\bk=(0,0)$, resulting in $n$ Dirac cones related by the $n$-fold rotation. To see this, let us consider the Hamiltonian (\ref{HS}) (without any mass term), which just describes two copies of a gapless Dirac Hamiltonian. Adding the term $m_1 \tau_z$ shifts the two cones in energy by $2m_1$ so that they intersect at a ring at zero energy. The Hamiltonian of this nodal ring is given by
\beq
\label{Hring}
h_\bk = (2m_1 - k_x^2 - k_y^2) \mu_z
\eeq
where $\mu_{x,y,z}$ denote the Pauli matrices in the space of the two bands intersecting at zero (the lower band from the Dirac cone shifted upwards and the upper band from the one shifted downwards). Adding the mass term $m_\bk \sigma_z \tau_x$ for $n=2, 6$ or $m_\bk \sigma_z \tau_y$ for $n=4$ and projecting it to the $\mu$ basis, we find that it has the form $m_\bk \mu_{x}$ or $m_\bk \mu_{y}$, which will gap-out the nodal ring except at $n$ points where it necessarily vanishes due to $m_{C_n \bk} = -m_\bk$.

\subsubsection{Multiple mirror/glide planes}
If the normal to the tangent plane lies in the intersection of multiple mirror planes mirror, all in the ``$-$'' representation, we can repeat from the case of a single mirror for each mirror separately and conclude that any $\T$-symmetric mass term will necessarily vanish at $\bk = (0,0)$. Similar to the case of rotation, this does not necessarily imply that we cannot move the Dirac cones away from $(0,0)$, since we can add the term $m_1 \tau_z$ to shift them in energy and get a nodal ring given by the Hamiltonian (\ref{Hring}) which can be gapped by adding the mass term $m_\bk \sigma_z \tau_x$ or $m_\bk \sigma_z \tau_y$ (whose projection in the $\mu$ basis is $m_\bk \mu_x$ or $m_\bk \mu_y$). This mass term vanishes at all mirror invariant lines due to the condition $m_{M \cdot \bk} = -m_{\bk}$, e.g., for two perpendicular mirror $M_x$ and $M_y$, we get two Dirac cones at $(0,\pm k_1)$ and $(\pm k_1,0)$. The case for multiple glides or a glide and a mirror is very similar since the action of glide at the $\Gamma$ point is identical to the action of mirror. We note here that our results agree with those of Ref.~\onlinecite{Wieder17} which showed we can only pin two Dirac points in the vicinity of a TRIM other than $\Gamma$ using two perpendicular glides.

Note that, despite the similarity between hinge phases protected by rotation and screw symmetries, without fine tuning the latter are not detectable by studying the dispersion at any plane on the surface. In this sense, they are similar to hinge states protected by inversion or rotoinversion in that they can only be observed if the surface is considered as a whole.

\section{Bulk-surface correspondence}
\label{Sec:BulkSurface}
In Sec.\ \ref{Sec:Indicators}, we have discussed how the symmetry-based indicators for time-reversal symmetric systems with strong spin-orbit coupling can mostly be reconciled with either a $\mathbb Z_2$ TI index or a mirror Chern number. However, as  listed in Table~\ref{tab:XBS}, there are new invariants which indicate that certain stacks of strong TIs remain nontrivial despite the lack of a conventional index.
These unconventional phases are expected to feature hinge-like anomalous surface states, as is discussed for specific cases in Refs.\ \onlinecite{Song17, Schindler17, Langbehn17,  Benalcazar17Aug, Fang17} and analyzed thoroughly in Sec.\ \ref{Sec:Surface}. In this section, we tie the bulk and surface perspectives  together and describe a concrete bulk-surface correspondence.

This is achieved by making an explicit connection between the specification of bulk symmetry representations and the surface mass-term analysis. In particular, we will show that all $X_{\rm BS}$-nontrivial phases have anomalous surface states (with suitably chosen sample geometry). However, the $X_{\rm BS}$ diagnosis and our surface states classification are not generally in one-to-one correspondence; rather, for most SG, different sTCIs become indistinguishable when one focuses only on symmetry representations, and therefore the same $X_{\rm BS}$ class is identified with multiple patterns of surface states. We will discuss the general structure behind such identification ambiguity, and in Table \ref{tab:SurfaceXBS_Full} in Appendix \ref{app:XBSRep} we provide the corresponding results for all SGs.

As the relation between $X^{\rm (w)}_{\rm BS}$ and the weak TI or mirror Chern indices has already been addressed in Sec.\ \ref{Sec:Indicators}, in the following we focus on the relation between $X^{\rm (s)}_{\rm BS}$ and $S^{\rm (s)}_{\rm SG}$.
To proceed, we make a simplifying assumption: we suppose the topological properties of the strong TI of interest can always be analyzed through a $\vec k \cdot \vec p$ (bulk) Dirac Hamiltonian about $\Gamma$ \cite{Molenkamp, Schnyder09, Ryu10, Witten16}:
\begin{equation}\begin{split}\label{eq:DiracH}
\mathcal H_{\vec k } = M \gamma_z +  \vec k \cdot \vec \sigma \, \gamma_x,
\end{split}\end{equation}
where we assume the mass term changes from $m<0$ in the bulk to $m>0$ outside the system, leading to a Dirac point localized to the surface where $m=0$ (Appendix \ref{app:surfaceD}).
Here, $\gamma$ is a set of Pauli matrices describing the orbital degrees of freedom.

We start by considering the action of a spatial symmetry $g$, defined as in Sec.~\ref{Sec:DSTI}, on the bulk Dirac Hamiltonian (\ref{eq:DiracH}), given by $g \cdot \H_{\bk} \equiv U_g \H_{\bk} U^\dagger_g  =  \H_{R_g \vec k} $. Here, $U_g \equiv u_g \oplus (\det R_g u_g )$, with $u_g$ defined as in  Sec.~\ref{Sec:DSTI}. Next, we connect the bulk characterization with the surface theory. As detailed in Appendix \ref{app:surfaceD}, for a surface with normal $\vechat n_{\vec r}$, the surface Hamiltonian is given by $h_{\vec r, \vec k} = (\vec k \times \vechat n_{\vec r})\cdot \vec \sigma$, which transforms under symmetry as $g \cdot h_{\br,\bk} \equiv u_g h_{\br,\bk} u_g^\dagger =  h_{g\cdot \vec r, R_g \vec k} $. Recall that, in Sec.\ \ref{Sec:Surface}, we showed that when two strong TIs are stacked together, the pattern of surface modes (if any) is determined by the signature $s_g$ of the symmetry representation, which can be expressed in terms of $\det R_g$ and the relative sign of the symmetry representation matrices across the two copies (cf.~Eq.~\ref{sg}). This means that, by specifying the (signed) symmetry representation matrices $u_g$ of the two TIs {\it in the bulk}, we can predict the pattern of the surface modes, thereby establishing the anticipated bulk-surface correspondence.

Such correspondence, however, utilizes more information than that available from the symmetry representations alone. To make connection with $X_{\rm BS}$, the symmetry-based indicators of band topology \cite{Po17}, we have to study what information is lost when we focus only on the symmetry representations. 
In other words, generally the identification of the surface states associated with a nontrivial class $[\vec b] \in X_{\rm BS}$ is not unique. 
Such nonuniqueness has two origins: either the choices  $\eta_g =\pm1$ correspond to the same representation, or certain nontrivial stacking patterns of strong TIs have symmetry representations compatible with an atomic insulator, and hence evades the $X_{\rm BS}$ diagnosis.

We illustrate these ideas from two concrete examples. Consider the space group $P2$ (SG 3), which is generated by lattice translation and a $C_2$ rotation about the $z$-axis. As discussed in Ref.\ \onlinecite{Fang17}, when we stack two strong TIs with opposite helicities, described respectively by the surface Hamiltonians $h_{\pm} =\pm k_y \sigma_x  -k_x \sigma_y $ and with $C_2$ represented by $u_{\pm, C_2} = i \sigma_z$, the resulting DSTI is nontrivial and features hinge modes when subjected to the $C_2$ symmetric, spherical open boundary conditions in Sec.\ \ref{Sec:Surface}. 
To reconcile with the analysis based on Eq.\ \eqref{eq:DiracH}, we can perform a basis rotation 
\begin{equation}\begin{split}\label{eq:}
\left\{
\begin{array}{rl}
h_- \mapsto&  \sigma_y ( - k_y \sigma_x  - k_x \sigma_y )  \sigma_y = k_y \sigma_x  - k_x \sigma_y  \\
u_{-, C_2}  \mapsto & \sigma_y (i \sigma_z)  \sigma_y =  - i \sigma_z
\end{array}
\right.,
\end{split}\end{equation}
which implies $\eta_{\pm, C_2} = \pm 1$ in our present framework. Eq.\ \eqref{sg} then correctly predicts that the DSTI $h_{+} \oplus h_-$ features anomalous surface states. However, as $\text{tr}(\pm i \sigma_z)=0$, the distinction between $h_{+} \oplus h_-$ and the trivial phase $h_{+} \oplus h_+$ cannot be diagnosed from symmetry representations alone. This is consistent with the result $X_{\rm BS} = 0$ found in Ref.\ \onlinecite{Po17} for $P2$ (SG 3), see also Table \ref{tab:XBS}.

As a second example, we consider $P4/m$ (SG 83) generated by the lattice translations and the point-group $C_{4 h}$. In Ref.\ \onlinecite{Po17}, it was found that the strong TI generates the subgroup $\mathbb Z_{8} < X_{\rm BS}$, and in the following we chart out the surface states consistent with the DSTIs, which are described by the even subgroup in $\mathbb Z_8$.
As $C_{4 h}$ is generated by the four-fold rotation $C_4$ and inversion $\mathcal I$, both of them having non-vanishing trace, the nonuniqueness in the surface signature of $X_{\rm BS}$ is more subtle. 
Recall from Sec.\ \ref{Sec:Surface} that, for $C_{4 h }$, the surface classification is given by $\mathbb Z \times \mathbb Z_2$, where the $\mathbb Z$ factor is generated by a DSTI with a mirror Chern mode protected by the mirror $M_z = \mathcal I C_4^2$, and the $\mathbb Z_2$ factor generated by a DSTI with only hinge modes. Let $\mathcal H_{ \eta_{\mathcal I}, \eta_{C_4}}$ be a  strong TI described by Eq.\ \eqref{eq:DiracH}.
From Eq.\ \eqref{sg}, we see that $\mathcal H_{-,-} \oplus \mathcal H_{-,-}$ has surface states described by $(1,0) \in \mathbb Z \times \mathbb Z_2$, whereas the surface of $\mathcal H_{-,-} \oplus \mathcal H_{-,+}$ is described by $(1,1)$. Note that, here, the helicity of the mirror Chern mode is fixed by $\eta_{M_Z} = \eta_{\mathcal I} \eta_{C_4}^2 = \eta_{\mathcal I}$.

Now, consider stacking four strong TIs, $\mathcal H_K = \mathcal H_{-,-} \oplus \mathcal H_{-,-} \oplus \mathcal H_{-,-} \oplus \mathcal H_{-,+}$. The surface remains nontrivial, and is classified by $(1,0)+(1,1)=( 2, 1)\in \mathbb Z \times \mathbb Z_2$, i.e., it features two helical modes at the equator, protected by $M_z$, together with the $C_4$-hinge mode. However, the symmetry representation of $\mathcal H_K$ happens to coincide with an atomic insulator (Appendix \ref{app:XBSRep}), and therefore the nontriviality of $\mathcal H_K$ is undetected from symmetry eigenvalues alone, i.e., $\mathcal H_K$ belongs to the trivial element in $X_{\rm BS}$. In other words, $X_{\rm BS}$ could at best resolve the surface signature in $\mathbb Z \times \mathbb Z_2$ modulo the subgroup generated by $\langle (2,1) \rangle$. From the line of arguments presented in Appendix \ref{app:XBSRep}, one can further show that $\langle (2,1) \rangle$ precisely captures the nonuniqueness in identifying the surface states associated with any $X_{\rm BS}$ class. For our present problem, a band insulator described by $2$ in the strong factor $\mathbb Z_8$ can have surface states described by, for instance, either $(1,0)$ or $(1,0) - (2,1) = (-1,1)$. Now, if we can stack these two systems together, we arrive at a bulk with symmetry indicator $4 \in \mathbb Z_8$ and a surface described by $(1,0) + (-1,1) = (0,1) \in \mathbb Z \times \mathbb Z_2$, which corresponds to a system with only hinge modes but not equatorial mirror Chern modes. This can be reconciled with the more detailed analysis in Ref.\ \onlinecite{Song17} and Sec.\ \ref{Sec:Surface}.

By a similar analysis, one can map out the surface-state ambiguity of $X_{\rm BS}^{\rm (s)}$ for any space group. We perform this study for the key space groups listed in Table \ref{tab:XBS}, and the results are tabulated in Table \ref{tab:XBS_BS}. In particular, by identifying the surface signature of the minimal DSTI in any SG, we see explicitly that a $X_{\rm BS}^{\rm (s)}$-nontrivial phase is either a strong TI or a sTCI. Together with the conventional indices we identified for the weak factors, this proves that all $X_{\rm BS}$-nontrivial phases possess anomalous surface states.

\begin{table}[h]
\begin{center}
\caption{{\bf Bulk-surface correspondence for the key space groups} $X^{\rm (s)}_{\rm BS}$ denotes the symmetry-based indicator generated by a strong topological insulator; $\mathcal S$ denotes the surface classification; 
$\zeta_0$ denotes the generators of the subgroup of $\mathcal S$ which is consistent with the trivial entry in  $X^{\rm (s)}_{\rm BS}$; $\zeta_2$ indicates representative surface signatures for the minimal doubled strong topological insulator, corresponding to the entry $2 \in X^{\rm (s)}_{\rm BS}$
\label{tab:XBS_BS}}
\begin{tabular}{c|c|c|c|c}\hline \hline
SG & $X_{\rm BS}^{\rm (s)}$ & $\mathcal S$ &  $\zeta_0 \in \mathcal S$ & $ \zeta_2 \in  \mathcal S$\\
\hline
{2}  ($P\bar{1}$) & $\mathbb Z_4$ & $\mathbb Z_2 $& $( 0)$ & $(1) $
\\
{81} ($P\bar{4}$) & $\mathbb Z_2$ & $\mathbb Z_2$ & $( 1)$ & $(0) \simeq (1)$
\\
{82}  ($I\bar{4}$) & $\mathbb Z_2$ & $\mathbb Z_2$& $( 1)$ & $(0) \simeq (1)$ 
\\
{83}  ($P4/m$) & $\mathbb Z_8$ & $\mathbb Z \times \mathbb Z_2 $ & $ (2,1) $ & $(1,0) \simeq (-1,1)$
\\
{87} ($I4/m$) & $\mathbb Z_8$ &  $\mathbb Z \times \mathbb Z_2 $ & $ (2,1) $ & $(1,0) \simeq (-1,1)$
\\
{174} ($P\bar 6$) & $\mathbb Z_{3}$
\footnote{
The same entry can correspond to either a strong or doubled strong topological insulator; the surface classification $\S$ is only applicable to the latter} & $\mathbb Z $ & $(3) $ & $(1) \simeq (-2)$
\\
{175} ($P6/m$) & $\mathbb Z_{12}$ & $\mathbb Z \times \mathbb Z_2 $ & $ (3,1) $ & $(1,0) \simeq (-2,1)$
\\
{176} ($P6_3/m$)\footnote{Nonsymmorphic; surface states on cylinder geometry periodic in the direction parallel to the screw axis} & $\mathbb Z_{12}$ & $\mathbb Z \times \mathbb Z_2 $ & $ (3,1)  $ & $(1,0) \simeq (-2,1)$
\\
\hline \hline
\end{tabular}
\end{center}
\end{table}

Finally, we comment on the generality of the present analysis, which relies on the Dirac model Eq.\ \eqref{eq:DiracH}.
In the above, we have restricted our attention to symmetry representations of the form $ u_g =\eta_g v_g$. This does not represent the general case, since, for instance, the valence bands at a TRIM could furnish 4D irreps \cite{Bradley}. 
Alternatively, certain symmetry representations lead to surface Dirac cones dispersing as $\sim k^3$ \cite{Fang17}, and therefore also falling outside of the Dirac description.
Strictly speaking, the Dirac Hamiltonian approach does not immediately inform the surface properties of such strong TIs. However, from the analysis discussed in Appendix \ref{app:XBSRep}, we found that, for any class in the strong factor $X_{\rm BS}^{\rm (s)}$, one can choose a representative whose symmetry representations agree with some atomic insulator everywhere in the Brillouin zone, except for the exchange of some irreps of the form $\eta_g v_g$ at $\Gamma$, i.e., it can be viewed as a stack of strong TIs  satisfying the simplifying assumptions adopted in the present analysis. In view of this, it is likely that our results are sufficient for establishing the essential connection between the $X_{\rm BS}^{\rm (s)}$ diagnosis and the (stable) surface signatures of DSTIs, but we will leave a more elaborated justification of this completeness conjecture to future works.

\section{Discussion}
\label{Sec:Discussion}

\begin{figure}
\center
\includegraphics[width=0.35\columnwidth]{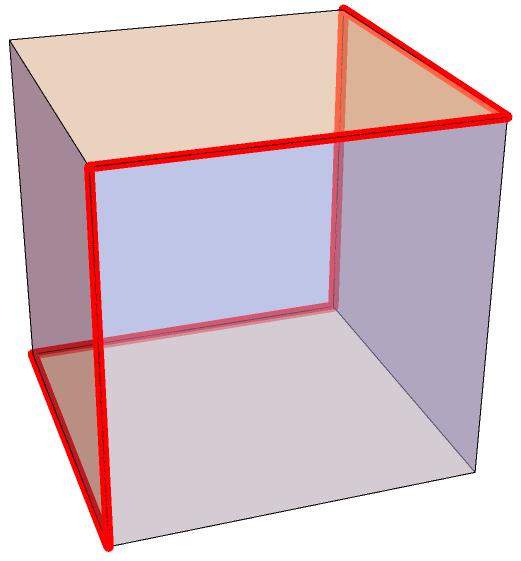}
\caption{Surface ``hinge'' mode protected by inversion symmetry on a cubic geometry, assuming no other spatial symmetries are present. The mode is confined to the edges of the cube since the surface mass term cannot change on a flat face.}
\label{Cube}
\end{figure}

In closing, we make several remarks concerning the connection between our present work and other recent ideas on the study of topological phases protected by crystalline symmetries.

We first note that the ``hinge'' surface states discussed in Sec.~\ref{Sec:Surface} are expected to be robust against weak perturbations which break their protecting crystalline symmetries, as suggested by Ref.~\onlinecite{Langbehn17}. The reason is that the local stability of the hinge modes relies only on TRS. As a result, the only way to remove them, assuming TRS is preserved, is to deform them to a point, which is only possible if the symmetry-breaking perturbation is large enough (compared to the scale of the surface energy gap). We also note that, when placed on a surface with planar faces e.g. a cube, the hinge modes are expected to be localized to the edges of the sample as shown in Fig.~\ref{Cube}, since the mass term is expected to remain constant on any flat surface.

\begin{figure}[h]
\begin{center}
{\includegraphics[width=0.45 \textwidth]{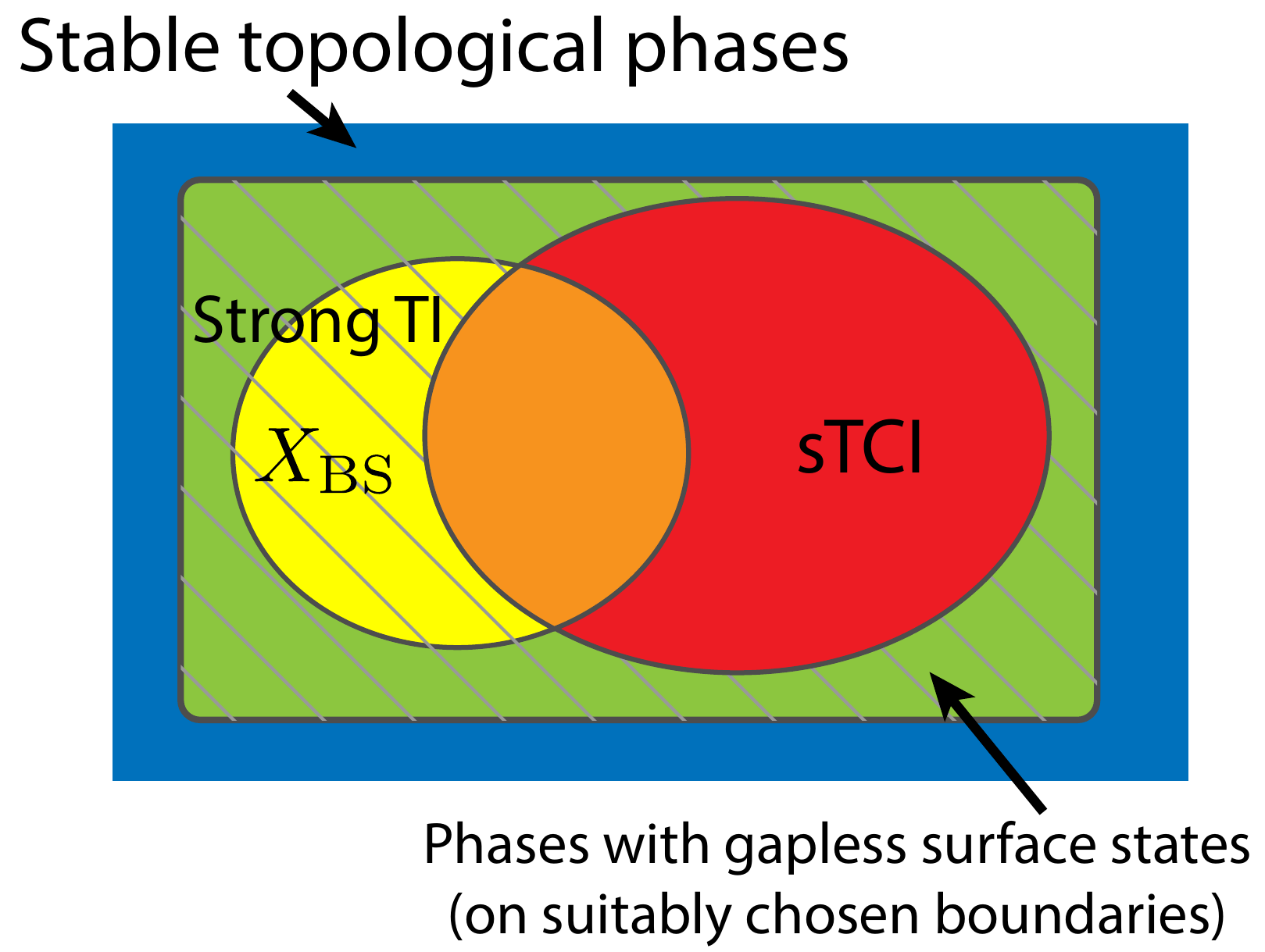}} 
\caption{
Hierarchy of crystalline topological insulators with time-reversal symmetry and strong spin-orbit coupling (class AII) in 3D. All nontrivial entries in the symmetry indicator, $X_{\rm BS}$, are compatible with gapped topological band structures with gapless surface states on suitably chosen sample geometries.
Excluding the strong topological insulators (TIs) from the $X_{\rm BS}$-nontrivial phases, we found that all the remaining phases are anomalous surface topological crystalline insulators (sTCIs), defined as nontrivial insulators with anomalous surface states but can be trivialized upon the breaking of some crystalline symmetries. sTCIs include all gapped topological phases with gapless surface states but are not strong TIs (hatched region).
These also include phases not diagnosed by $X_{\rm BS}$, i.e., the topological nature is not exposed by the symmetry eigenvalues in the bulk.
\label{fig:Venn}
 }
\end{center}
\end{figure}

It is worth reiterating that our classification results rely on a conjectured correspondence between sTCIs in a given space group and the signatures of the elements of this group. The resulting classification captures all known sTCIs in addition to predicting many new ones. Our results imply that the set of sTCIs strictly contains the set of phases which are $X_{\rm BS}$-nontrivial {\it but not} strong TIs (Fig.\ \ref{fig:Venn}). That is to say, all the non-trivial $X_{\rm BS}$ phases exhibit anomalous gapless surface states, which are either of the form of a surface Dirac cone, as for strong TIs, or are described by our sTCI surface classification. The set of sTCIs also strictly contains the set of conventional TCI phases with 2D surface Dirac cones \cite{Hsieh12, Tanaka12,   Wang16, Ezawa16, Varjas17, Wieder17,  Fang17}.
sTCIs without such 2D gapless surface states then feature only 1D helical hinge modes protected by inversion, rotoinversion or screw symmetries, whose existence can only be observed by considering the surface as a whole.

Next, we make connections between our results and more general ideas concerning crystalline SPT classifications.
Weak TIs, which can be understood in terms of spatially stacking 2D TIs using lattice translation symmetry,  exemplify the notion of  topological crystalline insulators. Curiously, on more general grounds it has been suggested that many, if not all, crystalline symmetry-protected topological phases can be understood in terms of a similar construction, upon the replacement of lattice translation by other crystalline symmetries \cite{Varjas17, Thorngren16,Fang17,Hermele17, Huang17,XiongGlide}. Here, we remark that a similar construction is also applicable to the DSTIs. To illustrate this, consider a space group with only lattice translation and a mirror symmetry $z \leftrightarrow - z$. The $z \in  \mathbb Z$ and $z\in (1/2 + \mathbb Z)$ planes are mirror-invariant, and if we pin a quantum spin Hall insulator to each of these planes, we will arrive at a band insulator which does not immediately yield an atomic limit. However, such an insulator does not possess a weak index; rather, we should view it as interlacing two weak TIs built from stacking 2D TIs with mirror-Chern number $+1$, one from pinning them to the $z=0$ plane and the other at the $z=1/2$ planes. Computing the mirror Chern numbers, one finds that the described band insulator has mirror Chern number $2$ on the $k_z=0$ plane and $0$ on the $k_z = \pi$ plane, i.e., it can also be understood as a DSTI. Schematically, we may write ``DSTI = weak-$0$ + weak-$1/2$.''

The above picture is, in fact, quite general: it applies to a large class of DSTIs arising in settings where the real-space deformation of ``weak-$0$'' to ``weak-${1/2}$'' is forbidden by symmetry, as in the case when inversion \cite{Po17} and/ or nonsymmorphic \cite{Wang16, Ezawa16, Varjas17, XiongGlide} symmetries are present. In particular, we note that the glide-protected TCI showcasing the ``hourglass fermion'' surface states also falls into this category \cite{Wang16, Ezawa16, XiongGlide}. 

It is also conceptually revealing to abstract our sTCI surface classification from the Dirac equation approach we have employed. Let $P$ be the group of spatial symmetries respected by the open boundary condition, e.g., in the filled-sphere geometry $P$ is the subgroup of the point-group which leaves the origin invariant. 
In essence, for every spatial symmetry $g \in P$ we assign a signature $s_g = \pm 1$, which conforms to the group multiplication $s_{g g'} = s_g s_{g'}$. 
Different assignment of signature can then be viewed simply as different homomorphisms from $P \rightarrow \mathbb Z_2$, which are classified by the first cohomology group of $P$ with $\mathbb Z_2$ coefficients, $H^{1}(P, \mathbb Z_2)$.

This interpretation is similar to the dimension-reduction approach advocated in Refs.\ \onlinecite{Hermele17,Huang17}, except that, due to the presence of additional internal symmetries (time-reversal for the case of DSTI), we do not fully reduce the classification to the symmetry charges at the point fixed by $P$; rather, we picture the construction of 3D phase by interlacing 2D SPTs \cite{Fang17}, and therefore the 2D classification ($\mathbb Z_2$ for class AII) enters into the coefficient of $H^1$. One can imagine a similar analysis for DSTI-like phases in a more general setting, where we replace $\mathbb Z_2$ by the appropriate lower-dimensional classification $X$ (an Abelian group), and incorporate a group action of $P$ on $X$ (orientation reversing elements in $P$ reverses the phase in $X$). We caution that, however, this provides only a partial understanding on the full classification. For instance, in the presence of mirror symmetries, the DSTIs can feature $\mathbb Z$-valued mirror Chern modes (neglecting interactions) at the surface, whereas the $H^1$ description captures only the parity of the number of modes. Such discrepancy arises because the mirror symmetry is effectively reduced to an onsite unitary symmetry on the mirror-invariant plane, which enriches the classification \cite{Hermele17, Huang17}. We leave the analysis of the general structure of such classifications as an interesting open problem.

Lastly, we make a few remarks about the spin-orbit-coupling-free case (class AI), whose indicators were also obtained in Ref.~\onlinecite{Po17}. 
In the absence of spin-orbit coupling, a new possibility that may be mandated by a nontrivial indicator is a gapless (semimetallic) phase (dubbed representation-enforced semimetals in Ref.\ \onlinecite{Po17}) \cite{KaneMele1, KaneMele2, Kim15, Yu15, Bian16}, and we provide an explicit example of such this scenario in the following.

An example of a semimetallic phase diagnosed by $X_{\rm BS}$ in class AI is provided by SG $P\bar 1$ (No.\ 2), where the only point-group symmetry is inversion, whose ``strong factor'' is $\Z_4$. The generator of such a factor has the same inversion eigenvalues at the TRIMs as the strong TI, and it can be thought of as the limit of a strong TI when the spin-orbit coupling is adiabatically switched off. The corresponding phase is a Dirac nodal ring semimetal with so-called drumhead surface states \cite{Kim15, Yu15, Bian16}. Its low-energy physics is captured by the effective spinless two-band Hamiltonian $\H_\bk = k_z \gamma_y + (k_x^2 + k_y^2 - k_0^2) \gamma_z$, which describes a nodal ring centered at $\bk = 0$ with radius $k_0$. Here, $\gamma$ denotes the Pauli matrices describing the orbital degree of freedom. $\H_\bk$ is invariant under spinless TRS, represented by complex conjugation, and inversion symmetry, represented by $\gamma_z$. A system constructed by stacking two copies of this Hamiltonian can be gapped out in the bulk by a mass term of the form $m_\bk \gamma_x \tau_y$. Inversion, however, would require $m_\bk$ to satisfy $m_{-\bk} = -m_\bk$, which implies that the mass has to vanish at least twice along the nodal ring, leaving behind two bulk Dirac nodes at generic, but inversion-related, momenta. This pair of Dirac nodes cannot be removed without breaking the symmetries \cite{ Ari, Hughes11} or changing the symmetry class (say by introducing spin-orbit coupling), and hence we conclude the $2 \in \Z_4$ entry in the $X_{\rm BS}$ diagnosis is enforced to be semimetallic by the symmetry representations. 
We will leave the general question of whether any of the $X_{\rm BS}$ entries for any space group realizes a gapped topological phase in class AI (with or without surface states) for future works.

\begin{acknowledgements}
The authors would like to acknowledge stimulating discussions with P.~W.~Brouwer, M.~Geier, A.~P.~Schnyder. 
In particular, we thank R.~Quieroz for helpful discussions on surface states protected by multiple mirrors and/or glides.
We also thank C.~Fang for informing us of their recent works \cite{SongFang17, FangSong17}, which also study the physical interpretation of phases with nontrivial symmetry indicators.  AV and HCP were supported by NSF DMR-1411343. AV acknowledges support from a Simons Investigator Award and ARO MURI program W911NF-12-1-0461.
HW acknowledges support from JSPS KAKENHI Grant Number JP17K17678.
\end{acknowledgements}

\appendix

\section{Relation between $\kappa_4$ and $\nu_0$}
\label{appk4n0}
In order to establish the connection between $\kappa_4$ and $\nu_0$, let us focus on the primitive lattice case (the body-centered case can be discussed the same way) and introduce the 1D Berry phase for each occupied band:
\begin{equation}
z_{(k_x,k_y)}\equiv\frac{1}{2\pi i}\int_{-\pi}^{\pi} dk_z u_{(k_x,k_y,k_z)}^*\partial_{k_z}u_{(k_x,k_y,k_z)},
\end{equation}
which is only well-defined modulo 1.  The $S_4$ symmetry imposes the relation $z_{(k_x,k_y)}=-z_{(-k_y,k_x)}$. This, in particular, implies that $z_{(k_x,k_y)}$ is quantized to either $0$ or $1/2$ at $(k_x,k_y)=(0,0)$ and $(\pi,\pi)$.  If (and only if) the two quantized values do not agree, the Berry phase $z_{(k_x,k_y)}$ has an odd winding as the 2D momentum changes along the loop $(0,0)\rightarrow(\pi,0)\rightarrow(\pi,\pi)\rightarrow(0,\pi)\rightarrow(0,0)$. This nontrivial winding indicates the strong index $(-1)^{\nu_0}=-1$ \cite{Soluyanov11}. On the other hand, the quantized value of $z_{(k_x,k_y)}$ at $(0,0)$ and $(\pi,\pi)$ can be diagnosed by the ratio of $S_4$-eigenvalues at $k_z=0$ and $k_z=\pi$, suggesting that $(-1)^{\nu_0}$ can be basically given by the product of $S_4$-eigenvalues of occupied bands over all high-symmetry momenta in $K_4$.  However, in this product form, the $S_4$-eigenvalues have to be restricted to those squaring into $+i$ (or, equivalently, $-i$) to avoid double-counting of TR pairs, just as in the case of the original Fu-Kane formula. All in all, we find
\begin{equation}
(-1)^{\nu_0}=(-1)^{\frac{n}{2}}\prod_{K}\prod_{\alpha=1,5}[e^{\frac{\alpha\pi}{4}i}]^{n_K^\alpha}=(-1)^{\kappa_4}.
\end{equation}
Again, $n=\sum_{\alpha=1,3,5,7}n_K^\alpha$ is the total number of occupied bands. The last equality can be verified from the definition Eq.~\eqref{defk4}, in the same way as we did for $\kappa_1$.

\section{Surface theory \label{app:surfaceD}}
In this appendix, we elaborate on the derivation of the surface theory starting from the bulk Dirac Hamiltonian in Eq.\ \eqref{eq:DiracH}, with the aim of explicitly connecting the symmetry properties of the two.

Recall the Dirac Hamiltonian in Eq.\ \eqref{eq:DiracH},
\begin{equation}\begin{split}\label{eq:}
\mathcal H_{\vec k } = M \gamma_z +   \vec k \cdot \vec \sigma \, \gamma_x.
\end{split}\end{equation}
Note that, in principle, there is a sign ambiguity for the term $\propto \vec k \cdot \vec \sigma \gamma_x$. However, one can show that this sign choice does not affect the surface analysis in any way, and therefore we simply set it to $+1$ in our treatment.
This should be contrasted with the helicity of the surface Dirac cone, which, in the presence of certain symmetries, plays a key role to the existence of anomalous surface states in a DSTI \cite{Fang17}.

To expose a boundary, we let the mass term $M$ be spatially dependent, such that the surface region is defined by $M(0)=0$, and $M(\lambda) \rightarrow {\rm sign}(\lambda)$ quickly away from the boundary region.
Let $\vechat n$ be the surface normal, and decompose $\vec k = \vec k_{\parallel} +  k_{\perp} \vechat n$ such that $ \vec k_{\parallel} \cdot \vechat n = 0$. In the Dirac Hamiltonian, we then replace $k_{\perp}  \rightarrow - i \partial_{\lambda}$, arriving at
\begin{equation}\begin{split}\label{eq:}
\mathcal H_{\vec k_{\parallel}, \lambda } = M(\lambda) \gamma_z +  \vec k_{\parallel} \cdot \vec \sigma \, \gamma_x -i   \vechat n \cdot \vec \sigma \, \gamma_x \partial_\lambda.
\end{split}\end{equation}
Our goal is to find eigenstates of $\mathcal H_{\vec k_{\parallel}, \lambda}$ which are exponentially localized to the surface region. This can be achieved through the ansatz
\begin{equation}\begin{split}\label{eq:}
\Psi(\vec k_{\parallel}, \lambda) = e^{- \int^\lambda_0 d \lambda' M(\lambda ')} \psi(\vec k_{\parallel}), 
\end{split}\end{equation}
which gives
\begin{equation}\begin{split}\label{eq:}
&\mathcal H_{\vec k_{\parallel}, \lambda } \Psi( \vec k_{\parallel, \lambda})\\
=&
\left (  \vec k_{\parallel} \cdot \vec \sigma \, \gamma_x + M(\lambda) \gamma_z
\left(  1-     \vechat n \cdot \vec \sigma \, \gamma_y \right) \right) \Psi( \vec k_{\parallel, \lambda}),
 \end{split}\end{equation}
and can be solved by 
\begin{equation}\begin{split}\label{eq:}
\left\{
\begin{array}{rl}
 \left(  1-     \vechat n \cdot \vec \sigma \, \gamma_y \right) \psi(\vec k_{\parallel}) =& 0\\
  \vec k_{\parallel} \cdot \vec \sigma \, \gamma_x  \psi(\vec k_{\parallel}) =& E_{\vec k_\parallel} \psi(\vec k_{\parallel})
\end{array}
\right.
.
\end{split}\end{equation}
The first condition implies $ P_{+} \psi(\vec k_{\parallel}) = \psi(\vec k_{\parallel})$, where $ P_{+}$ is the projector $P_{+} \equiv \frac{1}{2} \left(  1 +    \vechat n \cdot \vec \sigma \, \gamma_y \right)$ satisfying $[P_+ ,  \vec k_{\parallel} \cdot \vec \sigma \, \gamma_x]=0$. The $2$-dimensional surface Hamiltonian can then be found by restricting $  \vec k_{\parallel} \cdot \vec \sigma \, \gamma_x  $ to the subspace defined by $P_+$.

This is most easily achieved by introducing a rotation of basis. Let $V =\exp\left(  i  \frac{\pi}{4}
\vechat n \cdot \vec \sigma \,  \gamma_x \ \right)$, such that 
\begin{equation}\begin{split}\label{eq:}
V P_{+} V^\dagger =& \frac{1}{2}(1-  \gamma_z); \\
V \left(   \vec k_{\parallel} \cdot \vec \sigma \, \gamma_x \right)V^\dagger =& (\vec k \times \vechat n ) \cdot \vec \sigma,
\end{split}\end{equation}
where we used $ \vec k \times \vechat n = \vec k_{\parallel} \times \vechat n$. This gives explicitly a $2 \times 2$ surface Dirac-cone Hamiltonian $h_{\vec k} \equiv  (\vec k \times \vechat n ) \cdot \vec \sigma$.

Before proceeding, we make a few comments on the validity of our approach. The curvature of the space enters into our equation only through the (slow) spatial dependence of the surface normal $\vechat n$. This might not be the only effect resulting from the background curvature in the Dirac equation, since we generally expect both the local and global curvature to introduce extra terms in the equation \cite{WittenRMP}. We will, however, argue below that such terms will not be relevant for the physics we are considering. To see why, first note that the surface states associated to a DSTI are confined to regions in space corresponding to the domain walls of the mass term in the Dirac equation. As the presence of such domain walls is a topological property of the bulk, it is unaffected by the surface curvature. Now, we can imagine smoothly deforming any given surface such that all the nontrivial curvature is concentrated to isolated regions where the mass term is non-zero (in a symmetry-respecting manner), such that the gapless electronic theory at the domain walls resembles that defined assuming vanishing curvature. We then recover the surface theory described throughout, and hence the sTCI classification. Note that, such ability to decouple the possibly nontrivial curvature from the gapless surface states relies on having gapped regions on the surface of an sTCI, which prevents the sTCI surface states from exploring the global topology of the surface; this should be contrasted with the Dirac theory on the surface of a regular strong TI, which is gapless everywhere and is hence more susceptible to the effect of a nontrivial global curvature.

Next, we connect the symmetry representations furnished by the bulk and surface degrees of freedom.
Let $p = (\openone, 0)^T$ be the $4\times 2$ dimensional matrix projecting into the subspace defined by $P_+$, and we generalize $\vechat n$ to a slowly-varying function $ \vechat n_{\vec r}$, where $\vec r$ denotes a point on the boundary.
Then the surface Hamiltonian can be obtained from the bulk through $h_{\vec r, \vec k} \equiv p^T V_{\vec r} \mathcal H_{\vec k} V_{\vec r}^\dagger p $.
We assume the surface is symmetric under the (unitary) spatial symmetry $g = \{ R_g ~|~ \vec \gamma_g \}$, i.e., for any $\vec r$ on the surface, $g \cdot \vec r$ is also on the surface and their surface normals are related by $\hat n_{g \cdot \vec r} = R_g \hat n_{\vec r}$.
Recall that in the bulk, $g$ is represented by a unitary matrix $U_g=  u_g \oplus (\det R_g u_g )$, where $u_g $ equals to the standard spin-1/2 representation of $R_g$ up to a sign $\eta_g = \pm 1$, and that $\mathcal H_{\vec k}  =U_{g}^\dagger  \mathcal H_{R_g \vec k}U_{g}  $.
We can then deduce the transformation of the surface Hamiltonian through
\begin{equation}\begin{split}\label{eq:}
h_{\vec r, \vec k} = & p^T V_{\vec r} \mathcal H_{\vec k} V_{\vec r}^\dagger p \\
=&
p^T V_{\vec r} U_g^\dagger (V_{g \vec r}^\dagger V_{g \vec r}) \mathcal H_{R_g \vec k}  (V_{g \vec r}^\dagger V_{g \vec r}) U_g V_{\vec r}^\dagger p\\
= & \left( p^T V_{\vec r} U_g^\dagger V_{g  \cdot \vec r}^\dagger p \right) h_{g  \cdot \vec r, R_g \vec k}   \left( p^T V_{g  \cdot \vec r}  U_g V_{\vec r}^\dagger p \right),
\end{split}\end{equation}
where we used $[ V_{g  \cdot \vec r}  U_g V_{\vec r}^\dagger, p p^T] = 0$, and one can check that $p^T V_{g  \cdot \vec r}  U_g V_{\vec r}^\dagger p = u_g$. Therefore, $g \cdot h_{\vec r, \vec k} \equiv u_g h_{\vec r, \vec k}  u_g^\dagger =   h_{g \cdot  \vec r, R_g \vec k}  $, i.e., the surface Dirac cone furnishes the same symmetry representation as the valence bands, as one would expect.

Next, we study the general mass-term structure of stacked strong TIs.
Consider the surface theory for $n>1$ copies of the DSTI, with $k_n>1$ symmetry-allowed independent mass terms. The mass terms can be represented as $M_{i,\br} = m_{i,\br} \Gamma_i$, $i=1,\dots,k_n$, with $\Gamma_i$ chosen to satisfy $\Gamma_i^2 = 1$. By ``independent,'' we mean that the different mass terms anti-commute with each other, thus satisfying $\{\Gamma_i, \Gamma_j\} = 2 \delta_{ij}$. In addition, we assume all mass terms anticommute with the surface Hamiltonian so that the Hamiltonian is gapped when any of the mass terms in non-zero. We define the mass vector $\bm_\br=(m_{1,\br}, \dots, m_{k_n,\br})$ whose length gives the values of the gap at point $\br$. The transformation properties of the mass vector can be deduced from the requirement of the invariance of $\sum_i M_{i,\br}$ under symmetry
\begin{align}\begin{split}
g \cdot \sum_i M_{i,\br} =& \sum_{i} M_{i, g\cdot \br}\\
 \sum_i  m_{i,  \br} U_g   \Gamma_i  U_g^\dagger =&  \sum_i  m_{i,  g\cdot \br}     \Gamma_i  \\
 \sum_{ij}  O^g_{ji}  m_{i, \br} \Gamma_j =& \sum_i  m_{i,  g\cdot \br}     \Gamma_i ,
\end{split}
\end{align}
which implies that $g\cdot \bm_{\vec r} \equiv \bm_{g \cdot \vec r} = O^g \bm_{\vec r}$ with $O^g_{ij}$ an orthogonal matrix given by
\beq
O^g_{ij} = \frac{1}{4n} \mathrm{tr} \left( \Gamma_i U_g \Gamma_j U_g^{\dagger} \right)
\eeq
Thus, given a unitary representation acting on the Hamiltonian for the stacked DSTI, we get an induced orthogonal representation acting on the $(k-1)$-sphere.

\section{Bulk-surface correspondence for all space groups\label{app:XBSRep}}
In this appendix, we discuss some subtleties regarding bulk-surface correspondence of sTCI phases. In particular, we will focus on the relation between the sTCI classification and the information contained in the symmetry indicators. As noted in the main text, since the symmetry indicators only utilize data encoded in the symmetry representations, a given nontrivial indicator is generally compatible with more than one sTCI phases, i.e., knowing the indicator alone does not uniquely identify the surface signature of the nontrivial bulk. We will refer to this as the ``surface-state identification ambiguity.'' In the following, we will describe how such ambiguity can be systematically mapped out.

Let us first consider the simpler problem concerning ``weak'' phases defined by having at least one lattice translation taking a nontrivial signature, i.e, weak TIs and its mirror-enrichments.
As discussed in the main text, the relation between $X_{\rm BS}^{\rm (w)}$ and $\S_{\rm SG}^{\rm (w)}$, which concern sTCIs with either a weak TI or weak mirror Chern index, is readily understood from the existing discussion on the familiar topological band invariants \cite{Fu07, Fang12}. The ambiguity can be summarized as follows: First, for every $\Z_2$ factor (i.e., weak TI index) in $\S_{\rm SG}^{\rm (w)}$, $X_{\rm BS}^{\rm (w)}$ contains a corresponding $\Z_2$ if and only if the SG is centrosymmetric; Second, for every $\Z$ factor (i.e., weak mirror Chern number) in $\S_{\rm SG}^{\rm (w)}$, $X_{\rm BS}^{\rm (w)}$ contains a $\Z_{n}$ factor if and only if the SG contains an $n$-fold rotation about an axis normal to the associated mirror plane. This completely maps out the surface-state identification ambiguity for $\S_{\rm SG}^{\rm (w)}$.

It remains to study the case of $X_{\rm BS}^{\rm (s)}$ vs.\ $\S_{\rm SG}^{\rm (s)}$. 
This is far more complicated, as we detail below.

\subsection{Justification for the simplifying assumption}
In this subsection, we will first justify the simplifying assumption in establishing the bulk-boundary correspondence for the $X_{\rm BS}^{\rm (s)}$ phases, namely, we consider strong TIs with only band inversions at $\Gamma$ involving (signed) representations of the form $u_g = \eta_g v_g = \pm e^{- i \theta_g \vechat n_g \cdot \vec \sigma/2}$. For brevity, in the following we will refer to such representation as being ``quasi-standard,'' although this terminology is by no means standard outside of our present context.

A priori, our simplifying assumption is nontrivial as the symmetry representations of the non-relativistic electrons in a crystal are not ``quasi-standard'' in general, and that the band structures are also subjected to global connectivity constraints arising from, say, nonsymmorphic symmetries. 
Establishing the validity of the assumption requires a more technical discussion, which we will undertake below. Before that, we restate our main conclusion from this analysis: For any space group in class AII, one can choose a basis such that the ``strong factor'' in $X_{\rm BS}$ is generated by a strong TI which satisfies the simplifying assumption behind the Dirac analysis.

We now proceed to show this in the language of Ref.\ \onlinecite{Po17}, which we briefly review below. Insofar as symmetry representations, but not the detailed energetics, are concerned, any band structures isolated from above and below by band gaps can be represented by  a $D$-dimensional ``vector'' (more accurately, collection of $D$ integers) formed by the irrep multiplicities at different high-symmetry momenta. The symmetry-respecting ``vectors'' naturally  form an abelian group $\{ {\rm BS}\}$, with group addition corresponding physically to the stacking of the underlying systems. We then denote the subgroup of $\{ {\rm BS}\}$ which can arise from atomic insulators by $\{ {\rm AI}\}$. The symmetry-based indicator of band topology is defined as the mismatch between $\{ {\rm BS} \}$ and $\{ {\rm AI}\}$, which is mathematically captured by the quotient  $X_{\rm BS} \equiv \{ {\rm BS} \}/\{ {\rm AI} \} $.

Quasi-standard representations at $\Gamma$ play a special role in our Dirac analysis. The multiplicities of these irreps are encoded in certain components of the ``vectors'' in the description above. Let $\Pi$ be a projection which projects away these components, and suppose $\vec b_1, \vec b_2 \in \{ {\rm BS} \}$, $\vec b_1 \neq \vec b_2$, such that $\Pi ( \vec b_1 - \vec b_2) = 0$. Physically, this means while $\vec b_1$ and $\vec b_2$ are distinguishable in terms of symmetry representations, their only distinction arises solely from irrep exchanges involving only the quasi-standard representations at $\Gamma$. Further suppose $\vec b_1$ is nontrivial in $X_{\rm BS}$  but $\vec b_2$ is trivial, then one can choose a representative for the nontrivial class $[\vec b_1] \in X_{\rm BS}$ which satisfies the simplifying assumption behind the Dirac analysis, insofar as symmetry representations are concerned.

From the discussion above, we see that the kernel of the map $\Pi$, $\ker_{\rm BS} \Pi \equiv \{ \vec b \in \{ {\rm BS} \} ~:~ \Pi \vec b = \vec 0 \}$, plays a key role in the analysis. We can similarly define $\ker_{\rm AI} \Pi$ by replacing $\{ {\rm BS } \} \rightarrow \{ {\rm AI} \}$ in the definition.
The mismatch between $\ker_{\rm BS} \Pi$ and $\ker_{\rm AI} \Pi$ then describes band inversions of quasi-standard irreps at $\Gamma$ which lead to a $X_{\rm BS}$-nontrivial band structure. Again, this mismatch can be captured by a quotient, which has to be a subgroup of $X_{\rm BS}$. Through an explicit computation, we found
\begin{equation}\begin{split}\label{eq:KerBS}
X_{\rm BS}^{\rm (s)} = \frac{\ker_{\rm BS} \Pi}{\ker_{\rm AI} \Pi}
\end{split}\end{equation}
holds for all space groups in class AII.
This provides a formal justification of our simplifying assumption.

More concretely, the validity of Eq.\ \eqref{eq:KerBS} for any given SG can be established in three steps: first, we compute an explicit basis for $\{ {\rm AI} \}$ and  $\{ {\rm BS} \}$ following the recipe detailed in Ref.\ \onlinecite{Po17}; second, we identify the quasi-standard representations at $\Gamma$ and construct the projection $\Pi$; third, we compute the null spaces of $\Pi (\{{\rm BS}\})$ and $\Pi (\{{\rm AI}\})$, and evaluate the quotient to check if it agrees with the strong factor $X_{\rm BS}^{\rm (s)}$. 

Let us sketch out the described procedure for an explicit example. Consider SG 81 ($P\bar 4$), whose point group $S_4$ is generated by the rotoinversion $S_{4z}$.
There are four maximal-symmetry Wyckoff positions with $S_4$ as the site-symmetry group. As discussed in Sec.\ \ref{sec:k4} of the main text, the representations of $S_4$ furnished by spinful electrons can be described by the eigenvalue of $S_{4z}$, which takes the form $e^{i \alpha \pi/4}$ with $\alpha = \pm 1, \pm 3$. TRS pairs the representations with $\alpha = \pm 1$, and similarly for $\pm 3$.
On each of the four maximal-symmetry Wyckoff positions, we can construct an AI by having two electrons occupying the two orbitals labeled by $\alpha = \pm 1$. Similarly, we can construct another AI by putting the two electrons into the $\alpha = \pm 3$ orbitals, or by localizing them to the other Wyckoff positions. Upon stacking, these AIs generate all the possible AIs in our symmetry setting.

In momentum space, the symmetry representations are again given by the multiplicities of the $\alpha = \pm 1$ and $\pm 3$ representations at the four $S_4$-symmetric momenta: $\Gamma \equiv (0,0,0)$, M $ \equiv (\pi,\pi,0)$, Z$ \equiv (0,0,\pi)$, and A$\equiv (\pi,\pi,\pi)$. The representation data is encoded in the eight integers $\vec n \equiv (n^{\pm 1}_{\Gamma}, n^{\pm 3}_{\Gamma}, n^{\pm 1}_{\rm M}, n^{\pm 3}_{\rm M}, n^{\pm 1}_{\rm Z}, n^{\pm 3}_{\rm Z}, n^{\pm 1}_{\rm A}, n^{\pm 3}_{\rm A})$, where the subscript indicates the value of $\alpha$. In this notation, one can check that a legit choice of basis for $\{ {\rm AI} \}$ is
\begin{equation}\begin{split}\label{eq:}
\vec a_{1} =& (1, 0, 1, 0, 1, 0, 1, 0);\\
\vec a_{2} =& (0, 1, 0, 1, 0, 1, 0, 1);\\
\vec a_{3} =& (1, 0, 0, 1, 1, 0, 0, 1);\\
\vec a_{4} =& (1, 0, 1, 0, 0, 1, 0, 1);\\
\vec a_{5} =& (1, 0, 0, 1, 0, 1, 1, 0),
\end{split}\end{equation}
which arises by performing Fourier transform on five out of the eight AIs we constructed earlier. Note that we used only five of them as the remaining ones give $\vec n$'s that are linearly dependent to the ones above, i.e., the ``dimension'' of $\{ {\rm AI} \}$ is $d_{\rm AI}=5$, as was computed in Ref.\ \onlinecite{Po17}.

The next task is to compute $\{ {\rm BS}\}$. More systematically, this can be done via the use of the Smith normal form; here we simply note that the combination
\begin{equation}\begin{split}\label{eq:}
\frac{1}{2}(\vec a_1 + \vec a_3 + \vec a_4 + \vec a_5) = (2, 0, 1, 1, 1, 1, 1, 1)
\end{split}\end{equation}
corresponds to a band insulator which cannot be atomic. We further assert this is the only nontrivial combination one needs to consider.
As such, we infer a possible basis for $\{ {\rm BS} \}$ to be 
\begin{equation}\begin{split}\label{eq:}
\vec b_{i} =& \vec a_{i}  \text{ for } i = 1,2,3, 4; \\
\vec b_{5} =&\frac{1}{2}(\vec a_1 + \vec a_3 + \vec a_4 + \vec a_5),
\end{split}\end{equation}
and that $X_{\rm BS}^{\rm (s)} = \mathbb Z_2$.
This concludes the first step of the analysis.

Next, we identify the quasi-standard representations at $\Gamma$. 
For $S_4$, we have $v_{S_{4z}} = e^{- i  (2 \pi /4) \vechat z \cdot \vec \sigma/2} = e^{- i \pi \sigma_z/4}$, which corresponds simply to the $\alpha = \pm 1$ representation.
Furthermore, as  $ e^{ \pm i 3\pi /4}= - e^{ \mp i\pi/4} $, we conclude all the representations at $\Gamma$ are quasi-standard. The projection $\Pi$ can therefore be constructed explicitly as
\begin{equation}\begin{split}\label{eq:}
\Pi: \vec n \mapsto
(n^{\pm 1}_{\rm M}, n^{\pm 3}_{\rm M}, n^{\pm 1}_{\rm Z}, n^{\pm 3}_{\rm Z}, n^{\pm 1}_{\rm A}, n^{\pm 3}_{\rm A}).
\end{split}\end{equation}

It remains to compute the null spaces of $\Pi (\{{\rm BS}\})$ and $\Pi (\{{\rm AI}\})$. One finds
\begin{equation}\begin{split}\label{eq:}
 \vec b_1 + \vec b_2 - \vec b_5 =& (-1, 1, 0, 0, 0, 0, 0, 0);\\
 \vec a_1 + 2 \vec a_2 -  \vec a_3- \vec a_4- \vec a_5 =& (-2, 2, 0, 0, 0, 0, 0, 0),\\
\end{split}\end{equation}
which respectively generate $\ker_{\rm BS} \Pi$ and $\ker_{\rm AI} \Pi$. We thus conclude ${\ker_{\rm BS} \Pi}/{\ker_{\rm AI} \Pi} = \mathbb Z_2$, i.e., Eq.\ \eqref{eq:KerBS} is verified.

While we have provided a detailed example using SG 81, it is conceptually more revealing to understand why Eq.\ \eqref{eq:KerBS} should hold in general. To achieve that, we first translate Eq.\ \eqref{eq:KerBS} into a more physical language: {\it every nontrivial class in $X_{\rm BS}^{\rm (s)}$ can be represented by a BS which differs from an AI only by the exchange of some quasi-standard representations at $\Gamma$.} To establish  Eq.\ \eqref{eq:KerBS}  on physical grounds, it suffices to show that this defining property holds for the generator of $X_{\rm BS}^{\rm (s)}$; once a representative $\vec b_0 \in \{ {\rm BS} \}$ with the desired band-inversion interpretation is found for the generator, all the other classes can be represented by copies of $\vec b_0$, which will automatically enjoy the desired property.

To this end, first note that, by definition, the generator of $X_{\rm BS}^{\rm (s)}$ can be chosen to be a strong TI.
Given any SG with a symmetry-diagnosable strong TI (i.e., through the indices defined in Sec.\ \ref{Sec:Indicators}), one can imagine building a strong TI $\vec b_0$ by first starting with an AI, and then exchange a pair of quasi-standard representations at $\Gamma$, e.g., in a centrosymmetric SG, one can start with an AI with a quasi-standard representation at $\Gamma$ where  $u_{\I} = \openone$, and then upon a band inversion with $u_{\I} = -\openone$ one arrives at a strong TI as detected by the Fu-Kane criterion. 

It then remains to show that such a strong TI $\vec b_0$ can be constructed for any SG. This can be achieved in two steps: first, we note that one can always find an AI $\vec a_0$ whose representation content contains each of the quasi-standard representations at $\Gamma$. 
Such an AI arise from the generic Wyckoff position, as was established in a corollary in Ref.\ \onlinecite{Po17}.
Second, we can always perform a band inversion on $\vec a_0$ involving only the quasi-standard representations at $\Gamma$ and arrive at a strong TI. The only concern here is that one might violate some compatibility relations in performing the desired band inversion and end up with a semimetal. As shown in Sec.\ \ref{Sec:Indicators}, however, the key indices for the diagnosis of a strong TI use either inversion or $S_4$ eigenvalues, neither of which is subjected to any compatibility relations as they only leave isolated points (but not lines) invariant in the momentum space.

\subsection{Surface-state ambiguity}
Having established the formal framework, we turn to deriving the surface-state ambiguity in the $X_{\rm BS}^{\rm (s)}$ diagnosis. 
The main result here is reported in Table \ref{tab:SurfaceXBS_Full}, where we chart out such ambiguities for all $230$ SGs in class AII. The remaining of the section is devoted to an explanation on how these results are obtained. As shown in Sec.\ \ref{Sec:Surface} in the main text, the surface states of DSTIs can be classified by two types of indices: $\nu_M \in \mathbb Z$ for every independent mirror (note, not glide) plane $M$, and $\nu_\ell \in \mathbb Z_2$ for every other independent generator $\ell$ of the point group satisfying the conditions outlined in the main text. 
Suppose we have a band insulator with the quasi-standard irreps $\chi^{(i)}$, $i=1,\dots, 2 N_{\chi}$ and not necessarily distinct, exchanged at $\Gamma$, where we \emph{assume} each exchange can be modeled using the Dirac approach.
Let the symmetry $g$  be represented by $u_g^{(i)} = \eta_g^{(i)} v_g$ in $\chi^{(i)}$.
The surface indices are then given by
\begin{equation}\begin{split}\label{eq:SurfaceIndEx}
\nu_{M} = \frac{1}{2} \sum_{i=1}^{2 N_{\chi}} \eta_{M}^{(i)};~~~~
(-1)^{\nu_{\ell} } = (\det \ell)^{N_{\chi}} \prod_{i=1}^{2 N_{\chi}} \eta_{\ell}^{(i)}.
\end{split}\end{equation}
This can be seen by first grouping the $\{ \chi^{(i)} ~:~ i =1,\dots, 2N_{\chi} \}$ pair-wise and apply the bulk-surface correspondence in Sec.\ \ref{Sec:BulkSurface}, and then adding the resulting $N_{\chi}$ sets of surface indices. 
The resulting indices, Eq.\ \eqref{eq:SurfaceIndEx}, are insensitive to the initial, arbitrary choice on the pair-wise grouping, and are therefore well-defined.

To make connection to $X_{\rm BS}^{\rm (s)}$, we now restrict ourselves to information available from symmetry representations alone. 
As described in the main text, this introduces surface-state ambiguity as the symmetry representations contain less data than the Dirac description.
The first source of ambiguity is when the choices $\eta_g = \pm 1$ lead to the same irrep.
For instance, consider the mirror symmetry about a plane with surface normal $\vechat n_M$, which can be represented by the traceless unitary matrix $v_M = i \vechat n_M \cdot \vec \sigma$. Usually, this implies $\eta_M = \pm 1$ is unconstrained by the specification of the symmetry representation, and hence $X_{\rm BS}^{\rm (s)}$ does not constrain $\nu_M$, which leads to a surface-state ambiguity.

Yet, if the point-group contain $C_4$ or $C_6$ rotation together with inversion $\mathcal I$, then we may have relations like $\eta_{M} = \eta_{\mathcal I} \eta_{C_4}^2$ and hence $\eta_M$ becomes tied to the representations which can detect $\eta_{\mathcal I}$ and $\eta_{C_{4,6}}$. Such is the case of $P4/m$ (SG 83) discussed in the main text. This observation can also be viewed as a manifestation of the detection of mirror Chern numbers through $C_{4,6}$-symmetry eigenvalues \cite{Fang12}. (For $C_2$ together with $\mathcal I$, we can only detect the parity of the mirror Chern number, but the mirror-Chern parity is always even in a DSTI.)

However, there is a second source of surface-state ambiguity which can be present. This originates from nontrivial stacks of strong TIs which, when restricted to symmetry irreps alone, look indistinguishable from an atomic insulator. In other words, if we are just given a band insulator which differs from a reference atomic insulator only by the exchange of some quasi-standard irreps at $\Gamma$, it is not automatically justified that the exchange arose from a series of consecutive exchanges, each producing a strong TI amenable to a Dirac analysis. Such is the case when the irrep exchange pattern can be understood as the difference of two atomic insulators, which is captured precisely by $\ker_{\rm AI} \Pi$ in our framework.
Therefore, if an entry in $\ker_{\rm AI} \Pi$ gives rise to a nontrivial surface index according to Eq.\ \eqref{eq:SurfaceIndEx}, it implies, from representation alone, we cannot tell if we have at hand an atomic insulator or a DSTI with nontrivial surface states. The evaluation of the surface indices on $\ker_{\rm AI} \Pi$  then gives a subgroup of the surface classification corresponding to possible surfaces of $0 \in X_{\rm BS}^{\rm (s)}$, i.e., this subgroup captures the surface-state identification ambiguity.

While the two mentioned sources of ambiguity have rather different physical origins, in practice they may be intertwined with each other. As an illustrative example, consider the symmorphic SG $P2/m$ (SG 10), which is generated by lattice translations, inversion $\mathcal I$ and a two-fold rotation $C_{2z}$. Note that this implies a mirror symmetry $M_z \equiv \mathcal I C_{2z}$.
As shown in Table \ref{Classification}, the surface states of the DSTIs in this setting are classified by $\mathbb Z \times \mathbb Z_2$, where the first factor corresponds to equatorial states pinned to the mirror plane and protected by a $\mathbb Z$-valued mirror Chern number, and the $\mathbb Z_2$ entry is generated by a hinge mode, protected by $C_{2z}$, which passes through the north and south poles (and hence giving rise to 2D gapless surface states on appropriate faces \cite{Fang17}). The quasi-standard representations for both the $C_{2z}$ and $M_{z}$ are traceless, and therefore, from the discussion above, $X_{\rm BS}^{\rm (s)}$  should be blind to the existence of the mentioned surface states. Yet, the inversion symmetry $\mathcal I$ is not traceless, and using its symmetry eigenvalues one should be able to detect the total parity of the 1D modes on the surface. This small dilemma is resolved by noticing that $\eta_{\mathcal I} = \eta_{M_z} \eta_{C_{2z}}$, and therefore although we cannot detect the values of  $\eta_{M_z}$ and $ \eta_{C_{2z}}$ individually, their product become detectable. 
At a more technical level, in deriving the surface-state identification ambiguity of an $X_{\rm BS}^{\rm (s)}$-nontrivial phase, one has to first study the ``pre-image'' of the consistent assignments on the  $\eta_{g}$ of the traceless elements (in a quasi-standard representation), which implies each element in $\ker_{\rm AI} \Pi$ is now enhanced correspondingly. One then evaluates the surface indices on this enhanced version of $\ker_{\rm AI} \Pi$, which would reveal the subgroup of surface signatures that are consistent with the symmetry representations of $0 \in X_{\rm BS}^{\rm (s)}$.

\subsection{Table for the sTCI classification and surface-state ambiguity}
The full results of $\S_{\rm SG}$ and its surface-state identification ambiguity in $X_{\rm BS}$ are tabulated in Table \ref{tab:SurfaceXBS_Full}.
Let us illustrate how to read it through a few examples. First, consider the weak phases in SG 1 and 2. Both of the SGs have three independent weak TIs corresponding to the three independent lattice translations, and hence we expect $\S_{\rm SG}^{\rm (w)} = \mathbb Z_2 \times \mathbb Z_2 \times \mathbb Z_2 $ for both of them. However, while the weak TI indices are detected through the Fu-Kane parity criterion in SG 2, they are undetectable using symmetry eigenvalues in SG 1. This implies there are no surface-state identification ambiguity for $\S_{\rm SG}^{\rm (w)}$ in SG 2, whereas the entire $\S_{\rm SG}^{\rm (w)}$ is consistent with the identity in $X_{\rm BS}^{(w)} = \mathbb Z_1$ for SG 1. In Table \ref{tab:SurfaceXBS_Full}, the latter case is denoted by a quotient with $K^{\rm (w)}  = \langle \dots \rangle$, while the former is denoted by the absence of a quotient.

While the previous discussion focused on the detectability of the weak phases using the symmetry indicators, we recall that spatial symmetries can also restrict the possible set of weak phases, as discussed in the main text. For example, consider the SGs 3 and 4, which have the same point-group generated by $C_{2z}$, but the former is symmorphic and has $C_{2z}$ as a group element, whereas the latter is nonsymmorphic in which the two-fold rotation is extended into a $2_1$ screw. Correspondingly, the two SGs have different weak sTCI classifications, with  $\S_{\rm SG}^{\rm (w)} = \mathbb Z_2 \times \mathbb Z_2 \times \mathbb Z_2 $ for SG 3 but $\mathbb Z_2 \times \mathbb Z_2 $ for SG 4. The reduction of one of the $\mathbb Z_2$ factors in SG 4 arises as the square of $2_1$ is a lattice translation, and this translation cannot take a nontrivial $s_g$, as discussed in Sec.\ \ref{sec:transX} of the main text. Nonetheless, for both cases the symmetry indicators are unable to detect any of the weak TI indices, and hence we have $K^{\rm (w)} = \langle \dots \rangle$, denoting the full group, for both of them.

As a more nontrivial example, consider SG 123, which contains multiple mirrors extending the weak TI phases to weak mirror Chern insulators. Besides, the $C_4$ rotation about (say) the $z$-axis relate the weak index along the $x$ and $y$ direction, which leads to $\S_{\rm SG}^{\rm (w)} = \mathbb Z \times \mathbb Z$. Since the $\mathbb Z$-valued mirror Chern numbers can at best be detected $\mod n$ using an $n$-fold rotation \cite{Fang12}, the surface-state identification ambiguity will encode this modulo structure. This is seen in $K^{\rm (w)}  = \langle (4,0), (0,2) \rangle$ in Table \ref{tab:SurfaceXBS_Full}, which denotes the subgroup generated by the elements $(4,0)$ and $(0,2)$ in $\mathbb Z \times \mathbb Z$. Physically, this means that the first weak mirror Chern index (for phases deformable to stack of 2D phases along $z$) can be detected $\mod 4$ using $C_4$ about $z$, whereas the second weak mirror Chern index can be detected $\mod 2$ using (say) a $C_2$ rotation about the $x$ axis.

Next, we move on to the strong part of the sTCI classification. Recall that the strong part is tightly tied to the generators of the point group of the SG.
The chosen generators of the point groups are listed again in Table \ref{tab:SurfaceXBS_Full}. Note that, for a couple of SGs, the generating set differs slightly from that in Table \ref{Classification}. The factors in $\S_{\rm SG}^{\rm (s)}$ are listed in correspondence with the generators listed in the left-most column. 
An exception is made for three-fold rotations, which do not lead to any factor in $\S_{\rm SG}^{\rm (s)}$. They are always listed at the end of the generating set, and when they are present the number of factors in $\S_{\rm SG}^{\rm (s)}$ will be less than the number of independent generators. For instance, the SGs with point group $D_3$, generated by $C_{2x}$ and $C_{3z}$, all have $\S_{\rm SG}^{\rm (s)} = \mathbb Z_2$ with the only factor corresponding to $C_{2x}$.

However, depending on the SG, a mirror in the point group may correspond to a glide in the SG, with the former leading to a $\mathbb Z$ factor whereas the latter a $\mathbb Z_2$ factor.
To see this more concretely, compare the $\S_{\rm SG}^{\rm (s)}$ of SGs 83 and 85. 
Their point group, $C_{4h}$, is generated by a mirror $M_z$ and a four-fold rotation $C_{4z}$. 
Yet, as shown in the table, in tabulating the sTCI classification we instead chose $S_{4z}$ as a generator in lieu of $C_{4z}$.
First, focus on SG 83.
While $M_z$ leads to mirror Chern insulators with a $\mathbb Z$ classification, $S_{4z}$ protects sTCIs with a nontrivial $\mathbb Z_2$ index. As $M_z$ and $S_{4z}$ are independent, they lead to independent factors in the sTCI classification, giving $\S_{\rm SG}^{\rm (s)} = \mathbb Z \times \mathbb Z_2$.
Note that the indices associated with the other symmetries are fixed by the group structure, e.g., as $C_{4z} = M_z (S_{4z})^{-1}$, the $\mathbb Z_2$ index associated to the phase $(1,0) \in \mathbb Z \times \mathbb Z_2$ is $1 \in \mathbb Z_2$, but that to the phase $(1,1)$ is $0 \in \mathbb Z_2$.
In contrast, although sharing the same point group, SG 85 has $\S_{\rm SG}^{\rm (s)} = \mathbb Z_2 \times \mathbb Z_2$. This is because the would-be mirror in SG 85 has been extended into a glide, and hence the replacement $\mathbb Z \mapsto \mathbb Z_2$.

Finally, we describe how to extract the surface-state identification ambiguity from Table \ref{tab:SurfaceXBS_Full}. Consider SG 83 again.
As explained in the main text, the symmetry indicators cannot differentiate between a phase with a mirror Chern index of $2$ from one with the nontrivial ${C}_{4z}$ index, i.e., $(2,0) \simeq (0,1)$. 
(Here, $(0,1)$ is a phase a nontrivial $S_{4z}$ index but a trivial index for $\I$, as $\I =  M_z S_{4z}^2$, and hence it has a nontrivial index for $C_{4z} = \I S_{4z}$.)
This is represented in Table \ref{tab:SurfaceXBS_Full} by $K^{\rm (s)}  = \langle (2,1) \rangle$. For SG 85, however, one finds $K^{\rm (s)}  = \langle (0,1) \rangle$. This implies the index associated to $S_{4z}$ is undetectable from symmetry representations, and that an sTCI with a nontrivial indicator of $2 \in \mathbb Z_4$ in the strong part of $X_{\rm BS}$ will always have a nontrivial glide index.

As discussed above, we specify $K^{\rm (s)}$ by providing its generators. In writing down the group elements we have implicitly assumed a preferred basis. To clarify, we reiterate our choice here: each factor corresponds to a generator of the associated point group, as tabulated in the leftmost column, with the exception of $C_3$ rotations as they do not protect nontrivial phases. With this choice in mind, the data on $K^{\rm (s)}$ can be readily converted into information about the possible physical surface states of a given element in $X_{\rm BS}^{\rm (s)}$. 

Let us illustrate this using a nontrivial example. 
Consider SG 194. As listed in Table \ref{tab:SurfaceXBS_Full}, its point group is $D_{6h}$ and we choose the generators to be the three independent mirrors $M_z$, $M_{x}$, and $M_{\frac{\sqrt{3}x+y}{2}}$. 
Furthermore, we have  $\S_{\rm SG}^{\rm (s)} / K^{\rm (s)}= \mathbb Z \times \mathbb Z \times  \mathbb Z_{2}/ ~\langle ~(1, 0, 1), (0, 3, 1) ~\rangle$. The first and second $\mathbb Z$ factors in $\S_{\rm SG}^{\rm (s)}$ correspond to ``strong'' mirror-Chern phases protected by $M_z$ and $M_x$ respectively, whereas the last $\mathbb Z_2$ factor implies the mirror $M_{\frac{\sqrt{3}x+y}{2}}$ has been extended into a glide in this SG. The next step is to identify the generator of the group $ \S_{\rm SG}^{\rm (s)} / K^{\rm (s)}$, which can be chosen to be one of the three basis vectors $(1,0,0)$, $(0,1,0)$ or $(0,0,1)$. Given $K^{\rm (s)}=\langle ~(1, 0, 1), (0, 3, 1) ~\rangle$ and that the generators should have maximal order, we see that we can choose the generator to be $(0,1,0)$, which has order $6$. This gives $ \S_{\rm SG}^{\rm (s)} / K^{\rm (s)} = \mathbb Z_6$, which, as explained in the main text, corresponds to the even subgroup of $X_{\rm BS}^{\rm (s)} = \mathbb Z_{12}$.

The preceeding analysis also informs the possible surface states of any given entry of $X_{\rm BS}^{\rm (s)}$. For instance, suppose a material has the symmetry indicator of $2 \in \mathbb Z_{12}$. Then we know that it could be a sTCI classified by $(0,1,0) \in \mathbb Z \times \mathbb Z \times \mathbb Z_2$, i.e., it has $M_x$-protected surface states but not those protected by $M_z$ or the glide. However, due the ambiguity $(0,1,0) \simeq (0,-2,1) \simeq (1,-2,0)\simeq \dots$, if it has a higher $M_x$ mirror-Chern number it could also feature the coexistence of other surface states protected by the other mirror and/ or the glide.

In closing, we note that results concerning the physical interpretations of $X_{\rm BS}$ and its associated surface-state identification ambiguity has also been reported in Refs.\ \onlinecite{SongFang17}. In particular, a more explicit tabulation of the index association can be found there.

\onecolumngrid

\begin{center}
\begin{table}[h!]
\caption{
{\bf sTCI classification and symmetry indicators for all space groups.} The space groups are grouped by their associated point groups. $\mathcal S_{\rm SG}^{\rm (w)}$ ($\mathcal S_{\rm SG}^{\rm (s)}$) denotes the sTCI classification for phases with (without) a weak index.
$K^{\rm (w)}$ ($K^{\rm (s)}$) denotes the subgroup of $\mathcal S_{\rm SG}^{\rm (w)}$ ($\mathcal S_{\rm SG}^{\rm (s)}$) corresponding to a trivial symmetry indicator, i.e., it encodes the surface-state identification ambiguity. 
In each column of $\S_{\rm SG} / K$ (i.e. with superscript (w) or (s)), we write out $\S_{\rm SG}  $ and $K$ explicitly for most SGs.
 The exceptions are SGs with a trivial $K$, for which we list only $\S_{\rm SG}$ (say SG 2); besides, when $\S_{\rm SG}$ is trivial, we denote it by --. For simplicity, we denote $K$ by $\langle \dots \rangle$ when $K  = \S_{\rm SG}$, i.e., when the symmetry indicators cannot detect any phase in $\S_{\rm SG}$.
\label{tab:SurfaceXBS_Full}}
\begin{tabular}{c|ccc}
\hline \hline
Point Group (generators) ~&~ Space Group ~&~ $\S_{\rm SG}^{\rm (w)} / K^{\rm (w)}$ ~&~ $\S_{\rm SG}^{\rm (s)} / K^{\rm (s)}$\\
\hline
\multirow{1}{*}{$C_1$ ($1$)} & ~1~ & ~$ \mathbb Z_{2} \times \mathbb Z_{2} \times \mathbb Z_{2}/ \langle \dots \rangle$~ & ~--~\\
\hline
\multirow{1}{*}{$C_i$ ($\I$)} & ~2~ & ~$ \mathbb Z_{2} \times \mathbb Z_{2} \times \mathbb Z_{2}$~ & ~$ \mathbb Z_{2}$~\\
\hline
\multirow{3}{*}{$C_2$ ($C_{2z}$)} & ~3~ & ~$ \mathbb Z_{2} \times \mathbb Z_{2} \times \mathbb Z_{2}/ \langle \dots \rangle$~ & ~$ \mathbb Z_{2}/ \langle \dots \rangle$~\\
~ & ~4~ & ~$ \mathbb Z_{2} \times \mathbb Z_{2}/ \langle \dots \rangle$~ & ~$ \mathbb Z_{2}/ \langle \dots \rangle$~\\
~ & ~5~ & ~$ \mathbb Z_{2} \times \mathbb Z_{2}/ \langle \dots \rangle$~ & ~$ \mathbb Z_{2}/ \langle \dots \rangle$~\\
\hline
\multirow{4}{*}{$C_s$ ($M_z$)} & ~6~ & ~$ \mathbb Z \times  \mathbb Z_{2} \times \mathbb Z_{2}/ \langle \dots \rangle$~ & ~$ \mathbb Z/ \langle \dots \rangle$~\\
~ & ~7~ & ~$ \mathbb Z_{2} \times \mathbb Z_{2}/ \langle \dots \rangle$~ & ~$ \mathbb Z_{2}/ \langle \dots \rangle$~\\
~ & ~8~ & ~$ \mathbb Z_{2} \times \mathbb Z_{2}/ \langle \dots \rangle$~ & ~$ \mathbb Z/ \langle \dots \rangle$~\\
~ & ~9~ & ~$ \mathbb Z_{2}/ \langle \dots \rangle$~ & ~$ \mathbb Z_{2}/ \langle \dots \rangle$~\\
\hline
\multirow{6}{*}{$C_{2h}$ ($M_z, C_{2z}$)} & ~10~ & ~$ \mathbb Z \times  \mathbb Z_{2} \times \mathbb Z_{2}/ ~\langle ~(2, 0, 0) ~\rangle$~ & ~$ \mathbb Z \times  \mathbb Z_{2}/ ~\langle ~(1, 1) ~\rangle$~\\
~ & ~11~ & ~$ \mathbb Z_{2} \times \mathbb Z_{2}$~ & ~$ \mathbb Z \times  \mathbb Z_{2}/ ~\langle ~(1, 1) ~\rangle$~\\
~ & ~12~ & ~$ \mathbb Z_{2} \times \mathbb Z_{2}$~ & ~$ \mathbb Z \times  \mathbb Z_{2}/ ~\langle ~(1, 1) ~\rangle$~\\
~ & ~13~ & ~$ \mathbb Z_{2} \times \mathbb Z_{2}$~ & ~$ \mathbb Z_{2} \times \mathbb Z_{2}/ ~\langle ~(1, 1) ~\rangle$~\\
~ & ~14~ & ~$ \mathbb Z_{2}$~ & ~$ \mathbb Z_{2} \times \mathbb Z_{2}/ ~\langle ~(1, 1) ~\rangle$~\\
~ & ~15~ & ~$ \mathbb Z_{2}$~ & ~$ \mathbb Z_{2} \times \mathbb Z_{2}/ ~\langle ~(1, 1) ~\rangle$~\\
\hline
\multirow{9}{*}{$D_2$ ($C_{2z}, C_{2x}$)} & ~16~ & ~$ \mathbb Z_{2} \times \mathbb Z_{2} \times \mathbb Z_{2}/ \langle \dots \rangle$~ & ~$ \mathbb Z_{2} \times \mathbb Z_{2}/ \langle \dots \rangle$~\\
~ & ~17~ & ~$ \mathbb Z_{2} \times \mathbb Z_{2}/ \langle \dots \rangle$~ & ~$ \mathbb Z_{2} \times \mathbb Z_{2}/ \langle \dots \rangle$~\\
~ & ~18~ & ~$ \mathbb Z_{2}/ \langle \dots \rangle$~ & ~$ \mathbb Z_{2} \times \mathbb Z_{2}/ \langle \dots \rangle$~\\
~ & ~19~ & ~--~ & ~$ \mathbb Z_{2} \times \mathbb Z_{2}/ \langle \dots \rangle$~\\
~ & ~20~ & ~$ \mathbb Z_{2}/ \langle \dots \rangle$~ & ~$ \mathbb Z_{2} \times \mathbb Z_{2}/ \langle \dots \rangle$~\\
~ & ~21~ & ~$ \mathbb Z_{2} \times \mathbb Z_{2}/ \langle \dots \rangle$~ & ~$ \mathbb Z_{2} \times \mathbb Z_{2}/ \langle \dots \rangle$~\\
~ & ~22~ & ~$ \mathbb Z_{2} \times \mathbb Z_{2}/ \langle \dots \rangle$~ & ~$ \mathbb Z_{2} \times \mathbb Z_{2}/ \langle \dots \rangle$~\\
~ & ~23~ & ~$ \mathbb Z_{2}/ \langle \dots \rangle$~ & ~$ \mathbb Z_{2} \times \mathbb Z_{2}/ \langle \dots \rangle$~\\
~ & ~24~ & ~$ \mathbb Z_{2}/ \langle \dots \rangle$~ & ~$ \mathbb Z_{2} \times \mathbb Z_{2}/ \langle \dots \rangle$~\\
\hline
\multirow{22}{*}{$C_{2v}$ ($M_x, M_y$)} & ~25~ & ~$ \mathbb Z \times \mathbb Z \times  \mathbb Z_{2}/ \langle \dots \rangle$~ & ~$ \mathbb Z \times \mathbb Z/ \langle \dots \rangle$~\\
~ & ~26~ & ~$ \mathbb Z \times  \mathbb Z_{2}/ \langle \dots \rangle$~ & ~$ \mathbb Z \times  \mathbb Z_{2}/ \langle \dots \rangle$~\\
~ & ~27~ & ~$ \mathbb Z_{2} \times \mathbb Z_{2}/ \langle \dots \rangle$~ & ~$ \mathbb Z_{2} \times \mathbb Z_{2}/ \langle \dots \rangle$~\\
~ & ~28~ & ~$ \mathbb Z_{2} \times \mathbb Z_{2}/ \langle \dots \rangle$~ & ~$ \mathbb Z \times  \mathbb Z_{2}/ \langle \dots \rangle$~\\
~ & ~29~ & ~$ \mathbb Z_{2}/ \langle \dots \rangle$~ & ~$ \mathbb Z_{2} \times \mathbb Z_{2}/ \langle \dots \rangle$~\\
~ & ~30~ & ~$ \mathbb Z_{2}/ \langle \dots \rangle$~ & ~$ \mathbb Z_{2} \times \mathbb Z_{2}/ \langle \dots \rangle$~\\
~ & ~31~ & ~$ \mathbb Z_{2}/ \langle \dots \rangle$~ & ~$ \mathbb Z \times  \mathbb Z_{2}/ \langle \dots \rangle$~\\
~ & ~32~ & ~$ \mathbb Z_{2}/ \langle \dots \rangle$~ & ~$ \mathbb Z_{2} \times \mathbb Z_{2}/ \langle \dots \rangle$~\\
~ & ~33~ & ~--~ & ~$ \mathbb Z_{2} \times \mathbb Z_{2}/ \langle \dots \rangle$~\\
~ & ~34~ & ~$ \mathbb Z_{2}/ \langle \dots \rangle$~ & ~$ \mathbb Z_{2} \times \mathbb Z_{2}/ \langle \dots \rangle$~\\
~ & ~35~ & ~$ \mathbb Z_{2} \times \mathbb Z_{2}/ \langle \dots \rangle$~ & ~$ \mathbb Z \times \mathbb Z/ \langle \dots \rangle$~\\
~ & ~36~ & ~$ \mathbb Z_{2}/ \langle \dots \rangle$~ & ~$ \mathbb Z \times  \mathbb Z_{2}/ \langle \dots \rangle$~\\
~ & ~37~ & ~$ \mathbb Z_{2}/ \langle \dots \rangle$~ & ~$ \mathbb Z_{2} \times \mathbb Z_{2}/ \langle \dots \rangle$~\\
~ & ~38~ & ~$ \mathbb Z \times  \mathbb Z_{2}/ \langle \dots \rangle$~ & ~$ \mathbb Z \times \mathbb Z/ \langle \dots \rangle$~\\
~ & ~39~ & ~$ \mathbb Z_{2} \times \mathbb Z_{2}/ \langle \dots \rangle$~ & ~$ \mathbb Z \times  \mathbb Z_{2}/ \langle \dots \rangle$~\\
~ & ~40~ & ~$ \mathbb Z_{2}/ \langle \dots \rangle$~ & ~$ \mathbb Z \times  \mathbb Z_{2}/ \langle \dots \rangle$~\\
~ & ~41~ & ~$ \mathbb Z_{2}/ \langle \dots \rangle$~ & ~$ \mathbb Z_{2} \times \mathbb Z_{2}/ \langle \dots \rangle$~\\
~ & ~42~ & ~$ \mathbb Z_{2} \times \mathbb Z_{2}/ \langle \dots \rangle$~ & ~$ \mathbb Z \times \mathbb Z/ \langle \dots \rangle$~\\
~ & ~43~ & ~--~ & ~$ \mathbb Z_{2} \times \mathbb Z_{2}/ \langle \dots \rangle$~\\
~ & ~44~ & ~$ \mathbb Z_{2}/ \langle \dots \rangle$~ & ~$ \mathbb Z \times \mathbb Z/ \langle \dots \rangle$~\\
~ & ~45~ & ~$ \mathbb Z_{2}/ \langle \dots \rangle$~ & ~$ \mathbb Z_{2} \times \mathbb Z_{2}/ \langle \dots \rangle$~\\
~ & ~46~ & ~$ \mathbb Z_{2}/ \langle \dots \rangle$~ & ~$ \mathbb Z \times  \mathbb Z_{2}/ \langle \dots \rangle$~\\
\hline
\end{tabular}
\end{table}
\end{center}

\begin{center}
\begin{table}[h!]
(Continued from the previous page)\\
\begin{tabular}{c|ccc}
\hline
Point Group  (generators)~&~ Space Group ~&~ $\S_{\rm SG}^{\rm (w)} / K^{\rm (w)}$ ~&~ $\S_{\rm SG}^{\rm (s)} / K^{\rm (s)}$\\
\hline
\multirow{28}{*}{$D_{2h}$ ($M_x, M_y, M_z$)} & ~47~ & ~$ \mathbb Z \times \mathbb Z \times \mathbb Z/ ~\langle ~(2, 0, 0), (0, 2, 0), (0, 0, 2) ~\rangle$~ & ~$ \mathbb Z \times \mathbb Z \times \mathbb Z/ ~\langle ~(1, 1, 0), (1, 0, 1), (-1, 0, 1) ~\rangle$~\\
~ & ~48~ & ~$ \mathbb Z_{2}$~ & ~$ \mathbb Z_{2} \times \mathbb Z_{2} \times \mathbb Z_{2}/ ~\langle ~(1, 1, 0), (1, 0, 1) ~\rangle$~\\
~ & ~49~ & ~$ \mathbb Z_{2} \times \mathbb Z_{2}$~ & ~$ \mathbb Z \times  \mathbb Z_{2} \times \mathbb Z_{2}/ ~\langle ~(1, 1, 0), (0, 1, 1) ~\rangle$~\\
~ & ~50~ & ~$ \mathbb Z_{2}$~ & ~$ \mathbb Z_{2} \times \mathbb Z_{2} \times \mathbb Z_{2}/ ~\langle ~(1, 1, 0), (1, 0, 1) ~\rangle$~\\
~ & ~51~ & ~$ \mathbb Z \times  \mathbb Z_{2}/ ~\langle ~(2, 0) ~\rangle$~ & ~$ \mathbb Z \times \mathbb Z \times  \mathbb Z_{2}/ ~\langle ~(1, 0, 1), (0, 1, 1) ~\rangle$~\\
~ & ~52~ & ~--~ & ~$ \mathbb Z_{2} \times \mathbb Z_{2} \times \mathbb Z_{2}/ ~\langle ~(1, 1, 0), (1, 0, 1) ~\rangle$~\\
~ & ~53~ & ~$ \mathbb Z_{2}$~ & ~$ \mathbb Z \times  \mathbb Z_{2} \times \mathbb Z_{2}/ ~\langle ~(1, 1, 0), (0, 1, 1) ~\rangle$~\\
~ & ~54~ & ~$ \mathbb Z_{2}$~ & ~$ \mathbb Z_{2} \times \mathbb Z_{2} \times \mathbb Z_{2}/ ~\langle ~(1, 1, 0), (1, 0, 1) ~\rangle$~\\
~ & ~55~ & ~$ \mathbb Z/ ~\langle ~(2) ~\rangle$~ & ~$ \mathbb Z \times  \mathbb Z_{2} \times \mathbb Z_{2}/ ~\langle ~(1, 1, 0), (0, 1, 1) ~\rangle$~\\
~ & ~56~ & ~--~ & ~$ \mathbb Z_{2} \times \mathbb Z_{2} \times \mathbb Z_{2}/ ~\langle ~(1, 1, 0), (1, 0, 1) ~\rangle$~\\
~ & ~57~ & ~$ \mathbb Z_{2}$~ & ~$ \mathbb Z \times  \mathbb Z_{2} \times \mathbb Z_{2}/ ~\langle ~(1, 1, 0), (0, 1, 1) ~\rangle$~\\
~ & ~58~ & ~--~ & ~$ \mathbb Z \times  \mathbb Z_{2} \times \mathbb Z_{2}/ ~\langle ~(1, 1, 0), (0, 1, 1) ~\rangle$~\\
~ & ~59~ & ~$ \mathbb Z_{2}$~ & ~$ \mathbb Z \times \mathbb Z \times  \mathbb Z_{2}/ ~\langle ~(1, 0, 1), (0, 1, 1) ~\rangle$~\\
~ & ~60~ & ~--~ & ~$ \mathbb Z_{2} \times \mathbb Z_{2} \times \mathbb Z_{2}/ ~\langle ~(1, 1, 0), (1, 0, 1) ~\rangle$~\\
~ & ~61~ & ~--~ & ~$ \mathbb Z_{2} \times \mathbb Z_{2} \times \mathbb Z_{2}/ ~\langle ~(1, 1, 0), (1, 0, 1) ~\rangle$~\\
~ & ~62~ & ~--~ & ~$ \mathbb Z \times  \mathbb Z_{2} \times \mathbb Z_{2}/ ~\langle ~(1, 1, 0), (0, 1, 1) ~\rangle$~\\
~ & ~63~ & ~$ \mathbb Z_{2}$~ & ~$ \mathbb Z \times \mathbb Z \times  \mathbb Z_{2}/ ~\langle ~(1, 0, 1), (0, 1, 1) ~\rangle$~\\
~ & ~64~ & ~$ \mathbb Z_{2}$~ & ~$ \mathbb Z \times  \mathbb Z_{2} \times \mathbb Z_{2}/ ~\langle ~(1, 1, 0), (0, 1, 1) ~\rangle$~\\
~ & ~65~ & ~$ \mathbb Z \times  \mathbb Z_{2}/ ~\langle ~(2, 0) ~\rangle$~ & ~$ \mathbb Z \times \mathbb Z \times \mathbb Z/ ~\langle ~(1, 1, 0), (1, 0, 1), (-1, 0, 1) ~\rangle$~\\
~ & ~66~ & ~$ \mathbb Z_{2}$~ & ~$ \mathbb Z \times  \mathbb Z_{2} \times \mathbb Z_{2}/ ~\langle ~(1, 1, 0), (0, 1, 1) ~\rangle$~\\
~ & ~67~ & ~$ \mathbb Z_{2} \times \mathbb Z_{2}$~ & ~$ \mathbb Z \times \mathbb Z \times  \mathbb Z_{2}/ ~\langle ~(1, 0, 1), (0, 1, 1) ~\rangle$~\\
~ & ~68~ & ~$ \mathbb Z_{2}$~ & ~$ \mathbb Z_{2} \times \mathbb Z_{2} \times \mathbb Z_{2}/ ~\langle ~(1, 1, 0), (1, 0, 1) ~\rangle$~\\
~ & ~69~ & ~$ \mathbb Z_{2} \times \mathbb Z_{2}$~ & ~$ \mathbb Z \times \mathbb Z \times \mathbb Z/ ~\langle ~(1, 1, 0), (1, 0, 1), (-1, 0, 1) ~\rangle$~\\
~ & ~70~ & ~--~ & ~$ \mathbb Z_{2} \times \mathbb Z_{2} \times \mathbb Z_{2}/ ~\langle ~(1, 1, 0), (1, 0, 1) ~\rangle$~\\
~ & ~71~ & ~$ \mathbb Z_{2}$~ & ~$ \mathbb Z \times \mathbb Z \times \mathbb Z/ ~\langle ~(1, 1, 0), (1, 0, 1), (-1, 0, 1) ~\rangle$~\\
~ & ~72~ & ~$ \mathbb Z_{2}$~ & ~$ \mathbb Z \times  \mathbb Z_{2} \times \mathbb Z_{2}/ ~\langle ~(1, 1, 0), (0, 1, 1) ~\rangle$~\\
~ & ~73~ & ~$ \mathbb Z_{2}$~ & ~$ \mathbb Z_{2} \times \mathbb Z_{2} \times \mathbb Z_{2}/ ~\langle ~(1, 1, 0), (1, 0, 1) ~\rangle$~\\
~ & ~74~ & ~$ \mathbb Z_{2}$~ & ~$ \mathbb Z \times \mathbb Z \times  \mathbb Z_{2}/ ~\langle ~(1, 0, 1), (0, 1, 1) ~\rangle$~\\
\hline
\multirow{6}{*}{$C_4$ ($C_{4z}$)} & ~75~ & ~$ \mathbb Z_{2} \times \mathbb Z_{2}/ \langle \dots \rangle$~ & ~$ \mathbb Z_{2}/ \langle \dots \rangle$~\\
~ & ~76~ & ~$ \mathbb Z_{2}/ \langle \dots \rangle$~ & ~$ \mathbb Z_{2}/ \langle \dots \rangle$~\\
~ & ~77~ & ~$ \mathbb Z_{2} \times \mathbb Z_{2}/ \langle \dots \rangle$~ & ~$ \mathbb Z_{2}/ \langle \dots \rangle$~\\
~ & ~78~ & ~$ \mathbb Z_{2}/ \langle \dots \rangle$~ & ~$ \mathbb Z_{2}/ \langle \dots \rangle$~\\
~ & ~79~ & ~$ \mathbb Z_{2}/ \langle \dots \rangle$~ & ~$ \mathbb Z_{2}/ \langle \dots \rangle$~\\
~ & ~80~ & ~$ \mathbb Z_{2}/ \langle \dots \rangle$~ & ~$ \mathbb Z_{2}/ \langle \dots \rangle$~\\
\hline
\multirow{2}{*}{$S_4$ ($S_{4z}$)} & ~81~ & ~$ \mathbb Z_{2} \times \mathbb Z_{2}/ \langle \dots \rangle$~ & ~$ \mathbb Z_{2}/ \langle \dots \rangle$~\\
~ & ~82~ & ~$ \mathbb Z_{2}/ \langle \dots \rangle$~ & ~$ \mathbb Z_{2}/ \langle \dots \rangle$~\\
\hline
\multirow{6}{*}{$C_{4h}$ ($M_z,S_{4z}$)} & ~83~ & ~$ \mathbb Z \times  \mathbb Z_{2}/ ~\langle ~(4, 0) ~\rangle$~ & ~$ \mathbb Z \times  \mathbb Z_{2}/ ~\langle ~(2, 1) ~\rangle$~\\
~ & ~84~ & ~$ \mathbb Z_{2}$~ & ~$ \mathbb Z \times  \mathbb Z_{2}/ ~\langle ~(0, 1), (2, 0) ~\rangle$~\\
~ & ~85~ & ~$ \mathbb Z_{2}$~ & ~$ \mathbb Z_{2} \times \mathbb Z_{2}/ ~\langle ~(0, 1) ~\rangle$~\\
~ & ~86~ & ~$ \mathbb Z_{2}$~ & ~$ \mathbb Z_{2} \times \mathbb Z_{2}/ ~\langle ~(0, 1) ~\rangle$~\\
~ & ~87~ & ~$ \mathbb Z_{2}$~ & ~$ \mathbb Z \times  \mathbb Z_{2}/ ~\langle ~(2, 1) ~\rangle$~\\
~ & ~88~ & ~--~ & ~$ \mathbb Z_{2} \times \mathbb Z_{2}/ ~\langle ~(0, 1) ~\rangle$~\\
\hline
\multirow{10}{*}{$D_4$ ($C_{4z}, C_{2x}$)} & ~89~ & ~$ \mathbb Z_{2} \times \mathbb Z_{2}/ \langle \dots \rangle$~ & ~$ \mathbb Z_{2} \times \mathbb Z_{2}/ \langle \dots \rangle$~\\
~ & ~90~ & ~$ \mathbb Z_{2}/ \langle \dots \rangle$~ & ~$ \mathbb Z_{2} \times \mathbb Z_{2}/ \langle \dots \rangle$~\\
~ & ~91~ & ~$ \mathbb Z_{2}/ \langle \dots \rangle$~ & ~$ \mathbb Z_{2} \times \mathbb Z_{2}/ \langle \dots \rangle$~\\
~ & ~92~ & ~--~ & ~$ \mathbb Z_{2} \times \mathbb Z_{2}/ \langle \dots \rangle$~\\
~ & ~93~ & ~$ \mathbb Z_{2} \times \mathbb Z_{2}/ \langle \dots \rangle$~ & ~$ \mathbb Z_{2} \times \mathbb Z_{2}/ \langle \dots \rangle$~\\
~ & ~94~ & ~$ \mathbb Z_{2}/ \langle \dots \rangle$~ & ~$ \mathbb Z_{2} \times \mathbb Z_{2}/ \langle \dots \rangle$~\\
~ & ~95~ & ~$ \mathbb Z_{2}/ \langle \dots \rangle$~ & ~$ \mathbb Z_{2} \times \mathbb Z_{2}/ \langle \dots \rangle$~\\
~ & ~96~ & ~--~ & ~$ \mathbb Z_{2} \times \mathbb Z_{2}/ \langle \dots \rangle$~\\
~ & ~97~ & ~$ \mathbb Z_{2}/ \langle \dots \rangle$~ & ~$ \mathbb Z_{2} \times \mathbb Z_{2}/ \langle \dots \rangle$~\\
~ & ~98~ & ~$ \mathbb Z_{2}/ \langle \dots \rangle$~ & ~$ \mathbb Z_{2} \times \mathbb Z_{2}/ \langle \dots \rangle$~\\
\hline
\end{tabular}
\end{table}
\end{center}

\begin{center}
\begin{table*}[h!]
(Continued from the previous page)\\
\begin{tabular}{c|ccc}
\hline
Point Group  (generators)~&~ Space Group ~&~ $\S_{\rm SG}^{\rm (w)} / K^{\rm (w)}$ ~&~ $\S_{\rm SG}^{\rm (s)} / K^{\rm (s)}$\\
\hline
\multirow{12}{*}{$C_{4v}$ ($M_{x}, M_{xy}$)} & ~99~ & ~$ \mathbb Z \times  \mathbb Z_{2}/ \langle \dots \rangle$~ & ~$ \mathbb Z \times \mathbb Z/ \langle \dots \rangle$~\\
~ & ~100~ & ~$ \mathbb Z_{2}/ \langle \dots \rangle$~ & ~$ \mathbb Z \times  \mathbb Z_{2}/ \langle \dots \rangle$~\\
~ & ~101~ & ~$ \mathbb Z_{2}/ \langle \dots \rangle$~ & ~$ \mathbb Z \times  \mathbb Z_{2}/ \langle \dots \rangle$~\\
~ & ~102~ & ~$ \mathbb Z_{2}/ \langle \dots \rangle$~ & ~$ \mathbb Z \times  \mathbb Z_{2}/ \langle \dots \rangle$~\\
~ & ~103~ & ~$ \mathbb Z_{2}/ \langle \dots \rangle$~ & ~$ \mathbb Z_{2} \times \mathbb Z_{2}/ \langle \dots \rangle$~\\
~ & ~104~ & ~--~ & ~$ \mathbb Z_{2} \times \mathbb Z_{2}/ \langle \dots \rangle$~\\
~ & ~105~ & ~$ \mathbb Z/ \langle \dots \rangle$~ & ~$ \mathbb Z \times  \mathbb Z_{2}/ \langle \dots \rangle$~\\
~ & ~106~ & ~--~ & ~$ \mathbb Z_{2} \times \mathbb Z_{2}/ \langle \dots \rangle$~\\
~ & ~107~ & ~$ \mathbb Z_{2}/ \langle \dots \rangle$~ & ~$ \mathbb Z \times \mathbb Z/ \langle \dots \rangle$~\\
~ & ~108~ & ~$ \mathbb Z_{2}/ \langle \dots \rangle$~ & ~$ \mathbb Z \times  \mathbb Z_{2}/ \langle \dots \rangle$~\\
~ & ~109~ & ~--~ & ~$ \mathbb Z \times  \mathbb Z_{2}/ \langle \dots \rangle$~\\
~ & ~110~ & ~--~ & ~$ \mathbb Z_{2} \times \mathbb Z_{2}/ \langle \dots \rangle$~\\
\hline
\multirow{12}{*}{$D_{2d}$ ($M_{x},S_{4z}$)} & ~111~ & ~$ \mathbb Z_{2} \times \mathbb Z_{2}/ \langle \dots \rangle$~ & ~$ \mathbb Z \times  \mathbb Z_{2}/ \langle \dots \rangle$~\\
~ & ~112~ & ~$ \mathbb Z_{2}/ \langle \dots \rangle$~ & ~$ \mathbb Z_{2} \times \mathbb Z_{2}/ \langle \dots \rangle$~\\
~ & ~113~ & ~$ \mathbb Z_{2}/ \langle \dots \rangle$~ & ~$ \mathbb Z \times  \mathbb Z_{2}/ \langle \dots \rangle$~\\
~ & ~114~ & ~--~ & ~$ \mathbb Z_{2} \times \mathbb Z_{2}/ \langle \dots \rangle$~\\
~ & ~115~ & ~$ \mathbb Z \times  \mathbb Z_{2}/ \langle \dots \rangle$~ & ~$ \mathbb Z \times  \mathbb Z_{2}/ \langle \dots \rangle$~\\
~ & ~116~ & ~$ \mathbb Z_{2}/ \langle \dots \rangle$~ & ~$ \mathbb Z_{2} \times \mathbb Z_{2}/ \langle \dots \rangle$~\\
~ & ~117~ & ~$ \mathbb Z_{2}/ \langle \dots \rangle$~ & ~$ \mathbb Z_{2} \times \mathbb Z_{2}/ \langle \dots \rangle$~\\
~ & ~118~ & ~$ \mathbb Z_{2}/ \langle \dots \rangle$~ & ~$ \mathbb Z_{2} \times \mathbb Z_{2}/ \langle \dots \rangle$~\\
~ & ~119~ & ~$ \mathbb Z_{2}/ \langle \dots \rangle$~ & ~$ \mathbb Z \times  \mathbb Z_{2}/ \langle \dots \rangle$~\\
~ & ~120~ & ~$ \mathbb Z_{2}/ \langle \dots \rangle$~ & ~$ \mathbb Z_{2} \times \mathbb Z_{2}/ \langle \dots \rangle$~\\
~ & ~121~ & ~$ \mathbb Z_{2}/ \langle \dots \rangle$~ & ~$ \mathbb Z \times  \mathbb Z_{2}/ \langle \dots \rangle$~\\
~ & ~122~ & ~--~ & ~$ \mathbb Z_{2} \times \mathbb Z_{2}/ \langle \dots \rangle$~\\
\hline
\multirow{20}{*}{$D_{4h}$ ($M_{x}, M_{z}, M_{xy}$)} & ~123~ & ~$ \mathbb Z \times \mathbb Z/ ~\langle ~(4, 0), (0, 2) ~\rangle$~ & ~$ \mathbb Z \times \mathbb Z \times \mathbb Z/ ~\langle ~(1, 1, 0), (1, -1, 0), (0, 1, 2) ~\rangle$~\\
~ & ~124~ & ~$ \mathbb Z_{2}$~ & ~$ \mathbb Z \times  \mathbb Z_{2} \times \mathbb Z_{2}/ ~\langle ~(0, 1, 1), (2, 1, 0) ~\rangle$~\\
~ & ~125~ & ~$ \mathbb Z_{2}$~ & ~$ \mathbb Z \times  \mathbb Z_{2} \times \mathbb Z_{2}/ ~\langle ~(1, 0, 0), (0, 1, 0) ~\rangle$~\\
~ & ~126~ & ~--~ & ~$ \mathbb Z_{2} \times \mathbb Z_{2} \times \mathbb Z_{2}/ ~\langle ~(1, 0, 0), (0, 1, 0) ~\rangle$~\\
~ & ~127~ & ~$ \mathbb Z/ ~\langle ~(4) ~\rangle$~ & ~$ \mathbb Z \times \mathbb Z \times  \mathbb Z_{2}/ ~\langle ~(1, 0, 1), (0, 2, 1) ~\rangle$~\\
~ & ~128~ & ~--~ & ~$ \mathbb Z \times  \mathbb Z_{2} \times \mathbb Z_{2}/ ~\langle ~(0, 1, 1), (2, 1, 0) ~\rangle$~\\
~ & ~129~ & ~$ \mathbb Z_{2}$~ & ~$ \mathbb Z \times \mathbb Z \times  \mathbb Z_{2}/ ~\langle ~(1, 0, 0), (0, 1, 0) ~\rangle$~\\
~ & ~130~ & ~--~ & ~$ \mathbb Z_{2} \times \mathbb Z_{2} \times \mathbb Z_{2}/ ~\langle ~(1, 0, 0), (0, 1, 0) ~\rangle$~\\
~ & ~131~ & ~$ \mathbb Z/ ~\langle ~(2) ~\rangle$~ & ~$ \mathbb Z \times \mathbb Z \times  \mathbb Z_{2}/ ~\langle ~(1, 0, 0), (0, 0, 1), (0, 2, 0) ~\rangle$~\\
~ & ~132~ & ~$ \mathbb Z_{2}$~ & ~$ \mathbb Z \times \mathbb Z \times  \mathbb Z_{2}/ ~\langle ~(1, 0, 0), (0, 0, 1), (0, 2, 0) ~\rangle$~\\
~ & ~133~ & ~--~ & ~$ \mathbb Z_{2} \times \mathbb Z_{2} \times \mathbb Z_{2}/ ~\langle ~(1, 0, 0), (0, 1, 0) ~\rangle$~\\
~ & ~134~ & ~$ \mathbb Z_{2}$~ & ~$ \mathbb Z \times  \mathbb Z_{2} \times \mathbb Z_{2}/ ~\langle ~(1, 0, 0), (0, 1, 0) ~\rangle$~\\
~ & ~135~ & ~--~ & ~$ \mathbb Z \times  \mathbb Z_{2} \times \mathbb Z_{2}/ ~\langle ~(0, 1, 0), (0, 0, 1), (2, 0, 0) ~\rangle$~\\
~ & ~136~ & ~--~ & ~$ \mathbb Z \times \mathbb Z \times  \mathbb Z_{2}/ ~\langle ~(1, 0, 0), (0, 0, 1), (0, 2, 0) ~\rangle$~\\
~ & ~137~ & ~--~ & ~$ \mathbb Z \times  \mathbb Z_{2} \times \mathbb Z_{2}/ ~\langle ~(1, 0, 0), (0, 1, 0) ~\rangle$~\\
~ & ~138~ & ~--~ & ~$ \mathbb Z \times  \mathbb Z_{2} \times \mathbb Z_{2}/ ~\langle ~(1, 0, 0), (0, 1, 0) ~\rangle$~\\
~ & ~139~ & ~$ \mathbb Z_{2}$~ & ~$ \mathbb Z \times \mathbb Z \times \mathbb Z/ ~\langle ~(1, 1, 0), (1, -1, 0), (0, 1, 2) ~\rangle$~\\
~ & ~140~ & ~$ \mathbb Z_{2}$~ & ~$ \mathbb Z \times \mathbb Z \times  \mathbb Z_{2}/ ~\langle ~(1, 0, 1), (0, 2, 1) ~\rangle$~\\
~ & ~141~ & ~--~ & ~$ \mathbb Z \times  \mathbb Z_{2} \times \mathbb Z_{2}/ ~\langle ~(1, 0, 0), (0, 1, 0) ~\rangle$~\\
~ & ~142~ & ~--~ & ~$ \mathbb Z_{2} \times \mathbb Z_{2} \times \mathbb Z_{2}/ ~\langle ~(1, 0, 0), (0, 1, 0) ~\rangle$~\\
\hline
\multirow{4}{*}{$C_3$ ($C_{3z}$)} & ~143~ & ~$ \mathbb Z_{2}/ \langle \dots \rangle$~ & ~--~\\
~ & ~144~ & ~$ \mathbb Z_{2}/ \langle \dots \rangle$~ & ~--~\\
~ & ~145~ & ~$ \mathbb Z_{2}/ \langle \dots \rangle$~ & ~--~\\
~ & ~146~ & ~$ \mathbb Z_{2}/ \langle \dots \rangle$~ & ~--~\\
\hline
\multirow{2}{*}{$S_6$ ($S_{3z}$)} & ~147~ & ~$ \mathbb Z_{2}$~ & ~$ \mathbb Z_{2}$~\\
~ & ~148~ & ~$ \mathbb Z_{2}$~ & ~$ \mathbb Z_{2}$~\\
\hline
\multirow{7}{*}{$D_3$ ($ C_{2x},C_{3z}$)} & ~149~ & ~$ \mathbb Z_{2}/ \langle \dots \rangle$~ & ~$ \mathbb Z_{2}/ \langle \dots \rangle$~\\
~ & ~150~ & ~$ \mathbb Z_{2}/ \langle \dots \rangle$~ & ~$ \mathbb Z_{2}/ \langle \dots \rangle$~\\
~ & ~151~ & ~$ \mathbb Z_{2}/ \langle \dots \rangle$~ & ~$ \mathbb Z_{2}/ \langle \dots \rangle$~\\
~ & ~152~ & ~$ \mathbb Z_{2}/ \langle \dots \rangle$~ & ~$ \mathbb Z_{2}/ \langle \dots \rangle$~\\
~ & ~153~ & ~$ \mathbb Z_{2}/ \langle \dots \rangle$~ & ~$ \mathbb Z_{2}/ \langle \dots \rangle$~\\
~ & ~154~ & ~$ \mathbb Z_{2}/ \langle \dots \rangle$~ & ~$ \mathbb Z_{2}/ \langle \dots \rangle$~\\
~ & ~155~ & ~$ \mathbb Z_{2}/ \langle \dots \rangle$~ & ~$ \mathbb Z_{2}/ \langle \dots \rangle$~\\
\hline
\end{tabular}
\end{table*}
\end{center}

\begin{center}
\begin{table*}[h!]
(Continued from the previous page)\\
\begin{tabular}{c|ccc}
\hline
Point Group  (generators)~&~ Space Group ~&~ $\S_{\rm SG}^{\rm (w)} / K^{\rm (w)}$ ~&~ $\S_{\rm SG}^{\rm (s)} / K^{\rm (s)}$\\
\hline
\multirow{6}{*}{$C_{3v}$ ($M_{x}, C_{3z}$)} & ~156~ & ~$ \mathbb Z_{2}/ \langle \dots \rangle$~ & ~$ \mathbb Z/ \langle \dots \rangle$~\\
~ & ~157~ & ~$ \mathbb Z_{2}/ \langle \dots \rangle$~ & ~$ \mathbb Z/ \langle \dots \rangle$~\\
~ & ~158~ & ~--~ & ~$ \mathbb Z_{2}/ \langle \dots \rangle$~\\
~ & ~159~ & ~--~ & ~$ \mathbb Z_{2}/ \langle \dots \rangle$~\\
~ & ~160~ & ~$ \mathbb Z_{2}/ \langle \dots \rangle$~ & ~$ \mathbb Z/ \langle \dots \rangle$~\\
~ & ~161~ & ~--~ & ~$ \mathbb Z_{2}/ \langle \dots \rangle$~\\
\hline
\multirow{6}{*}{$D_{3d}$ ($M_{x}, C_{2x}, C_{3z}$)} & ~162~ & ~$ \mathbb Z_{2}$~ & ~$ \mathbb Z \times  \mathbb Z_{2}/ ~\langle ~(1, 1) ~\rangle$~\\
~ & ~163~ & ~--~ & ~$ \mathbb Z_{2} \times \mathbb Z_{2}/ ~\langle ~(1, 1) ~\rangle$~\\
~ & ~164~ & ~$ \mathbb Z_{2}$~ & ~$ \mathbb Z \times  \mathbb Z_{2}/ ~\langle ~(1, 1) ~\rangle$~\\
~ & ~165~ & ~--~ & ~$ \mathbb Z_{2} \times \mathbb Z_{2}/ ~\langle ~(1, 1) ~\rangle$~\\
~ & ~166~ & ~$ \mathbb Z_{2}$~ & ~$ \mathbb Z \times  \mathbb Z_{2}/ ~\langle ~(1, 1) ~\rangle$~\\
~ & ~167~ & ~--~ & ~$ \mathbb Z_{2} \times \mathbb Z_{2}/ ~\langle ~(1, 1) ~\rangle$~\\
\hline
\multirow{6}{*}{$C_6$ ($C_{6z}$)} & ~168~ & ~$ \mathbb Z_{2}/ \langle \dots \rangle$~ & ~$ \mathbb Z_{2}/ \langle \dots \rangle$~\\
~ & ~169~ & ~--~ & ~$ \mathbb Z_{2}/ \langle \dots \rangle$~\\
~ & ~170~ & ~--~ & ~$ \mathbb Z_{2}/ \langle \dots \rangle$~\\
~ & ~171~ & ~$ \mathbb Z_{2}/ \langle \dots \rangle$~ & ~$ \mathbb Z_{2}/ \langle \dots \rangle$~\\
~ & ~172~ & ~$ \mathbb Z_{2}/ \langle \dots \rangle$~ & ~$ \mathbb Z_{2}/ \langle \dots \rangle$~\\
~ & ~173~ & ~--~ & ~$ \mathbb Z_{2}/ \langle \dots \rangle$~\\
\hline
\multirow{1}{*}{$C_{3h}$ ($M_z, C_{3z}$)} & ~174~ & ~$ \mathbb Z/ ~\langle ~(3) ~\rangle$~ & ~$ \mathbb Z/ ~\langle ~(3) ~\rangle$~\\
\hline
\multirow{2}{*}{$C_{6h}$ ($M_z, C_{6z}$)} & ~175~ & ~$ \mathbb Z/ ~\langle ~(6) ~\rangle$~ & ~$ \mathbb Z \times  \mathbb Z_{2}/ ~\langle ~(3, 1) ~\rangle$~\\
~ & ~176~ & ~--~ & ~$ \mathbb Z \times  \mathbb Z_{2}/ ~\langle ~(3, 1) ~\rangle$~\\
\hline
\multirow{6}{*}{$D_6$ ($C_{6z}, C_{2x}$)} & ~177~ & ~$ \mathbb Z_{2}/ \langle \dots \rangle$~ & ~$ \mathbb Z_{2} \times \mathbb Z_{2}/ \langle \dots \rangle$~\\
~ & ~178~ & ~--~ & ~$ \mathbb Z_{2} \times \mathbb Z_{2}/ \langle \dots \rangle$~\\
~ & ~179~ & ~--~ & ~$ \mathbb Z_{2} \times \mathbb Z_{2}/ \langle \dots \rangle$~\\
~ & ~180~ & ~$ \mathbb Z_{2}/ \langle \dots \rangle$~ & ~$ \mathbb Z_{2} \times \mathbb Z_{2}/ \langle \dots \rangle$~\\
~ & ~181~ & ~$ \mathbb Z_{2}/ \langle \dots \rangle$~ & ~$ \mathbb Z_{2} \times \mathbb Z_{2}/ \langle \dots \rangle$~\\
~ & ~182~ & ~--~ & ~$ \mathbb Z_{2} \times \mathbb Z_{2}/ \langle \dots \rangle$~\\
\hline
\multirow{4}{*}{$C_{6v}$ ($M_{x}, M_{\frac{\sqrt{3}x+y}{2}}$)} & ~183~ & ~$ \mathbb Z_{2}/ \langle \dots \rangle$~ & ~$ \mathbb Z \times \mathbb Z/ \langle \dots \rangle$~\\
~ & ~184~ & ~--~ & ~$ \mathbb Z_{2} \times \mathbb Z_{2}/ \langle \dots \rangle$~\\
~ & ~185~ & ~--~ & ~$ \mathbb Z \times  \mathbb Z_{2}/ \langle \dots \rangle$~\\
~ & ~186~ & ~--~ & ~$ \mathbb Z \times  \mathbb Z_{2}/ \langle \dots \rangle$~\\
\hline
\multirow{4}{*}{$D_{3h}$ ($M_z, M_{x}, C_{3z}$)} & ~187~ & ~$ \mathbb Z/ ~\langle ~(3) ~\rangle$~ & ~$ \mathbb Z \times \mathbb Z/ ~\langle ~(1, 0), (0, 3) ~\rangle$~\\
~ & ~188~ & ~--~ & ~$ \mathbb Z \times  \mathbb Z_{2}/ ~\langle ~(0, 1), (3, 0) ~\rangle$~\\
~ & ~189~ & ~$ \mathbb Z/ ~\langle ~(3) ~\rangle$~ & ~$ \mathbb Z \times \mathbb Z/ ~\langle ~(1, 0), (0, 3) ~\rangle$~\\
~ & ~190~ & ~--~ & ~$ \mathbb Z \times  \mathbb Z_{2}/ ~\langle ~(0, 1), (3, 0) ~\rangle$~\\
\hline
\multirow{4}{*}{$D_{6h}$ ($M_z, M_{x}, M_{\frac{\sqrt{3}x+y}{2}}$)} & ~191~ & ~$ \mathbb Z/ ~\langle ~(6) ~\rangle$~ & ~$ \mathbb Z \times \mathbb Z \times \mathbb Z/ ~\langle ~(1, 1, 0), (1, -1, 0), (0, 1, 3) ~\rangle$~\\
~ & ~192~ & ~--~ & ~$ \mathbb Z \times  \mathbb Z_{2} \times \mathbb Z_{2}/ ~\langle ~(0, 1, 1), (3, 1, 0) ~\rangle$~\\
~ & ~193~ & ~--~ & ~$ \mathbb Z \times \mathbb Z \times  \mathbb Z_{2}/ ~\langle ~(1, 0, 1), (0, 3, 1) ~\rangle$~\\
~ & ~194~ & ~--~ & ~$ \mathbb Z \times \mathbb Z \times  \mathbb Z_{2}/ ~\langle ~(1, 0, 1), (0, 3, 1) ~\rangle$~\\
\hline
\multirow{5}{*}{$T$ ($C_{3,\bn_1}, C_{3,\bn_2}$)} & ~195~ & ~$ \mathbb Z_{2}/ \langle \dots \rangle$~ & ~--~\\
~ & ~196~ & ~--~ & ~--~\\
~ & ~197~ & ~$ \mathbb Z_{2}/ \langle \dots \rangle$~ & ~--~\\
~ & ~198~ & ~--~ & ~--~\\
~ & ~199~ & ~$ \mathbb Z_{2}/ \langle \dots \rangle$~ & ~--~\\
\hline
\multirow{7}{*}{$T_h$ ($M_z, C_{3,\bn_1}$)} & ~200~ & ~$ \mathbb Z/ ~\langle ~(2) ~\rangle$~ & ~$ \mathbb Z/ ~\langle ~(2) ~\rangle$~\\
~ & ~201~ & ~$ \mathbb Z_{2}$~ & ~$ \mathbb Z_{2}$~\\
~ & ~202~ & ~--~ & ~$ \mathbb Z/ ~\langle ~(2) ~\rangle$~\\
~ & ~203~ & ~--~ & ~$ \mathbb Z_{2}$~\\
~ & ~204~ & ~$ \mathbb Z_{2}$~ & ~$ \mathbb Z/ ~\langle ~(2) ~\rangle$~\\
~ & ~205~ & ~--~ & ~$ \mathbb Z_{2}$~\\
~ & ~206~ & ~$ \mathbb Z_{2}$~ & ~$ \mathbb Z_{2}$~\\
\hline
\end{tabular}
\end{table*}
\end{center}

\begin{center}
\begin{table*}[h!]
(Continued from the previous page)\\
\begin{tabular}{c|ccc}
\hline
Point Group  (generators)~&~ Space Group ~&~ $\S_{\rm SG}^{\rm (w)} / K^{\rm (w)}$ ~&~ $\S_{\rm SG}^{\rm (s)} / K^{\rm (s)}$\\
\hline
\multirow{8}{*}{$O$ ($C_{4,z}, C_{3,\bn_1}$)} & ~207~ & ~$ \mathbb Z_{2}/ \langle \dots \rangle$~ & ~$ \mathbb Z_{2}/ \langle \dots \rangle$~\\
~ & ~208~ & ~$ \mathbb Z_{2}/ \langle \dots \rangle$~ & ~$ \mathbb Z_{2}/ \langle \dots \rangle$~\\
~ & ~209~ & ~--~ & ~$ \mathbb Z_{2}/ \langle \dots \rangle$~\\
~ & ~210~ & ~--~ & ~$ \mathbb Z_{2}/ \langle \dots \rangle$~\\
~ & ~211~ & ~$ \mathbb Z_{2}/ \langle \dots \rangle$~ & ~$ \mathbb Z_{2}/ \langle \dots \rangle$~\\
~ & ~212~ & ~--~ & ~$ \mathbb Z_{2}/ \langle \dots \rangle$~\\
~ & ~213~ & ~--~ & ~$ \mathbb Z_{2}/ \langle \dots \rangle$~\\
~ & ~214~ & ~$ \mathbb Z_{2}/ \langle \dots \rangle$~ & ~$ \mathbb Z_{2}/ \langle \dots \rangle$~\\
\hline
\multirow{6}{*}{$T_d$ ($M_{\bn_1}, C_{3,\bn_1}$)} & ~215~ & ~$ \mathbb Z_{2}/ \langle \dots \rangle$~ & ~$ \mathbb Z/ \langle \dots \rangle$~\\
~ & ~216~ & ~--~ & ~$ \mathbb Z/ \langle \dots \rangle$~\\
~ & ~217~ & ~$ \mathbb Z_{2}/ \langle \dots \rangle$~ & ~$ \mathbb Z/ \langle \dots \rangle$~\\
~ & ~218~ & ~--~ & ~$ \mathbb Z_{2}/ \langle \dots \rangle$~\\
~ & ~219~ & ~--~ & ~$ \mathbb Z_{2}/ \langle \dots \rangle$~\\
~ & ~220~ & ~--~ & ~$ \mathbb Z_{2}/ \langle \dots \rangle$~\\
\hline
\multirow{10}{*}{$O_h$ ($M_{x},M_{xy}, C_{3,\bn_1}$)} & ~221~ & ~$ \mathbb Z/ ~\langle ~(4) ~\rangle$~ & ~$ \mathbb Z \times \mathbb Z/ ~\langle ~(2, 1), (0, 2) ~\rangle$~\\
~ & ~222~ & ~--~ & ~$ \mathbb Z_{2} \times \mathbb Z_{2}/ ~\langle ~(1, 0) ~\rangle$~\\
~ & ~223~ & ~--~ & ~$ \mathbb Z \times  \mathbb Z_{2}/ ~\langle ~(0, 1), (2, 0) ~\rangle$~\\
~ & ~224~ & ~$ \mathbb Z_{2}$~ & ~$ \mathbb Z \times  \mathbb Z_{2}/ ~\langle ~(1, 0) ~\rangle$~\\
~ & ~225~ & ~--~ & ~$ \mathbb Z \times \mathbb Z/ ~\langle ~(2, 1), (0, 2) ~\rangle$~\\
~ & ~226~ & ~--~ & ~$ \mathbb Z \times  \mathbb Z_{2}/ ~\langle ~(2, 1) ~\rangle$~\\
~ & ~227~ & ~--~ & ~$ \mathbb Z \times  \mathbb Z_{2}/ ~\langle ~(1, 0) ~\rangle$~\\
~ & ~228~ & ~--~ & ~$ \mathbb Z_{2} \times \mathbb Z_{2}/ ~\langle ~(1, 0) ~\rangle$~\\
~ & ~229~ & ~$ \mathbb Z_{2}$~ & ~$ \mathbb Z \times \mathbb Z/ ~\langle ~(2, 1), (0, 2) ~\rangle$~\\
~ & ~230~ & ~--~ & ~$ \mathbb Z_{2} \times \mathbb Z_{2}/ ~\langle ~(1, 0) ~\rangle$~\\
\hline
\hline
\end{tabular}
\end{table*}
\end{center}
\clearpage
\twocolumngrid

\bibliography{refs}

\end{document}